\documentclass[]{aa}
\usepackage{graphicx}
\usepackage{color}
\usepackage{colortbl}
\usepackage{soul}
\usepackage{natbib}
\bibpunct{(}{)}{;}{a}{}{,}
\graphicspath{{figures/}}
\usepackage{subfigure}
\usepackage{algorithm,algorithmicx}

\begin{document}
%
%
%%%%%%%%%%%%%%%%%%%%%%%%%%%%%%%%%%%%%%%%%%%%%%%%%%%%%%%%%%%%%%
%
% Title
%
%%%%%%%%%%%%%%%%%%%%%%%%%%%%%%%%%%%%%%%%%%%%%%%%%%%%%%%%%%%%%%
\title{Modelling the solar transition region using an adaptive conduction method}
  \author{C. D. Johnston\inst{1}
  \and P. J. Cargill\inst{1, 2}
  \and A. W. Hood\inst{1} 
  \and I. De Moortel\inst{1,3}
  \and S. J. Bradshaw\inst{4} 
  \and A. C. Vaseekar\inst{1}}
  \institute{School of Mathematics and Statistics, University 
  of St Andrews, St Andrews, Fife, KY16 9SS, UK.
  \and
  Space and Atmospheric Physics, The Blackett Laboratory,
  Imperial College, London, SW7 2BW, UK.
  \and
  Rosseland Centre for Solar Physics, University of Oslo, PO 
  Box 1029  Blindern, NO-0315 Oslo, Norway.
  \and
  Department of Physics and Astronomy, Rice University,
  Houston, TX 77005, USA.
  \\
  \email{cdj3@st-andrews.ac.uk}
  }
%
%
%%%%%%%%%%%%%%%%%%%%%%%%%%%%%%%%%%%%%%%%%%%%%%%%%%%%%%%%%%%%%%
%
% Abstract
%
%%%%%%%%%%%%%%%%%%%%%%%%%%%%%%%%%%%%%%%%%%%%%%%%%%%%%%%%%%%%%%
\abstract
  {
  Modelling the solar \textbf{T}ransition 
  \textbf{R}egion with the use of
  an \textbf{A}daptive \textbf{C}onduction (TRAC) method
  permits fast and accurate 
  numerical solutions of the field-aligned hydrodynamic 
  equations, capturing the
  enthalpy 
  exchange between the corona and transition region, 
  when the corona undergoes impulsive heating. 
  The TRAC method eliminates 
  the need for highly resolved numerical grids in the 
  transition region  and 
  the commensurate very short time steps that are required for 
  numerical stability. When employed with coarse spatial 
  resolutions, typically achieved in multi-dimensional 
  magnetohydrodynamic
  codes, the errors at peak density
  are less than 5\% and the 
  computation time is three orders of magnitude faster than 
  fully resolved field-aligned models. This paper presents 
  further examples that demonstrate the versatility and 
  robustness of the method over a range of heating events, 
  including impulsive and 
  quasi-steady footpoint heating. A detailed 
  analytical assessment of the TRAC method is also presented, 
  showing that the approach works through all phases of an 
  impulsive heating event because (i)~the 
  total radiative 
  losses and (ii) the total heating
  when integrated over the transition region
  are both preserved at 
  all temperatures under the
  broadening modifications of the method. The 
  results from the numerical simulations complement this 
  conclusion.
  }
  \titlerunning{Modelling the transition region using an adaptive conduction method}
  \maketitle
  %
  %
%%%%%%%%%%%%%%%%%%%%%%%%%%%%%%%%%%%%%%%%%%%%%%%%%%%%%%%%%%%%%%
%
% Introduction
%
%%%%%%%%%%%%%%%%%%%%%%%%%%%%%%%%%%%%%%%%%%%%%%%%%%%%%%%%%%%%%%
\section{Introduction}
  \label{Sect:Intro}
  \indent
  The computational modelling of the plasma response to either 
  impulsive or quasi-steady coronal heating
  has a long history 
  and it 
  has become an essential tool in understanding flares of 
  all sizes, as well as active region and quiet Sun heating
  \citep[see e.g.][for a review]{paper:Reale2014}. 
  A 
  widely used approach to investigate the
  energy release involves studying the 
  response of the plasma along a flux element (or field-line or 
  loop) to an imposed heating function. Such one-dimensional 
  (1D)
  hydrodynamic models have the benefit of being relatively 
  simple to implement and run and can readily generate 
  observables such as 
  emission line intensities and profiles. 
  However, 
  they suffer from major computational limitations brought 
  about by the very narrow transition region (TR), between 
  chromosphere and corona. 
  \\
  \indent
  \citet[][hereafter BC13]{paper:Bradshaw&Cargill2013}
  demonstrated that inadequate 
  TR spatial resolution leads to a potentially major 
  underestimate of the coronal density in numerical
  simulations. 
  The very steep temperature gradients in the TR are associated 
  with thermal conduction between
  the corona and chromosphere. 
  Resolving these gradients
  requires a highly resolved grid 
  which, in turn, acts as a major 
  constraint on the time step ($\Delta t$). 
  Stability in an explicit 
  numerical scheme that models thermal conduction requires,
  \begin{align}
    \Delta t < \textrm{min}(
    k_B n (\Delta s)^2 
    /
    \kappa_\parallel(T)
    ),
    \label{Eqn:conduction_CFL} 
  \end{align}
  where 
  $\kappa_\parallel(T) = \kappa_0 T^{5/2}$ 
  is the field-aligned Spitzer-H{\"a}rm~(SH)
  coefficient of thermal conduction
  \citep{paper:Spitzer1953}
  with $\kappa_0=10^{-11}$, 
  $T$ the 
  plasma temperature,
  $n$ the number
  density, $\Delta s$ the numerical cell width and the 
  time step is limited by the minimum of the bracketed quantity 
  across the entire grid.
  \\
  \indent
  For a uniform grid, this condition is 
  always set by the coronal properties, and so is grossly 
  inefficient. Thus, a highly non-uniform mesh that puts grid 
  points preferentially in the TR is of considerable benefit in 
  reducing the time a given simulation takes to run.
  However, an 
  additional problem is that the TR moves in response to 
  coronal heating and cooling, so ideally the 
  most highly resolved region 
  should be able to move upward or downward
  as required. 
  Adaptive grid codes have been developed that can address this 
  problem satisfactorily
  \citep[][BC13]{paper:Bettaetal1997,
  paper:Antiochosetal1999,
  paper:Bradshaw&Mason2003}
  so that for a single loop, a 
  brute force approach can be implemented, given adequate 
  computational facilities.  
  \\
  \indent
  Some authors also avoid the conduction stability 
  condition,  
  defined in Eq. \eqref{Eqn:conduction_CFL},
  by using either implicit methods
  \citep[e.g.][]{paper:Hansteen1993} or
  operator splitting methods 
  \citep[e.g.][]{paper:Bothetal2011,
  paper:Gudiksenetal2011}. 
  Such operator splitting methods enable the advection terms to 
  be integrated in time using an explicit numerical scheme 
  while treating thermal conduction separately with an implicit 
  method. However, with both approaches, spatial 
  convergence to the correct solution in the TR may still 
  require a very fine grid, as discussed further in Section 4.2 
  of BC13.
  \\
  \indent
  Therefore, 
  for studies of multiple loops forming an active 
  region
  \citep{paper:Bradshaw&Viall2016,paper:Barnesetal2019}
  and for long simulations 
  \citep{paper:Fromentetal2018,
  paper:Winebargeretal2018}, 
  it is desirable 
  to develop methods that mitigate
  the need for highly resolved numerical grids.
  Further, there is also 
  a need for a 1D 
  code that can be run quickly in order to 
  assess the viability of physical ideas. The same 
  considerations obviously also apply to the difficulties of 
  modelling thermal conduction in multi-dimensional
  magnetohydrodynamic (MHD) models, where machine 
  limitations dictate the number of grid points run in a 
  simulation
  \citep[e.g.][]{paper:Hoodetal2016,paper:Realeetal2016,
  paper:Warneckeetal2017,
  paper:Reidetal2018,
  paper:Sykoraetal2018,
  paper:Howsonetal2019,
  paper:Knizhniketal2019} 
  \\
  \indent
  In recent years, two approaches have been 
  proposed. 
  One, by 
  \cite{paper:Johnstonetal2017a,paper:Johnstonetal2017b,
  paper:Johnstonetal2019},
  was tested 
  in a field-aligned hydrodynamic code, and modelled the TR 
  as an unresolved discontinuity using a physically motivated 
  jump condition across it. Comparison with fully resolved 1D 
  simulations of impulsive heating and the development of 
  thermal non-equilibrium (TNE) showed good agreement 
  \citep{paper:Johnstonetal2019}.
  \\
  \indent
  A second approach is due, in its original form, to 
  \cite{paper:Linkeretal2001}
  and, subsequently
  \citet[][hereafter L09]{paper:Lionelloetal2009}
  and 
  \cite{paper:Mikicetal2013}.  
  In order to decrease the mesh resolution
  requirements (and so increase the 
  conduction time step), 
  they chose to broaden the TR by setting 
  the parallel thermal conductivity 
  ($\kappa_\parallel(T)$) to be 
  constant below a fixed temperature, defined herein as $T_c$, 
  with  $T_c = 2.5 \times 10^5$~K a typical value
  used.  
  At the same time, 
  they modified the optically thin radiative loss function 
  ($\Lambda(T)$) 
  below $T_c$ such that $\kappa_\parallel(T) \Lambda(T)$ 
  gave the same function of temperature as for $T\geq T_c$.
  The method is discussed in 
  more detail in Section \ref{Sect:TRAC_analytical_assessment}, 
  but for a static loop, L09 showed 
  that this approach 
  gave almost identical coronal conditions to 
  those obtained using  
  the classical SH heat flux formulation at all 
  temperatures. In addition, a range of tests on, 
  in particular,
  TNE 
  \citep{paper:Mikicetal2013},
  showed that the method worked 
  for dynamically evolving loops as well. 
  \\
  \indent
  However, further investigation has revealed that for a more 
  general class of problems, in particular involving
  strong impulsive 
  heating, the L09 method with a fixed $T_c$
  and coarse spatial resolution has 
  shortcomings similar to those identified by BC13 when the 
  entire TR was under-resolved.
  In a recent paper
  \citep[][hereafter JB19]{paper:Johnston&Bradshaw2019},  
  we 
  proposed an important modification to their approach,
  to
  model the 
  \textbf{T}ransition \textbf{R}egion 
  using an \textbf{A}daptive \textbf{C}onduction (TRAC)
  method.
  In TRAC,
  $T_c$ was allowed to vary throughout the simulation 
  in a way that adapted to the resolution requirements at any 
  time. 
  This permitted the modelling of very dynamic phenomenon 
  such as strong flares, and we found (i) excellent agreement 
  between our approach and a fully resolved 1D model
  and (ii)
  extremely significant 
  savings in computation time.
  \\
  \indent
  JB19 presented just a sample of results.
  In this paper, we examine 
  how well the TRAC method 
  works for a wider range of problems, why it works
  through all phases of an impulsive heating event, 
  and what other physics can be included in the approach.
  Section \ref{Sect:TRAC}
  outlines the TRAC method, 
  Section \ref{Sect:Model_and_experiments} describes
  the numerical model
  and
  an extensive analysis 
  of our test problems is presented in
  Section \ref{Sect:Results}. 
  We conclude with a discussion of the TRAC method in 
  Section \ref{Sect:Discussion}
  and a series of Appendices contain supplementary 
  material.
  %
  %
%%%%%%%%%%%%%%%%%%%%%%%%%%%%%%%%%%%%%%%%%%%%%%%%%%%%%%%%%%%%%%
%
% The TRAC Method
%
%%%%%%%%%%%%%%%%%%%%%%%%%%%%%%%%%%%%%%%%%%%%%%%%%%%%%%%%%%%%%%
\section{The TRAC method
  \label{Sect:TRAC}}
  \indent
  In the JB19 Letter, 
  we introduced the ideas behind the TRAC method but 
  space did not permit a full description. This is presented
  in the following subsections. 
  \\
  \indent
  We model the plasma response 
  to heating by considering the single fluid, 
  field-aligned hydrodynamic equations
  for a coronal loop, with uniform 
  cross-section,
  \begin{align}
    & 
    \frac{\partial\rho}{\partial t} + v\frac{\partial\rho}
    {\partial s} = - \rho\frac{\partial v}{\partial s}; 
    \label{eqn:1d_continuity}
    \\[1.5mm]
    & 
    \rho \frac{\partial v}{\partial t} + \rho v
    \frac{\partial 
    v}
    {\partial s}
    = -
    \frac{\partial P}
    {\partial s} - \rho g_{\parallel} 
    + 
    \frac{\partial}
    {\partial s}
    \!   
    \left(    
    \rho \nu \frac{\partial v}{\partial s}
    \right)
    \!   
    ;
    \label{eqn:1d_motion}
    \\[1.5mm]
    & 
    \rho \frac{\partial\epsilon}{\partial t} + \rho v
    \frac{\partial
    \epsilon}
    {\partial s}
    = -P\frac{\partial 
    v}
    {\partial s} - 
    	\frac{\partial F_c}{\partial s} + Q
    - \! n^2 \Lambda(T)
    \!
    + 
    \!
    \rho \nu 
    \!   
    \left(
    \frac{\partial v}{\partial s}
    \right)^{\! \! \!2} \! ; \!  \! 
    \label{eqn:1d_tee}
    \\[1.5mm]
    & 
    P = 2 \, k_B n T.
    \label{eqn:gas_law}
  \end{align} 
  Here, $s$ is the spatial coordinate along the magnetic 
  field, 
  $\rho$ is the mass density, $P$ is the gas pressure, $T$ is 
  the temperature,  $k_B$ is the Boltzmann 
  constant, 
  $\epsilon=P/(\gamma-1) \rho $ is the specific 
  internal energy density,
  $n$ is the number density ($n=\rho/1.2m_p$, $m_p$ is the
  proton mass),
  $v$ is the velocity parallel to the 
  magnetic field, $g_{\parallel}$ is the field-aligned 
  gravitational acceleration (for which we use a profile 
  that corresponds to a semi-circular 
  strand),
  $\nu$ is the dynamic viscosity,
  $F_c=-\kappa_\parallel(T)\partial T/\partial s$
  is the SH heat flux
  with $\kappa_\parallel(T)$ defined following Eq.
  \eqref{Eqn:conduction_CFL},
  $Q$ is a heating function,
  $\Lambda(T)$ is the 
  radiative loss function in an optically thin plasma, 
  which we approximate using the 
  piecewise continuous function defined in
  \cite{paper:Klimchuketal2008},
  and 
  $\rho \nu (\partial v / \partial s)^2$
  is a viscous heating term which is added to the heating
  function $Q$.
  \\
  \indent
  Throughout this paper, we assume equilibrium ionization 
  and use the SH conductivity.   
  We note that this form of thermal conduction 
  assumes that the mean free path of the electrons 
  remains small compared to the characteristic 
  scale lengths
  \citep[see e.g.][]{paper:Lie-Svendsenetal1999}.
  However,   
  a radiative loss function that depends on the history of 
  the plasma 
  \citep[e.g.][]{paper:Hansteen1993}
  and alternative conductivity models
  also can potentially be
  implemented with the TRAC method.
  %
  %
%%%%%%%%%%%%%%%%%%%%%%%%%%%%%%%%%%%%%%%%%%%%%%%%%%%%%%%%%%%%%%
%
% Identification
%
%%%%%%%%%%%%%%%%%%%%%%%%%%%%%%%%%%%%%%%%%%%%%%%%%%%%%%%%%%%%%%
\subsection{Identification of an adaptive cutoff temperature
  \label{Sect:TRAC_identification}}
  \indent
  For impulsive heating, the evolution of a loop can be divided 
  into three phases
  \citep[e.g.][]{paper:Cargill1994,paper:Klimchuk2006,
  paper:Cargilletal2015}.  
  As 
  the loop is heated, a strong enhancement of thermal 
  conduction from corona to chromosphere arises. This leads to 
  the TR moving downwards, and the excessive conductive flux 
  leads to an enthalpy flux upwards into the corona, 
  increasing 
  the coronal density
  \citep{paper:Antiochos&Sturrock1978,
  paper:Klimchuketal2008}. This phase has the most severe 
  requirements on TR numerical resolution. 
  Following the termination of 
  the heating, the coronal temperature declines, but the 
  density continues to increase until a balance between 
  downward conduction and 
  TR radiation is reached with 
  only small mass motions. This is the time of maximum coronal 
  density. 
  After this, as the corona cools further, its density 
  decreases, leading to a downward enthalpy flux whose 
  magnitude is determined by the TR radiation requirements
  \citep{paper:Bradshaw&Cargill2010a,
  paper:Bradshaw&Cargill2010b}. 
  These three phases are demonstrated very simply in the 
  approximate methods developed by
  \cite{paper:Klimchuketal2008} and
  \cite{paper:Cargilletal2012a,paper:Cargilletal2012b}.
  \\
  \indent	
  The essence of the TRAC method is to ensure that the 
  TR is resolved 
  at all times 
  in the most effective way 
  while ensuring
  that these three 
  phases,
  as well as intervening times,
  are modelled correctly. The first part
  of achieving this is 
  to identify the locations in the simulation where the 
  temperature profile is unresolved, and the second part 
  (described in subsequent subsections) is to 
  modify $\kappa_
  \parallel (T)$, $\Lambda(T)$ and 
  $Q(T)$
  in such a way that the temperature profile 
  becomes resolved at these locations.
  \\
  \indent
  We define the adaptive cutoff temperature 
  ($T_c$) such that $T_c$ is the 
  temperature associated with 
  the upper location on the grid of
  any unresolved grid cells 
  (i.e. the temperature is under-resolved 
  below $T_c$). [Here we 
  assume a symmetric loop but the method can 
  be adapted readily to consider asymmetry between the 
  footpoints by performing the following calculation at each 
  footpoint.]
  This is done using an algorithm based on the 
  method employed by 
  \cite{paper:Johnstonetal2017a,paper:Johnstonetal2017b}
  for locating the top of an unresolved transition region, 
  and is restated here for completeness.
  \\
  \indent
  The temperature length scale is defined as,
  \begin{align}
    L_T(s) = \dfrac{T}{|dT/ds|},
    \label{Eqn:L_T} 
  \end{align}
  and the local resolution in the simulation is given by,
  \begin{align}
    L_R(s) = \Delta s,
    \label{Eqn:L_R}
  \end{align}
  where $s$ is the spatial coordinate along the magnetic field
  and $\Delta s$ is the local grid cell width.  
  We note that these definitions can be used on 
  either a uniform
  or non-uniform grid.
  \\
  \indent
  Using these definitions, 
  the cutoff temperature is defined as 
  the maximum temperature that violates the resolution 
  criteria of
  \cite{paper:Johnstonetal2017a,paper:Johnstonetal2017b},
  \begin{align}
    T_c=\textrm{max}(T(s)) \,\,\textrm{such that}\,\,
    \dfrac{L_R(s)}{L_T(s)} > \delta=\dfrac{1}{2},
    \label{Eqn:T_c} 
  \end{align}
  which corresponds to having an
  insufficient number of grid cells across 
  the temperature length scale (i.e. unresolved 
  temperature gradients). The choice of 
  $\delta=1/2$ is the minimum resolution 
  criteria. However, we note that choosing $n$ grid points 
  across $L_T$ ($\delta=1/n$) will result in higher cutoff 
  temperatures for increasing values of $n>2$.
  \\
  \indent
  An upper bound for the cutoff temperature is set as 
  20\% of the peak 
  coronal temperature in the loop at that time, but for the
  simulations we have performed, 
  the results are only weakly dependent on this upper bound. A
  lower bound is set as the 
  temperature value of the isothermal
  chromosphere, taken as 
  $T_{\textrm{chrom}}=2 \times 10^4$~K. 
  Therefore, we  dynamically 
  adjust $T_c$ with the criteria that it should
  also satisfy,
  \begin{align}
    T_{\textrm{chrom}} \leq T_c \leq  0.2 \,\textrm{max}(T(s)).
    \label{Eqn:T_c_criteria} 
  \end{align}	
  \indent
  One important
  aspect of the model that could not be discussed in JB19 for 
  reasons of space is that $T_c$ can undergo sudden jumps when 
  the entire TRAC region becomes temporarily
  resolved. In that case, $T_c$ 
  briefly defaults to the minimum value, 
  introducing significant changes 
  in the radiative properties of the TRAC region 
  for a limited number of time steps.
  While this does not have any effect on the coronal 
  quantities, it is aesthetically unpleasing, and 
  is easily removed.
  This is discussed in Appendix
  \ref{App:A} 
  and in this paper we use the cutoff
  temperature limiter that is described there.
  %
  %
%%%%%%%%%%%%%%%%%%%%%%%%%%%%%%%%%%%%%%%%%%%%%%%%%%%%%%%%%%%%%%
%
% Analytical assessment of the TRAC region modifications
%
%%%%%%%%%%%%%%%%%%%%%%%%%%%%%%%%%%%%%%%%%%%%%%%%%%%%%%%%%%%%%%
\subsection{Analytical assessment of the TRAC region   
  modifications
  \label{Sect:TRAC_analytical_assessment}}
  \indent
  We now turn to an
  analytical assessment of the TRAC method
  in order to (i) 
  justify the broadening modifications that are
  used and 
  (ii) demonstrate
  how the approach works through all phases of an impulsive 
  heating event.
  \\
  \indent
  As noted in the introduction,
  \cite{paper:Linkeretal2001} introduced an 
  artificial TR broadening by modifying $\kappa_\parallel(T)$ 
  such that it was constant for temperatures below
  0.25~MK. 
  For static loops, they found that while the 
  coronal
  temperature
  of the loop was roughly the same when
  compared with a full SH solution, the density 
  was larger by up to a factor of two. 
  This difference in the coronal 
  density indicates that the approximation underestimates the 
  TR radiation when compared to SH conduction. 
  \\
  \indent
  Subsequently, L09 demonstrated that 
  modifying $\kappa_\parallel(T)$ such that 
  $\kappa_\parallel(T) \Lambda(T)$ 
  was the same function of temperature in both the SH 
  and modified models gave excellent agreement with the
  coronal 
  temperature and density. We now show that this result 
  has generality through all phases of the evolution of a 
  heated loop.
  \\
  \indent
  It is required that the radiative losses integrated across 
  the TRAC region be independent of the specific form of 
  $\kappa_\parallel(T)$, $\Lambda(T)$ and $Q(T)$. These 
  integrated losses are defined as,
  \begin{align}
    \mathcal{R}_{\textsc{trac}} = 
    \int_{s_b}^{s_c}
    n^2
    \Lambda(T)
    \, ds
    =
    \int_{T_b}^{T_c}
    \left(
    \dfrac{P}{2k_BT}
    \!
    \right)^{\!\!2}
    \!\!
    \Lambda(T)
    \dfrac{L_T}{T} 
    \, dT,
    \label{Eqn:R_TRAC}
  \end{align}
  where $s_b$ ($s_c$) is the
  spatial coordinate at the
  base (top) of the TRAC region and
  $L_T$ is as defined in  Eq.~\eqref{Eqn:L_T}.
  \\
  \indent
  To demonstrate this in a simple way, we 
  start by writing the 
  energy
  equation \eqref{eqn:1d_tee}
  in the TRAC region in the following conservative form,
  \begin{align}
    \frac{\partial E }{\partial t} 
    =-
    \frac{\partial}
    {\partial s} (Ev + Pv + F_c) + Q(s,t) -\! n^2 \Lambda(T)
    \! + \! \rho g_\parallel v, \!
    \label{Eqn:1d_tec}
  \end{align}
  where
  $E = P/(\gamma - 1) + \rho v^2/2$.
  We note that Eq. \eqref{Eqn:1d_tec} applies
  to the fluid in single species codes such as the
  Lagrangean remap code
  \citep{paper:Arber2001, 
  paper:Johnstonetal2017a,paper:Johnstonetal2017b}
  and
  electrons in multi-species codes such as HYDRAD
  \citep[][BC13]{paper:Bradshaw&Mason2003,
  paper:Bradshaw&Cargill2006}.
  \\
  \indent
  Next, following \cite{paper:Klimchuketal2008},
  we assume that the TRAC region is quasi-steady, 
  gravity is neglected and any flows are subsonic.
  These assumptions are made only to
  derive the analytical approximations that 
  follow. 
  They are not imposed on the TRAC region  
  in numerical simulations,
  as is discussed
  in Section \ref{Sect:TRAC_broadening}.
  Then we can solve for $L_T$ by writing
  Eq.~\eqref{Eqn:1d_tec} in the form,
  \begin{align}
    \dfrac{\kappa_\parallel(T)}{T}
    \!
    \!
    \left(
    \dfrac{T}{L_T}
    \right)^{\!\!2}
    \!
    \!
    -
    \! 
    5k_B J
    \left(
    \dfrac{T}{L_T}
    \right)
    \!
    \!
    - 
    \!
    \left[
    \!
    \left(
    \!
    \dfrac{P}{2k_BT}
    \!
    \right)^{\!\!2}
    \!\!
    \Lambda(T)  
    \! 
    - 
    \!
    Q
    \right] 
    \! 
    \!
    =  0,
    \label{Eqn:T_L_T_quadratic}
  \end{align}
  where $J = nv$ is the mass flux and here we 
  retain the 
  TR heating term. 
  $J$ is positive (negative) for an upflow 
  (downflow). 
  This can be solved for $T/L_T$ as,
  \begin{align}
    \dfrac{T}{L_T} = 
    \dfrac{ 
    \!
    5k_B J 
    \! 
    \pm
    \!
    \sqrt{25k_B^2J^2
    \! 
    +
    \!
    4\dfrac{\kappa_\parallel(T)}{T}
    \!
    \!
    \left[
    \!
    \left(
    \!
    \dfrac{P}{2k_BT}
    \!
    \right)^{\!\!2}
    \!\!
    \Lambda(T)
    \! 
    - 
    \!
    Q
    \right] 
    \!
    \!
    }
    }
    {2\dfrac{\kappa_\parallel(T)}{T}},
    \label{Eqn:T_L_T_quadratic_formula}
  \end{align}
  where the positive (negative) root corresponds to
  the increasing (decreasing) temperature gradient 
  found at the
  left-hand (right-hand) leg of the loop.
  Substituting this into 
  Eq. \eqref{Eqn:R_TRAC} 
  shows that only the combinations $
  \kappa_\parallel(T) \Lambda(T)$ 
  and $\kappa_\parallel(T) Q(T)$ occur 
  in the expression for
  the radiative losses integrated across
  the TRAC region ($\mathcal{R}_{\textsc{trac}}$). 
  Hence, so long as these 
  combinations are properly adjusted, the same
  integrated radiative losses, 
  $\mathcal{R}_{\textsc{trac}}$, arise for 
  both the 
  TRAC and SH conduction models. 
  \\
  \indent
  Adopting the approach of 
  \cite{paper:Klimchuketal2008}
  and neglecting TR 
  heating,
  we can further simplify Eq. 
  \eqref{Eqn:T_L_T_quadratic_formula}
  into three limits: 
  (i) strong evaporation (neglect radiation);
  (ii) peak 
  density (neglect dynamics); and (iii) radiative
  cooling (neglect thermal 
  conduction). 
  For these three phases we find:
  \begin{align}
    \textrm{(i)} \quad
    &
    \dfrac{T}{L_T} = \dfrac{5k_BJT}{\kappa_\parallel(T)};
    \\[2mm] 
    & 
    \mathcal{R}_{\textsc{trac}} = 
    \int_{T_b}^{T_c}
    \left(
    \dfrac{P}{2k_BT}
    \!
    \right)^{\!\!2}
    \dfrac{\kappa_\parallel(T)\Lambda(T)}{5k_BJT} 
    \, dT; 
    \label{Eqn:R_TRAC_strong_evaporation}
    \\[4mm] 
    \textrm{(ii)} \quad
    &
    \dfrac{T}{L_T} = 
    \dfrac{P}{2k_B}
    \sqrt{
    \dfrac{\Lambda(T)}{T\kappa_\parallel(T)}
    };
    \\[2mm] 
    & 
    \mathcal{R}_{\textsc{trac}} = 
    \int_{T_b}^{T_c}
    \dfrac{P}{2k_B}
    \sqrt{
    \dfrac{\kappa_\parallel(T)\Lambda(T)}{T^3}
    }
    \, dT; 
  \end{align}
  and,
   \begin{align}
    \textrm{(iii)} \quad
    &
    \dfrac{T}{L_T} = 
    \left(
    \dfrac{P}{2k_BT}
    \!
    \right)^{\!\!2}
    \dfrac{\Lambda(T)}{5k_B|J|};
    \\[2mm] 
    & 
    \mathcal{R}_{\textsc{trac}} = 
    \int_{T_b}^{T_c}
    5k_B|J|
    \, dT,
  \end{align}
  where $J$ is negative (a downflow) in the third regime. Since 
  the product $\kappa_\parallel(T) \Lambda(T)$ 
  is assumed to be the 
  same function of temperature for both models, then the 
  integrated 
  radiative losses at all temperatures in the TRAC region are 
  the same for both the SH and TRAC methods.
  \\
  \indent 
  Furthermore, a
  similar analysis also holds for the heating 
  integrated across the TRAC region,
  \begin{align}
    Q_{\textsc{trac}} = 
    \int_{s_b}^{s_c}
    Q(T)
    \, ds
    =
    \int_{T_b}^{T_c}
    Q(T)
    \dfrac{L_T}{T} 
    \, dT.
    \label{Eqn:Q_TRAC}
  \end{align}
  In particular,
  as long as 
  the combination $\kappa_\parallel(T)Q(T)$ is properly
  adjusted, then
  $Q_{\textsc{trac}}$
  is independent of the specific form of
  $\kappa_\parallel(T)$ and $Q(T)$,
  so that the integrated heating at all temperatures in
  the TRAC region is also the same for both the SH and TRAC 
  methods.
  %
  %
%%%%%%%%%%%%%%%%%%%%%%%%%%%%%%%%%%%%%%%%%%%%%%%%%%%%%%%%%%%%%%
%
% Broadening the TRAC region
%
%%%%%%%%%%%%%%%%%%%%%%%%%%%%%%%%%%%%%%%%%%%%%%%%%%%%%%%%%%%%%%
\subsection{Broadening the TRAC region
  \label{Sect:TRAC_broadening}}
  \indent
  Having identified the 
  adaptive cutoff temperature,
  the second part of the TRAC method 
  is to broaden the steep temperature and   
  density gradients in the TRAC region.
  This is achieved using
  an extension of the approach developed by 
  \cite{paper:Linkeretal2001},
  L09 and 
  \cite{paper:Mikicetal2013}
  as
  discussed in detail in the preceding section. 
  \\
  \indent
  Using 
  the results presented in Section 
  \ref{Sect:TRAC_analytical_assessment}, 
  below the cutoff temperature ($T_c$):
  \\
  \\
  (i)~the parallel thermal conductivity
  ($\kappa_\parallel(T)$)
  is set to a constant value $\kappa_\parallel^{\prime}(T)$, 
  so that,
  \begin{align}
    \kappa_\parallel(T) &= \kappa_0 T^{5/2},
    \quad \,\,\textrm{for all}\,\, T \geq T_c;
    \\
    \kappa_\parallel^{\prime}(T) &= \kappa_\parallel(T_c),
    \quad \,\,\,\textrm{for all}\,\, T < T_c,
  \end{align}
  \\
  (ii)~the radiative loss rate 
  ($\Lambda(T)$)
  is modified to $\Lambda^{\prime}(T)$ to preserve
  $\kappa_\parallel(T)\Lambda(T)=
  \kappa_\parallel^{\prime}(T)\Lambda^{\prime}(T)$,
  \begin{align}
    \Lambda(T) &= \Lambda(T),
    \quad \,\,\textrm{for all}\,\, T \geq T_c;
    \\
    \Lambda^{\prime}(T) &= \Lambda(T) 
    \left(\dfrac{T}{T_c}
    \right)^{5/2},
    \quad \,\,\textrm{for all}\,\, T < T_c,
  \end{align}
  \\
  and,
  \\
  \\
  (iii) the heating rate ($Q(T)$)  
  is modified to $Q^{\prime}(T)$ to preserve
  $\kappa_\parallel(T) Q(T)=
  \kappa_\parallel^{\prime}(T)Q^{\prime}(T)$,
  \begin{align}
    Q(T) &= Q(T),
    \quad \,\,\textrm{for all}\,\, T \geq T_c;
    \\
    Q^{\prime}(T) &= Q(T) 
    \left(\dfrac{T}{T_c}
    \right)^{5/2},
    \quad \,\,\textrm{for all}\,\, T < T_c.
  \end{align}
  When TRAC is employed in the numerical simulations in 
  Sections \ref{Sect:Model_and_experiments} and
  \ref{Sect:Results}, 
  we solve the full set of equations 
  \eqref{eqn:1d_continuity}--\eqref{eqn:gas_law}
  in the TRAC region 
  \citep[e.g.][]{paper:Johnstonetal2017b} and corona, 
  but with the use of the modified
  $\kappa_
  \parallel^{\prime}(T)$, $\Lambda^{\prime}(T)$ and 
  $Q^{\prime}(T)$ in the TRAC region. 
  Thus the assumptions, 
  such as subsonic flows
  made in the analytic solutions, are not 
  present, enabling us to assess their validity by a comparison 
  between the analytic and numerical solutions 
  (see Appendix \ref{App:B}).
  \\
  \indent
  As we show further on in this paper, 
  increasing the parallel thermal conductivity
  while decreasing the radiative loss and heating
  rates, at temperatures below $T_c$, has the
  desired effect of
  broadening the temperature
  length scales in the TRAC region.
  This helps TRAC prevent the
  heat flux jumping across any unresolved regions while 
  maintaining accuracy in the
  properly resolved parts of the atmosphere. 
  Further, we note that the formulation of TRAC: 
  (i) makes no assumptions about the spatial
  resolution in a simulation;
  (ii) reduces to the classical SH conduction model 
  when the TR is properly resolved (e.g. see JB19);
  and (iii) may still be implemented
  in both explicit and implicit numerical schemes
  that model thermal conduction.	
  %
  %
%%%%%%%%%%%%%%%%%%%%%%%%%%%%%%%%%%%%%%%%%%%%%%%%%%%%%%%%%%%%%%
%
% Numerical model and experiments
%
%%%%%%%%%%%%%%%%%%%%%%%%%%%%%%%%%%%%%%%%%%%%%%%%%%%%%%%%%%%%%%
\section{Numerical model and experiments
  \label{Sect:Model_and_experiments}}
  \indent
  In JB19, we demonstrated the viability of the TRAC method 
  with two examples, 
  namely a 600~s (long) and 60~s (short)
  heating pulse in a loop of 
  total length 100~Mm, including a 10~Mm 
  chromosphere at each end. The heating pulse was triangular, 
  uniformly distributed along the loop, with a peak value of 
  $2 \times 10^{-2}$~Jm$^{-3}$s$^{-1}$. 
  The 600 (60)~s pulse thus 
  had $4.8 \times 10^{8(7)}$~Jm$^{-2}$, 
  which for an aspect 
  ratio of 10, gives a total energy release of $10^{23(22)}$~J.
  \\
  \indent
  As in JB19, the TRAC results
  for these two examples are compared with the 
  SH results that are obtained using the adaptive 
  mesh refinement code HYDRAD 
  \citep[][BC13]{paper:Bradshaw&Mason2003,
  paper:Bradshaw&Cargill2006}. 
  HYDRAD has been extensively 
  described elsewhere (e.g. BC13) so we only restate the 
  details relevant to the results presented here. 
  \\
  \indent
  We run the HYDRAD code in 
  single fluid mode to solve
  equations
  \eqref{eqn:1d_continuity}--\eqref{eqn:gas_law}.
  The largest grid cell in all of
  our calculations has a width of $10^6$ m (1,000 km)
  and each successive refinement splits the cell into two.
  Thus, a refinement level of RL leads to cell widths decreased 
  by $1/2^{\textrm{RL}}$. 
  In this study, 
  the adaptive 
  mesh in HYDRAD
  is limited to 14 levels of refinement, defined as 
  RL = [0, 1, 2, \ldots , 13, 14]. 
  A mesh with RL = 14 has a 
  minimum grid cell size $2^{14}$ = 16384 times smaller than a 
  uniform grid with RL = 0,
  corresponding 
  to a grid cell width of 61 m
  in the most highly resolved parts of the TR.
  \\
  \indent
  BC13 demonstrated that the value of 
  RL needed for a \lq converged\rq\ 
  solution depended on the problem 
  being solved (Table 1 there), but here we work with RL = 14 
  as the benchmark for comparison.
  Hereafter we refer to the 
  (benchmark) HYDRAD solutions computed with RL=14 and the SH
  conduction method as the SH solutions. 
  When TRAC is implemented in HYDRAD, we use RL = 5 
  so that the 
  minimum grid size of 31.25~km 
  is a factor $2^9 = 512$ times larger
  than the corresponding SH solution. 
  These
  simulations are identical in all respects
  except for the value of RL and the conduction method used.
  However,
  this leads to TRAC run times for a typical problem being of 
  order $500 - 1000$ times faster. 
  We focus on the 
  TRAC solutions that are computed with RL=5 here
  because
  of the improvement in the accuracy of the 
  temperature evolution that is accessible with minimal 
  increase in computation time, when compared with the
  RL=3 simulations (see Section 
  \ref{Sect:Uniform_heating}).
  \\
  \indent
  The results of two uniform heating cases,
  namely the long and short heating pulses,
  are presented in Sections \ref{Sect:Long_pulse}
  \& \ref{Sect:Short_pulse},
  respectively.
  While these experiments 
  are representative of reasonably powerful 
  flares, it is also important to consider how the 
  TRAC method performs for a 
  wider range of uniform heating events
  and spatially non-uniform heating functions
  in order for future users to have confidence in the
  method. 
  The latter is addressed 
  in Section \ref{Sect:Footpoint_heating}
  through the consideration of both 
  impulsive and steady
  footpoint heating.   
  The former involves 
  a parameter study for uniform heating,
  which we present in Section \ref{Sect:Uniform_heating},
  that covers several orders
  of magnitude for the total energy released.
  %
  %
%%%%%%%%%%%%%%%%%%%%%%%%%%%%%%%%%%%%%%%%%%%%%%%%%%%%%%%%%%%%%%
%
% Results
%
%%%%%%%%%%%%%%%%%%%%%%%%%%%%%%%%%%%%%%%%%%%%%%%%%%%%%%%%%%%%%%
\section{Results
  \label{Sect:Results}}
  %
  %
%%%%%%%%%%%%%%%%%%%%%%%%%%%%%%%%%%%%%%%%%%%%%%%%%%%%%%%%%%%%%%
%
% Long pulse
%
%%%%%%%%%%%%%%%%%%%%%%%%%%%%%%%%%%%%%%%%%%%%%%%%%%%%%%%%%%%%%%
\subsection{Long pulse
  \label{Sect:Long_pulse}}
  \indent
  We first consider the details of the 600~s heating pulse
  simulations. 
  In particular, we present a 
  comprehensive description of certain aspects of the
  loop evolution
  in order to demonstrate and explain 
  why the TRAC simulations are so successful
  in describing the coronal response to heating while using
  such large grid cell widths.   
  Furthermore, a 
  detailed analysis of the global evolution of the 
  loop, 
  during the three key phases discussed in 
  Section \ref{Sect:TRAC_analytical_assessment}, 
  is also  presented in Appendix \ref{App:B}.
\subsubsection{Coronal response to heating}
  \indent
  Starting with the coronal response,
  the upper 
  two panels (row 1) 
  of Fig. \ref{Fig:600s_pulse_RL_5_time_evolution}
  show the coronal averaged 
  temperature and density
  as a function of time, where the averaging is
  calculated
  over the 50\% of the loop nearest the apex. 
  The three curves are 
  the SH solution with RL = 14 (solid red line), 
  the TRAC solution with RL = 5 
  (dashed blue line) 
  and the SH solution with RL = 5 (dashed red line).  
  In the time-dependent plots in Fig.
  \ref{Fig:600s_pulse_RL_5_time_evolution}
  all of the quantities are shown
  with 1~s temporal resolution.
  \\
  \indent
  As 
  expected, 
  the under-resolved SH solution shows major differences 
  in the density
  from the resolved one (BC13). 
  On the other hand,   
  the TRAC and resolved SH solutions 
  show excellent agreement and both show the familiar pattern 
  described in Section \ref{Sect:TRAC_identification}
  of a rapid temperature 
  increase, followed by a slower density increase, then a 
  cooling and draining
  \citep{paper:Bradshaw&Cargill2006,
  paper:Cargilletal2012a,paper:Cargilletal2012b,
  paper:Cargilletal2015,
  paper:Reale2016}. 
  We note that in 
  Fig. \ref{Fig:600s_pulse_RL_5_time_evolution},
  the TRAC and resolved SH temperatures
  lie on top of each other.
  \\
  \indent
  The second pair of panels (row 2) in 
  Fig. \ref{Fig:600s_pulse_RL_5_time_evolution}
  show the normalised difference  
  between the resolved SH and TRAC solutions
  (e.g. ($T_{\textrm{TRAC}}-T_{\textrm{SH}})/T_{\textrm{SH}}$). 
  The TRAC solution shows a slightly higher density 
  throughout the simulation, which will be discussed later,
  but the difference at any time is less than
  5\%. The rapid 
  increase in the percentage temperature difference towards the 
  end of the simulation reflects the slightly more rapid 
  decrease of $T_{\textrm{TRAC}}$ as the cooling tries to 
  return to its 
  initial equilibrium. 
  This is a common feature in the late radiative phase when 
  comparing accurate numerical solutions with approximate 
  methods
  \citep[e.g.][]{paper:Cargilletal2012a}. 
\subsubsection{Energy balance in the TRAC region}
  \indent
  We now turn to a consideration of how
  the TRAC modifications to the parallel thermal conductivity,
  radiative loss and heating rates affect the local energy
  balance and subsequent dynamics inside the TRAC region.
  Using the approach of 
  \cite{paper:Johnstonetal2017a,paper:Johnstonetal2017b},
  Eq. \eqref{Eqn:1d_tec} 
  can be rewritten to describe 
  the energy balance in the TRAC region as,
  \begin{align}
    \frac{\gamma}{\gamma - 1} P_c v_c
    + \frac{1}{2} \rho_c 
    v_c^3 
    + F_{c,c} =
    Q_{\textsc{trac}}
    - \mathcal{R}_{\textsc{trac}},
    \label{Eqn:1d_utr_jc}
  \end{align}
  where the subscripts 
  \lq c\rq\ and \lq TRAC\rq\
  indicate quantities 
  evaluated 
  at the 
  top of and integrated across the TRAC region,
  respectively.  
  This location is determined by the temperature 
  value of 
  $T_c$ and the temperature domain is the same for SH and TRAC 
  results, although may correspond to different spatial 
  locations (e.g. see Figs. 
  \ref{Fig:600s_pulse_RL_5_300s_snapshots} - 
  \ref{Fig:600s_pulse_RL_5_2000s_snapshots}).
  \\
  \indent 
  The fifth and sixth 
  panels (row 3)  in 
  Fig. \ref{Fig:600s_pulse_RL_5_time_evolution}
  show the dominant terms in the 
  integrated energy equation \eqref{Eqn:1d_utr_jc} 
  for the TRAC 
  solution (left) and SH (right).
  In these panels,
  the blue and orange 
  curves are the downward 
  heat flux, 
  ($F_{c,c}=-\kappa_\parallel(\partial T/\partial s)$
  evaluated at $T_c$)
  and downward or upward enthalpy flux 
  ($F_{e,c}=\gamma/(\gamma - 1) P_c v_c$)
  respectively,
  with dashed (solid) lines corresponding to a downflow 
  (upflow). 
  The red (green) 
  curve is the total radiative loss (heating) integrated across 
  the TRAC region. 
  \\
  \indent
  Both methods show the standard picture of an 
  initial phase where the downward heat flux is balanced by an 
  upward enthalpy flux (evaporation), a density maximum at 
  900~s when the heat flux is balanced by radiative
  losses and a decay phase when the radiation is driven by a 
  downward enthalpy flux. 
  These figures show that, even though there is
  good agreement between the coronal quantities, there are 
  significant fluctuations in, and at the top of, the TRAC 
  region, especially during the decay phase
  in the SH solution. These can be 
  attributed to the continual relocation of $T_c(s)$ as the top 
  of the TRAC region
  retreats back upwards.
\subsubsection{TRAC region cutoff temperature and thickness
  \label{Sect:Long_pulse_Tc_l}}
  \indent
  Next we focus on the dynamic evolution of the 
  cutoff temperature and the effect this has on the
  broadening of the TRAC region.
  The panels in row 4 of 
  Fig.~\ref{Fig:600s_pulse_RL_5_time_evolution}
  show $T_c$ and the
  thickness of the TRAC region ($\ell$) 
  as a function of time, with the 
  same colour coding as before. $T_c$ evolves in approximately 
  the same manner as the coronal
  temperature, beginning at 0.12~MK and 
  rising to 1.45~MK at peak heating. Thus, the ability to vary 
  $T_c$ is an important aspect of obtaining the correct coronal 
  properties: retaining this at a fixed value of 0.25~MK would 
  have led to a major under-resolution of the TR
  (e.g. see JB19).
  \\
  \indent
  The impact of 
  the TRAC approach can also be seen in the right-hand panel. 
  Here the TRAC region 
  thickness is increased by an order of magnitude 
  when the TRAC method is applied. The artificial broadening is 
  at a maximum at the time of peak heating, 
  which is associated with the 
  most extreme downward heat flux (see panel 5), and 
  subsequently settles down to being roughly a factor 10 larger 
  than the SH thickness. 
  We note that the spikes in the SH 
  model thickness are approximately the width
  of the minimum grid size in the TRAC simulation. They
  arise due to the motion of $T_c(s)$. 
\subsubsection{Integrated radiative losses
  \label{Sect:Long_pulse_RL}}
  \indent
  Finally,
  in order to test the
  analytical predictions made in 
  Section \ref{Sect:TRAC_analytical_assessment},
  we now consider the details of
  the integrated radiative losses
  from the SH and TRAC models.
  The 
  lower two panels (row 5) of Fig.
  \ref{Fig:600s_pulse_RL_5_time_evolution}
  show the integrated radiative losses in the 
  TRAC region (lower pair of curves) and the total
  over half of the loop (upper pair of curves), 
  and the ratio of 
  these quantities   
  in the left and right-hand panels, respectively.
  In Appendix \ref{App:B}, 
  we discuss in detail the agreement and discrepancies 
  between the results of Section
  \ref{Sect:TRAC_analytical_assessment} and these 
  numerical simulations.
  Here we note 
  the good agreement of the integrated losses after $600$~s
  (the end of the heating phase), and Appendix \ref{App:B}
  outlines the causes of the discrepancy prior to this time;
  in particular, amongst other 
  things, 
  the violation of the subsonic assumption made in 
  Section \ref{Sect:TRAC_analytical_assessment}. 
  The smaller 
  integrated TRAC
  region losses
  up to this time lead to a slightly higher coronal density due 
  to enhanced evaporation and accounts for the difference in 
  the coronal densities shown in the second and fourth
  panels.
  The spikes in 
  the SH radiation are
  due to the upward and downward motion 
  at the base of the TRAC region.
  %
  %
%%%%%%%%%%%%%%%%%%%%%%%%%%%%%%%%%%%%%%%%%%%%%%%%%%%%%%%%%%%%%%
%
% Short pulse
%
%%%%%%%%%%%%%%%%%%%%%%%%%%%%%%%%%%%%%%%%%%%%%%%%%%%%%%%%%%%%%%
\subsection{Short pulse
  \label{Sect:Short_pulse}}
  \indent
  Fig. \ref{Fig:60s_pulse_RL_5_time_evolution}
  shows the outcome of the 60~s heating pulse
  simulations in the same 
  format as 
  Fig. \ref{Fig:600s_pulse_RL_5_time_evolution}.
  For such a strongly impulsive
  heating event, the evolution is 
  much more dynamic
  than for
  the long pulse, with the coronal plasma sloshing
  to and fro  within the 
  loop \citep{paper:Reale2016}. 
  But the agreement between the $T$ and $n$ obtained with
  the SH and TRAC models remains good
  (the errors are less than 5\% throughout the simulation), 
  with even the oscillations seen in 
  the coronal density showing reasonable timing agreement.
  \\
  \indent 
  The plasma
  sloshing is also 
  reflected in the continual change of sign in the 
  enthalpy flux at the
  top of the TRAC region ($F_{e,c}$), superposed on the 
  fluctuations described earlier. 
  The agreement between 
  the integrated radiative losses
  is good, once again with the 
  exception of the
  times around peak heating,  which then
  leads to a slightly larger TRAC density. 
  This happens because the
  downflows at the base of the TR are supersonic
  during this period.
  However, even though the analytical model
  does not work as
  well during the heating phase, 
  it has little consequence for
  the coronal evolution because the radiative losses are
  small compared to the other terms in Eq. 
  \eqref{Eqn:1d_utr_jc} at that time.
  %
  %
%%%%%%%%%%%%%%%%%%%%%%%%%%%%%%%%%%%%%%%%%%%%%%%%%%%%%%%%%%%%%%
%
% Parameter Study for Uniform heating
%
%%%%%%%%%%%%%%%%%%%%%%%%%%%%%%%%%%%%%%%%%%%%%%%%%%%%%%%%%%%%%%
\subsection{Parameter study for uniform heating
  \label{Sect:Uniform_heating}}
  \indent
  We have
  also carried out a comparison between TRAC and the fully
  resolved SH simulations for a wider range of 
  uniform heating events,
  using the suite of twelve examples 
  first established by BC13 and further analysed in 
  \cite{paper:Johnstonetal2017a}. 
  These are in addition to the two examples 
  presented in detail above. 
  The parameter study focusses on short (60~Mm) and 
  long (180~Mm) loops with a range of heating functions.
  \\
  \indent 
  Table \ref{Table:uniform_simulations}
  summarises the results.
  The columns are, from left to 
  right: the case number and stage of evolution, 
  the heating event parameters 2L, $Q_H$ and
  $\tau_H$, the sample time 
  then the average
  temperature using SH(RL=14) and TRAC(RL=5,[3]), the 
  percentage difference between these, and the
  corresponding density 
  values at the same time. 
  The survey focuses on three 
  times: peak heating, peak density and during the decay phase. 
  The first two are readily identifiable from the simulations, 
  and 
  the third is chosen to be representative of a time when the 
  coronal part of the loop is cooling largely by radiation to 
  space and an enthalpy flux to the TR. 
  \\
  \indent
  In all cases, the discrepancy between the
  TRAC(RL=5) and SH(RL=14)
  models is 
  small. 
  For the temperature, there are
  just five occasions when 
  the normalised difference 
  is greater than 1\%, all 
  occurring in the decay phase. 
  The errors are larger for the density,
  with the majority over 1\%,
  but there are only
  two instances when the difference is greater than 
  5\%. 
  \\
  \indent
  The heating events in the longer loop show better 
  agreement than the shorter loop and the
  largest discrepancies are for the strongest heating
  events. 
  The first of these happens because 
  shorter loops require greater spatial resolution than longer 
  loops for a given peak temperature (BC13).
  The second arises because
  the heat flux that hits the TR and subsequent evaporation
  is systematically larger 
  for stronger heating events \citep{paper:Johnstonetal2017b},
  which in turn 
  increases the difficulty of 
  capturing accurately the corona and TR enthalpy exchange.
  It is also widely known
  \citep[e.g.][]{paper:Cargilletal2012b}
  that modelling the
  coronal density
  with approximate methods is much 
  more challenging than the temperature. 
  This can be attributed to the temperature being set initially 
  by the 
  direct in-situ heating, while the density 
  evolution
  relies on the 
  difficult 
  interplay in the TR between downward conduction and 
  upward enthalpy. 
  \\
  \indent
  We also show 
  the equivalent results 
  when the TRAC simulations are computed with 125 km grid cells 
  (RL=3) 
  in Table \ref{Table:uniform_simulations} 
  in square 
  brackets. 
  These are the simulations that correspond to 
  the same spatial 
  resolution as those
  presented in Fig. 3 of JB19.
  The density errors are of similar order to TRAC(RL=5).
  Once again, there are
  only two instances when the normalised difference is greater
  than 5\%. 
  On the other hand, the temperature errors do show
  significant variation, with the TRAC(RL=3) 
   differences
  consistently larger than TRAC(RL=5).  
  The latter is associated with a higher cutoff temperature
  having more influence on the coronal temperature evolution,
  while the former, in contrast, is indicative of the rapid 
  convergence of the TRAC method when modelling the enthalpy
  exchange.
  %
  %
%%%%%%%%%%%%%%%%%%%%%%%%%%%%%%%%%%%%%%%%%%%%%%%%%%%%%%%%%%%%%%
%
% Footpoint Heating
%
%%%%%%%%%%%%%%%%%%%%%%%%%%%%%%%%%%%%%%%%%%%%%%%%%%%%%%%%%%%%%%
\subsection{Footpoint heating
  \label{Sect:Footpoint_heating}}
  \indent
  Footpoint heating of coronal loops, either steady or 
  impulsive, is a topic of considerable importance. Steady 
  footpoint heating can be associated with phenomena such 
  as coronal rain
  \citep[e.g.][]{paper:Schrijver2001,
  paper:Antolinetal2010,paper:Antolinetal2015,
  paper:Antolin2020}
  and long-period 
  extreme ultra-violet (EUV) pulsations 
  \citep[e.g.][]{paper:Auchereetal2014,
  paper:Auchereetal2018,
  paper:Fromentetal2015,
  paper:Fromentetal2017,
  paper:Fromentetal2018,
  paper:Fromentetal2019,
  paper:Pelouzeetal2019}, while
  unsteady 
  footpoint heating can arise due to, for example, 
  precipitation of energetic particles during flares
  \citep{paper:Testaetal2014} and
  chromospheric reconnection during
  surface magnetic flux cancellation 
  \citep{paper:Chittaetal2018}.
  \\
  \indent
  In order
  to model the coronal response to footpoint heating 
  accurately,
  the TRAC method must include the modification
  of the heating rate ($Q(T)$) as described in  
  Section \ref{Sect:TRAC_broadening}. 
  This is an important extension to the
  technique developed by L09 for
  broadening the TR because when this modification
  is not included, there can be large discrepancies 
  in the total energy injected 
  into the loop 
  (and subsequent evolution)
  between
  the TRAC and SH models. 
\subsubsection{Impulsive energy release
  \label{Sect:Impulsive_footpoint_heating}}
  \indent 
  In \cite{paper:Johnstonetal2017b}, 
  we examined a number of 
  impulsive footpoint heating examples in the 
  context of the jump condition model. 
  The most challenging were those 
  involving heating at the base of the TR 
  (i.e. $s= s_b$ in the initial equilibrium, referred
  to as the \lq fp2\rq\ examples in that paper). 
  We consider two such examples here, namely 
  a loop with 2L =~60~Mm, 
  and a heating function comprised of
  a Gaussian pulse centred at the base of each TR with a
  half-width of 0.75~Mm, lasting for 
  600 and 
  60~s.
  The former (latter) has a peak heating rate 
  of 0.21 (2.1)~Jm$^{-3}$s$^{-1}$
  at the maximum of the Gaussian profile.
  \\
  \indent
  The results are summarised in Figs.
  \ref{Fig:fp2_Case_600s_RL_5_time_evolution} and
  \ref{Fig:fp2_Case_60s_RL_5_time_evolution},
  which 
  are of the same format as Figs. 
  \ref{Fig:600s_pulse_RL_5_time_evolution}
  and \ref{Fig:60s_pulse_RL_5_time_evolution}. 
  They show little 
  difference from the previous cases of 
  uniform coronal energy release, indicating 
  the robustness of the method. 
  In 
  particular,
  consistent with
  the analytical assessment of the
  TRAC method presented in 
  Section \ref{Sect:TRAC_analytical_assessment},
  both the integrated radiative losses and 
  integrated heating
  show good agreement 
  between the TRAC and SH models,
  but as before there is
  discrepancy between the radiative losses at times
  during the heating phase.
  We have also tested
  impulsive footpoint heating in loops of 
  total length 180~Mm. These simulations show the same 
  fundamental properties as the 60 Mm loop.
\subsubsection{Steady energy release
  \label{Sect:Steady_footpoint_heating}}
  \indent
  Thermal non-equilibrium (TNE) is a phenomenon that 
  can occur in coronal loops when the heating is quasi-steady 
  and concentrated towards the footpoints \citep[e.g.][]
  {paper:Mulleretal2003,paper:Antolinetal2010,
  paper:Peteretal2012,paper:Mikicetal2013,
  paper:Fromentetal2018}.
  The response of a loop to such heating conditions is to 
  undergo evaporation and condensation cycles with a period on 
  the timescale of hours.
  \\
  \indent
  For the case of steady footpoint heating,
  \cite{paper:Johnstonetal2019} demonstrated that
  with the SH conduction model
  inadequate TR resolution can lead to 
  significant discrepancies in TNE cycle behaviour, with TNE 
  being suppressed in under-resolved loops. 
  To compare this influence of numerical resolution on TNE
  (in coronal loops)
  with the TRAC method,
  we repeat  those
  steady footpoint heating simulations here. 
  \\
  \indent 
  Fig. \ref{Fig:TNE_f_vs_resolution} contrasts the 
  TRAC model with the SH results.
  The upper two rows show the coronal 
  averaged temperature ($T$) as a function of time,  
  for selected  values of RL,
  for the SH and TRAC methods, respectively.
  The TRAC solutions are significantly less 
  dependent on the 
  spatial resolution than the SH results,
  with cyclic TNE 
  \citep{paper:Fromentetal2018,
  paper:Winebargeretal2018,
  paper:Klimchuk&Luna}
  arising for all values of 
  RL, while maintaining high levels 
  of accuracy throughout. 
  The details of the time evolution of the temperature 
  as a function of position are
  representative of those described fully in 
  \cite{paper:Johnstonetal2019}. 
  \\
  \indent
  The lower two rows show 
  the dependence of TNE cycle frequency (row 3)
  and simulation computation time (row 4)
  on the minimum spatial 
  resolution.
  In these plots the blue (red) 
  lines correspond to the TRAC (SH) model.
  Convergence of the TNE cycle 
  period and thermodynamic evolution (i.e. the same temperature 
  extrema) is seen with TRAC for RL~$\geq 5$  
  (corresponding to a TR grid cell width of 31.25~km),
  while the SH model requires 
  a TR grid resolution of 1.95 km or better (RL~$\geq 9$).
  This 
  relaxation of the resolution requirements
  represents a substantial  
  saving in the computation time.
  The improvement in run time is comparable to that
  described in \cite{paper:Johnston&Bradshaw2019}, 
  which we note is
  achieved in all of the simulations presented
  in this paper.
  Even if computationally one can only
  achieve a TR resolution of 500 km (RL=1), then 
  with TRAC method,
  the error in the cycle period is just 10\%
  whereas without TRAC there is no cycle detected. 
  %
  %
%%%%%%%%%%%%%%%%%%%%%%%%%%%%%%%%%%%%%%%%%%%%%%%%%%%%%%%%%%%%%%
%
% Discussion
%
%%%%%%%%%%%%%%%%%%%%%%%%%%%%%%%%%%%%%%%%%%%%%%%%%%%%%%%%%%%%%%
\section{Discussion}
  \label{Sect:Discussion}
  \indent
  This paper extends the work of JB19 and demonstrates the 
  versatility and robustness of the TRAC method over a range of 
  impulsive and quasi-steady footpoint
  heating events, as well as the 
  theoretical underpinning of its success. 
  Furthermore, the method should prove amenable to extension in 
  multi-dimensional MHD
  simulations, though
  a more sophisticated treatment will be required;
  in particular, how the magnetic field evolution
  modifies the prescription of the
  cutoff temperature along a field-line. 
  Flux tube area expansion will also 
  form part of such an extension. 
  \\
  \indent
  The main consequence of not adequately
  resolving the transition region 
  (TR) in numerical simulations of impulsive heating is that 
  the resulting coronal density is artificially low (BC13). 
  This happens because the 
  downward heat flux is forced to \lq jump\rq\ 
  across an under-resolved TR to the 
  chromosphere, where the incoming energy is then strongly 
  radiated. 
  Hence, the integrated radiative losses are 
  significantly overestimated with a lack of spatial resolution
  on use of
  the SH conduction method 
  \citep[see e.g.][]{paper:Johnstonetal2017a}.
  \\
  \indent
  In contrast, when using the TRAC method, the
  integrated radiative losses are accurately 
  accounted for which 
  helps ensure that the energy balance across the TRAC 
  region is an accurate approximation  
  of the properly resolved SH solution.
  This is achieved by enforcing certain 
  conditions on the parallel thermal conductivity, radiative 
  loss and heating rates that are not met 
  physically (e.g. by the SH conduction method), but do enable 
  the TR to be broadened so that the steep gradients are 
  spatially resolved even when using coarse numerical grids.
  \\
  \indent
  The resulting
  accuracy of the coronal plasma evolution means that 
  simulations using TRAC can be used to follow coronal 
  observables with confidence. 
  However,
  despite the method conserving
  both the integrated radiative losses and 
  integrated heating across the TRAC region,
  caution is needed with any forward modelling below 
  the cutoff temperature~($T_c$). 
  This can be seen by examining 
  the dependence of the differential
  emission measure (DEM) on temperature. 
  Using the 
  expressions for $L_T$ from Section 
  \ref{Sect:TRAC_analytical_assessment}, we can
  calculate the 
  temperature distribution of the 
  DEM
  for each model,
  \begin{align}
    \textrm{DEM}(T) = n^2 
    \left(
    \dfrac{\partial T}{\partial s} 
    \right)^{\!\!-1}
     =     
     \left(
    \dfrac{P}{2k_BT}
    \!
    \right)^{\!\!2}
    \dfrac{T}{L_T}. 
    \label{Eqn:DEM} 
  \end{align}
  In the three limits, (i)
  strong evaporation, (ii) peak density and 
  (iii) decay, we find:
  \begin{align}
    \textrm{(i)} \quad
    &
    L_T \sim \kappa_\parallel(T); 
    \\[4mm] 
    \textrm{(ii)} \quad
    &
    L_T \sim   
    \dfrac{T^3 \kappa_\parallel(T)}{P^2\Lambda(T)}; 
  \end{align}
  and,
   \begin{align}
    \textrm{(iii)} \quad
    &
    L_T \sim \dfrac{T^3n|v|}{P^2\Lambda(T)}.
  \end{align}
  Therefore,
  the temperature dependence of the DEM now differs 
  significantly between TRAC and SH in all regimes since the 
  combination $\kappa_\parallel(T)\Lambda(T)$ 
  no longer appears. Thus, 
  observables cannot be calculated below $T_c$
  with confidence.
  However, 
  for all temperatures above $T_c$, 
  including the majority of the upper TR, the 
  emission can be synthesised
  accurately.
  \\
  \indent
  In summary, the TRAC method allows the highly
  efficient numerical integration of the hydrodynamic equations
  through the computationally demanding TR. 
  The outcome is an accurate and
  time-dependent \lq boundary condition\rq\ 
  for the domain of
  interest, which is comprised of all the plasma in the corona
  and above $T_c$.
  Below the cutoff temperature, the modifications 
  to the heat flux, heating and cooling rates 
  broaden the steep gradients in the TR
  while conserving key quantities in the energy equation.
  %
  %
%%%%%%%%%%%%%%%%%%%%%%%%%%%%%%%%%%%%%%%%%%%%%%%%%%%%%%%%%%%%%%
%
% Acknowledgements
%
%%%%%%%%%%%%%%%%%%%%%%%%%%%%%%%%%%%%%%%%%%%%%%%%%%%%%%%%%%%%%%
\begin{acknowledgements} 
  This work has received support from the European Union 
  Horizon 2020 research and innovation programme (grant 
  agreement No. 647214),
  from
  the
  UK Science and 
  Technology Facilities Council through the consolidated 
  grant ST/N000609/1
  and the Research Council of Norway 
  through its Centres of Excellence scheme, project number 
  262622.
  S.J.B. is grateful to the National Science Foundation 
  for supporting this work through CAREER award AGS-1450230. 
  C.D.J.
  acknowledges support from the International Space Science 
  Institute (ISSI), Bern, Switzerland to the International Team 
  401 \lq\lq
  Observed Multi-Scale Variability of Coronal Loops as a 
  Probe of Coronal Heating\rq\rq.
  We also 
  acknowledge useful discussions with Dr Z. Miki{\'c}
  and very constructive comments from the referee, 
  Dr Philip Judge.
\end{acknowledgements}
  %
  %
%%%%%%%%%%%%%%%%%%%%%%%%%%%%%%%%%%%%%%%%%%%%%%%%%%%%%%%%%%%%%%
%
% References
%
%%%%%%%%%%%%%%%%%%%%%%%%%%%%%%%%%%%%%%%%%%%%%%%%%%%%%%%%%%%%%%
\bibliographystyle{aa}
\bibliography{TRAC_Paper_2_Field_Aligned_HD}
  %
  %
  %%%%%%%%%%%%%%%%%%%%%%%%%%%%%%%%%%%%%%%%%%%%%%%%%%%%%%%%%%%%%    
  %
  % Fig:600s_pulse_RL_5_time_evolution
  %
\begin{figure*}
  \vspace*{-2mm}
  \hspace*{0.075\linewidth}
  \subfigure{\includegraphics[width=0.4\linewidth]
  {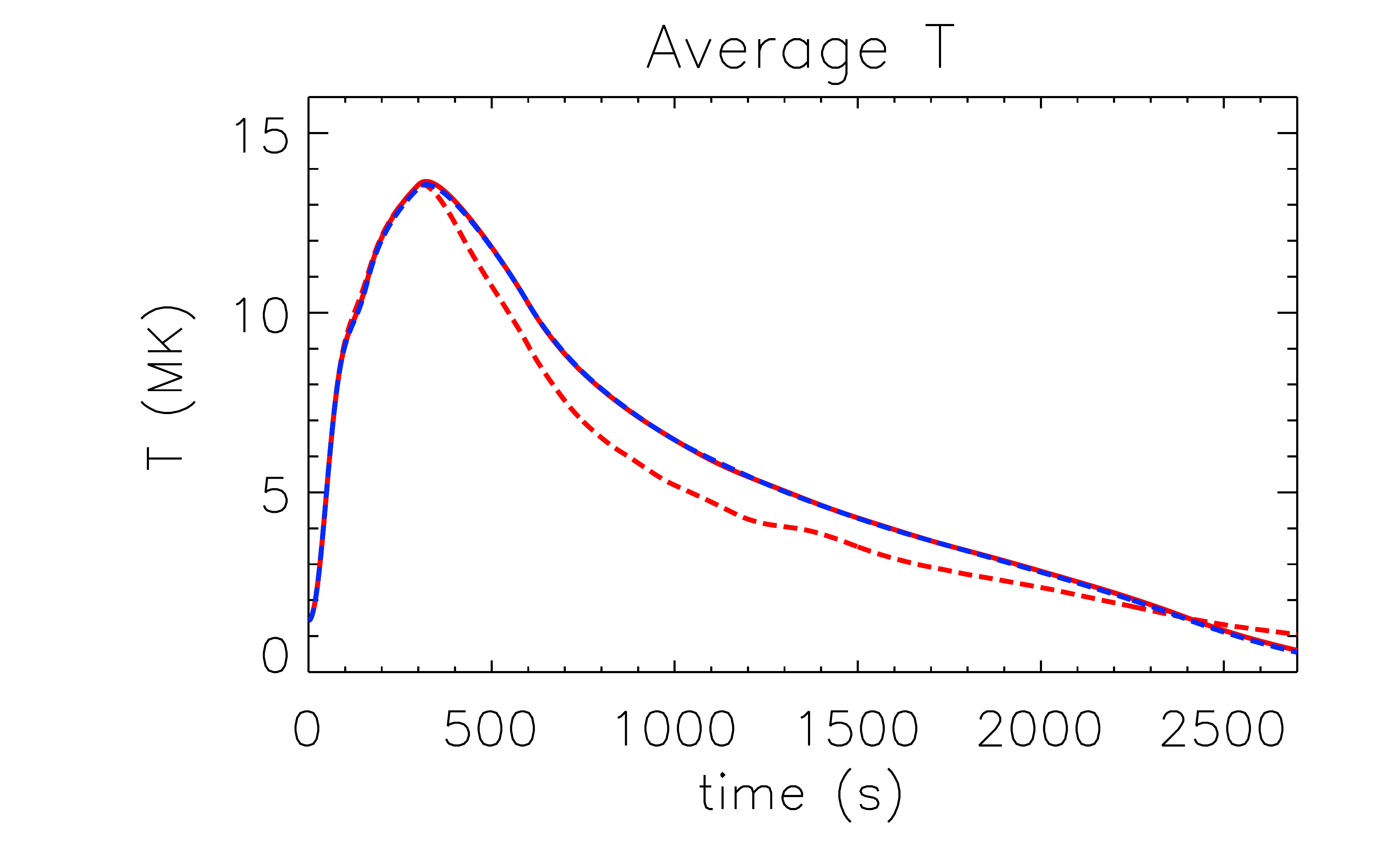}}
  \subfigure{\includegraphics[width=0.4\linewidth]
  {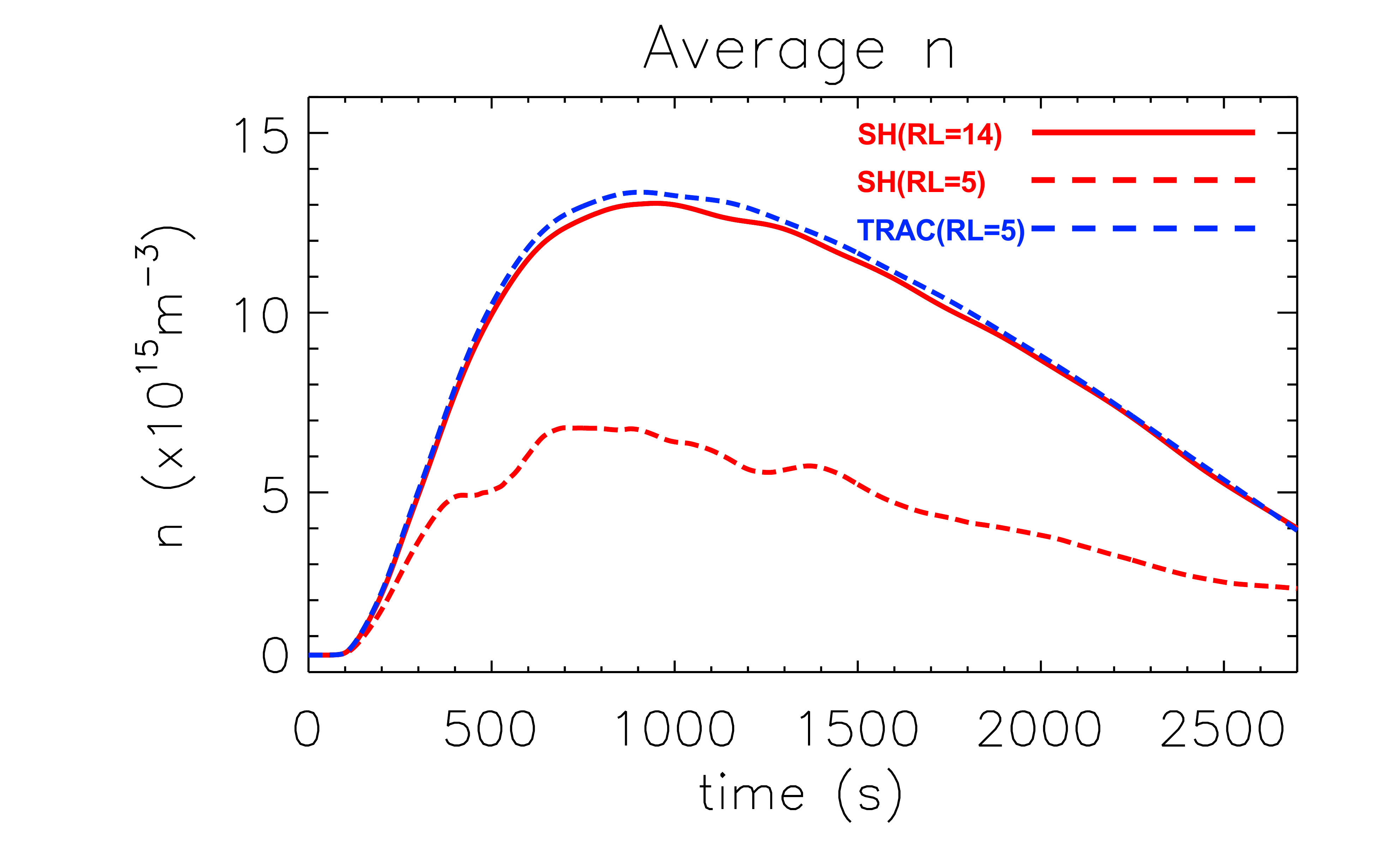}}
  \\[-4mm]
  \hspace*{0.075\linewidth}
  \subfigure{\includegraphics[width=0.4\linewidth]
  {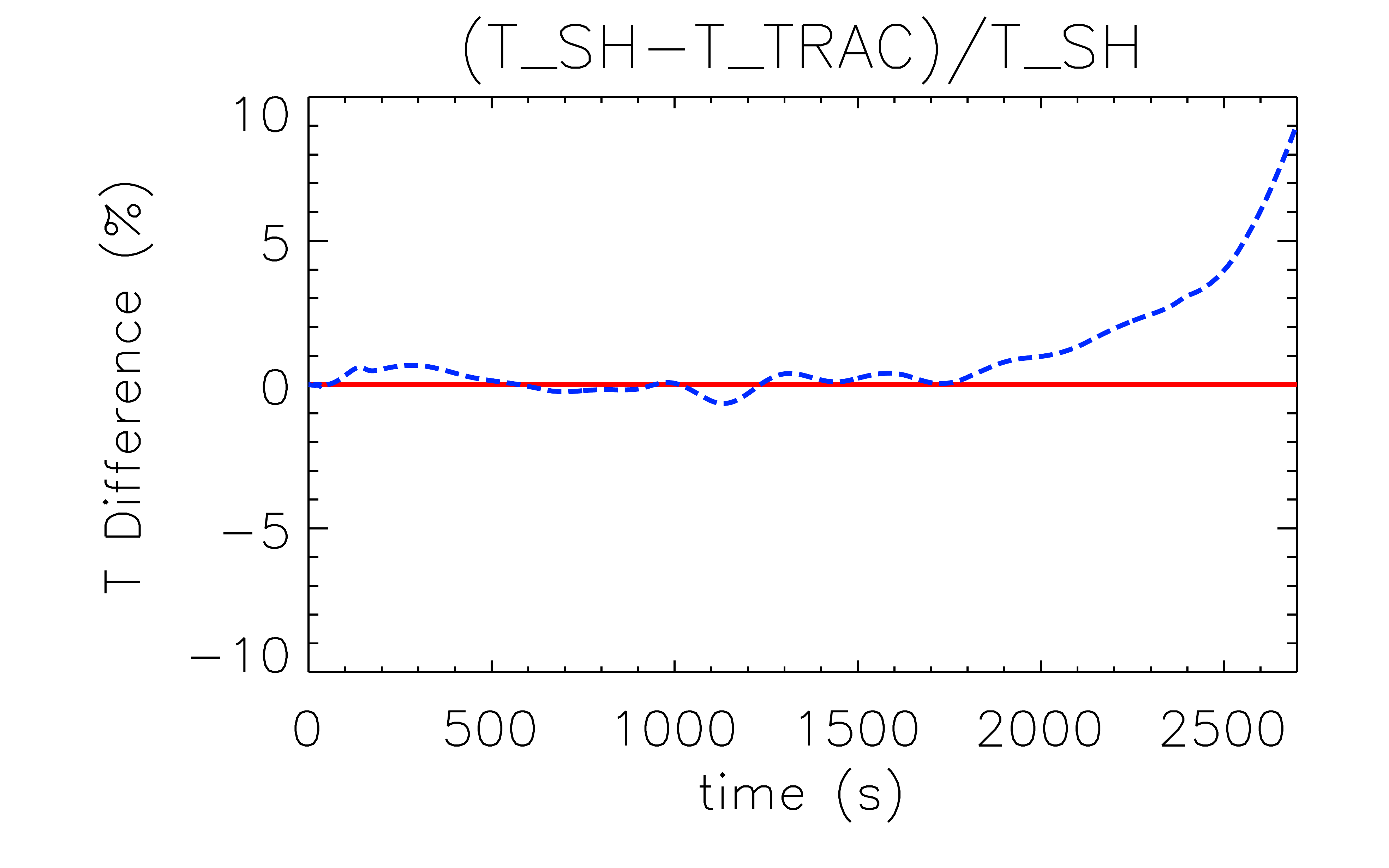}}
  \subfigure{\includegraphics[width=0.4\linewidth]
  {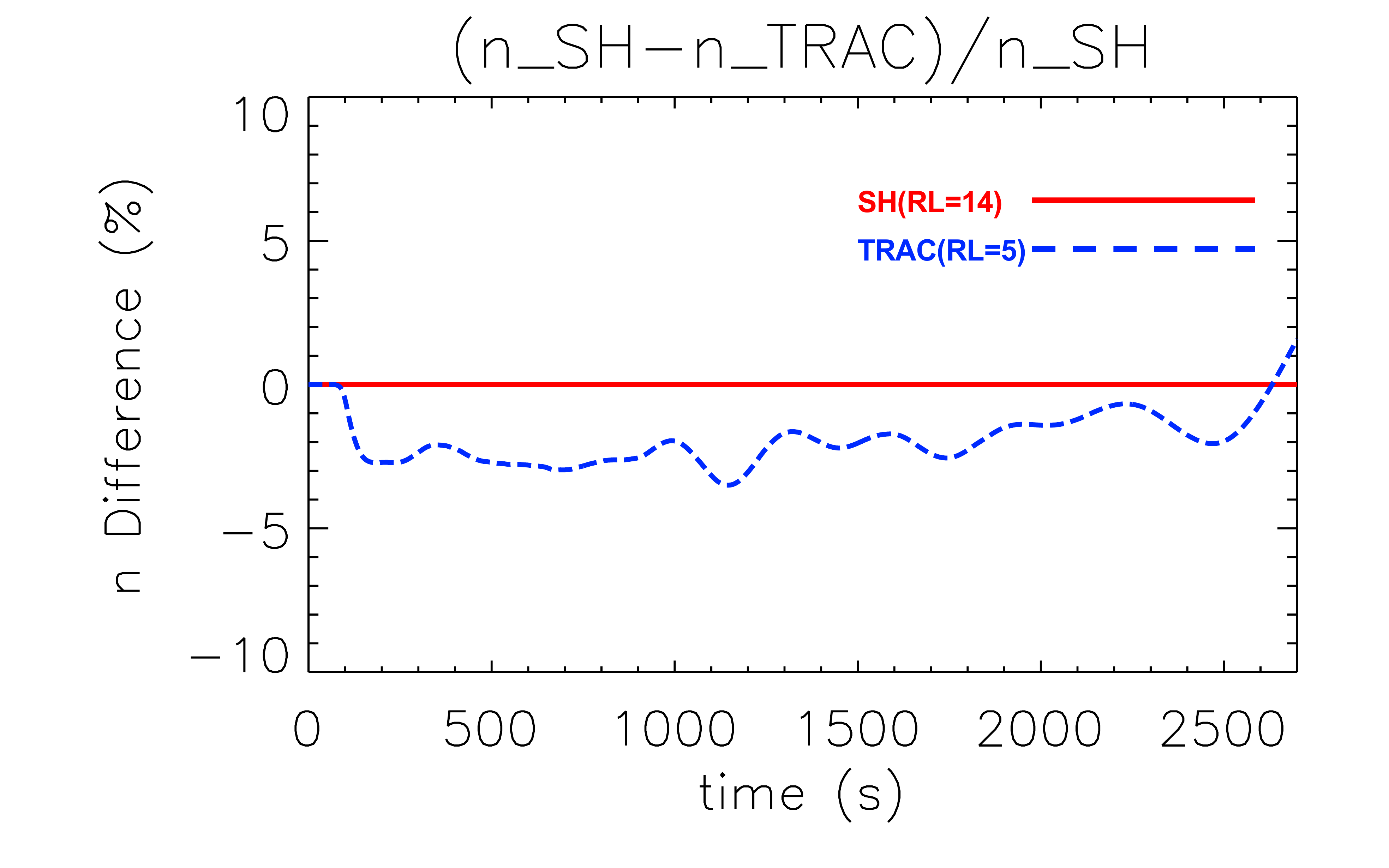}}
  \\[-4mm]
  \hspace*{0.075\linewidth}
  \subfigure{\includegraphics[width=0.4\linewidth]
  {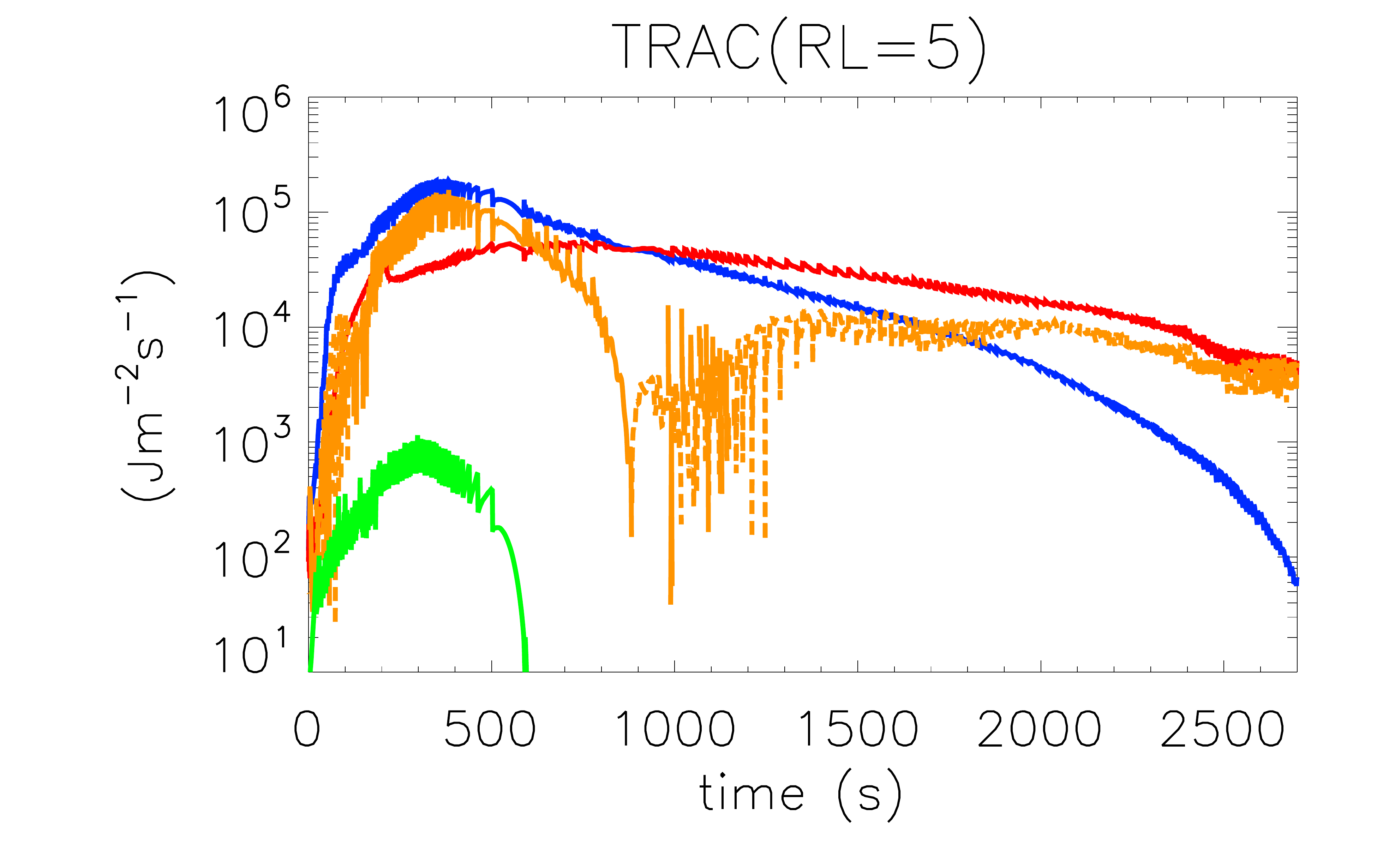}}
  \subfigure{\includegraphics[width=0.4\linewidth]
  {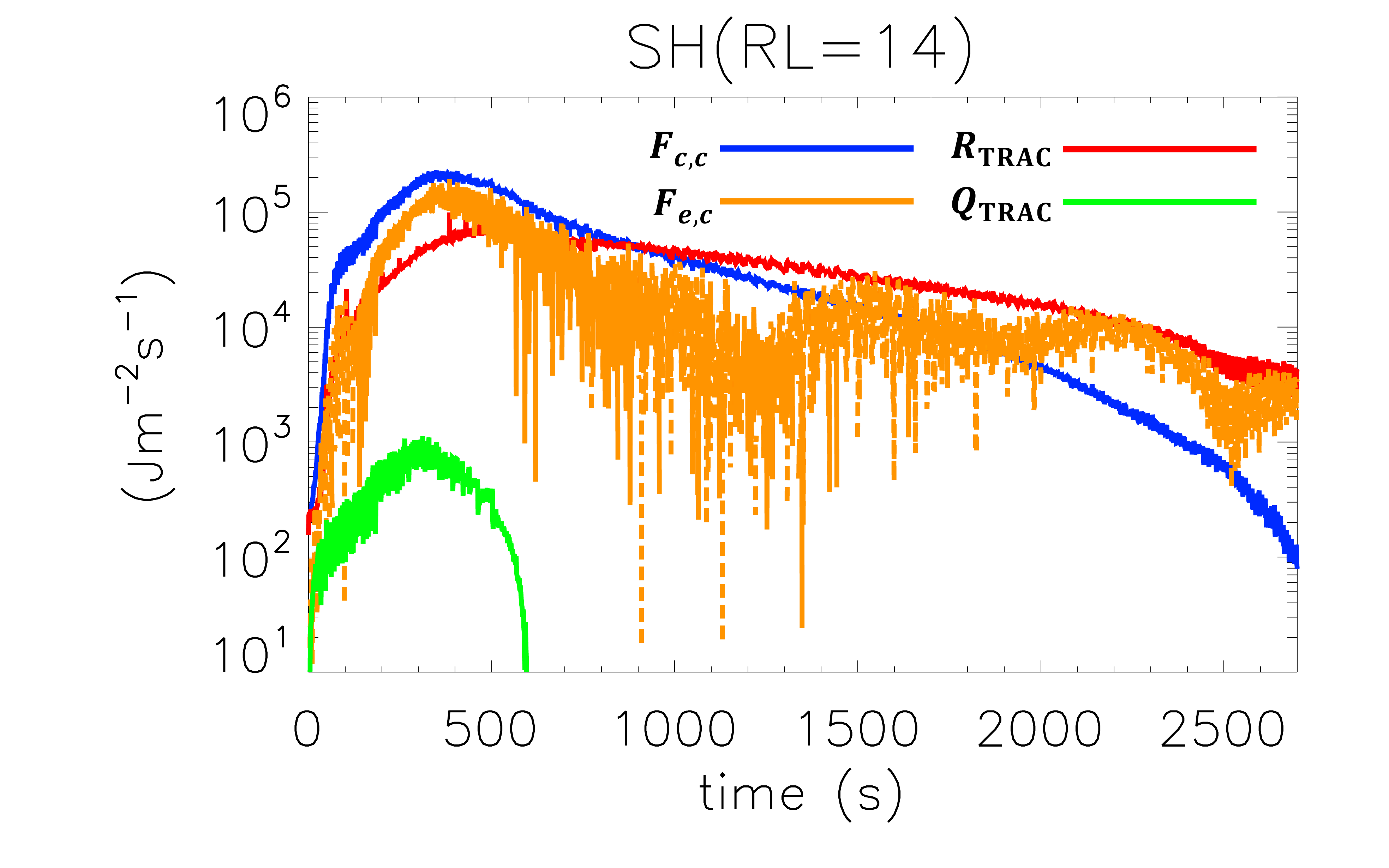}}
  \\[-4mm]
  \hspace*{0.075\linewidth}
  \subfigure{\includegraphics[width=0.4\linewidth]
  {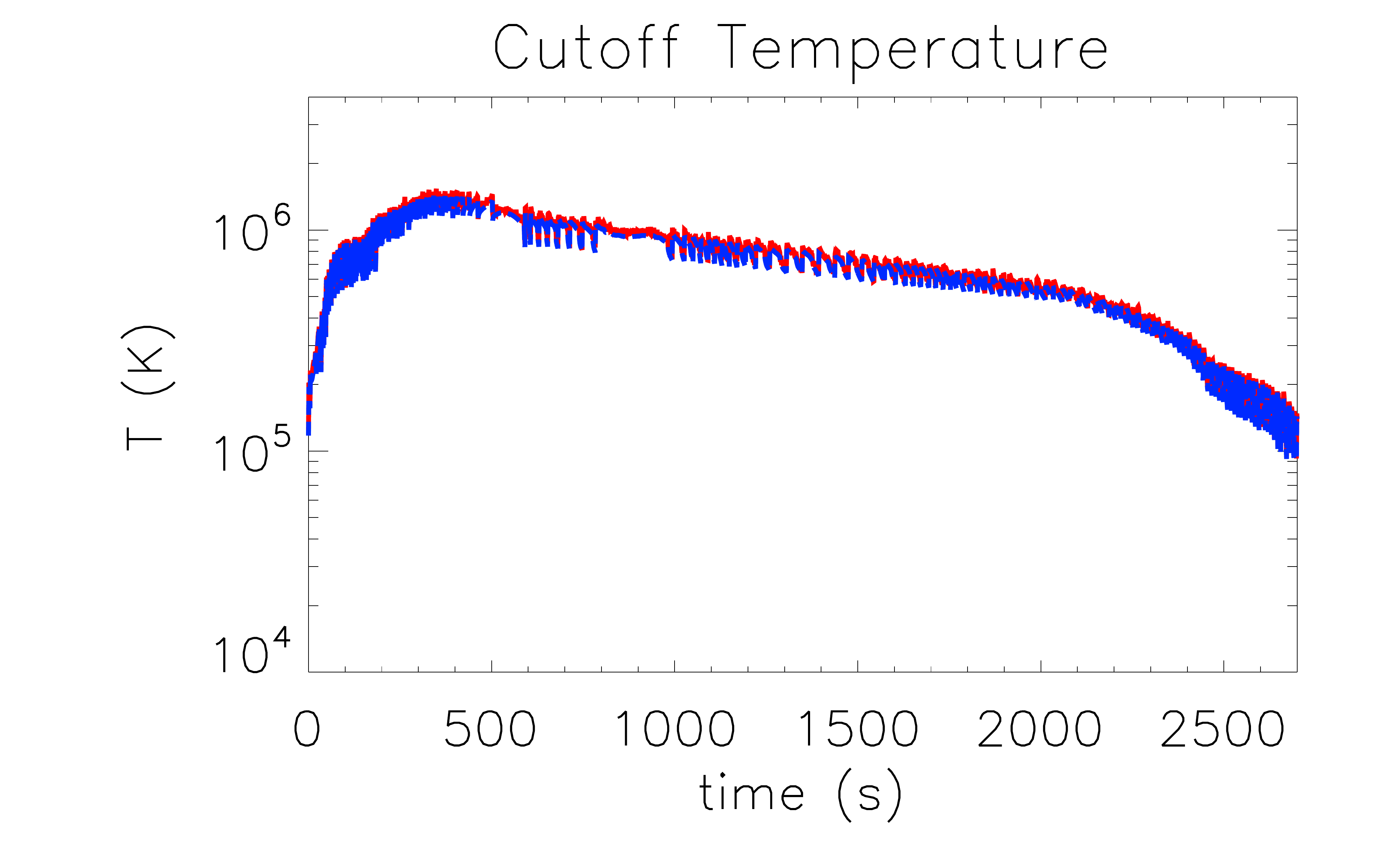}}
  \subfigure{\includegraphics[width=0.4\linewidth]
  {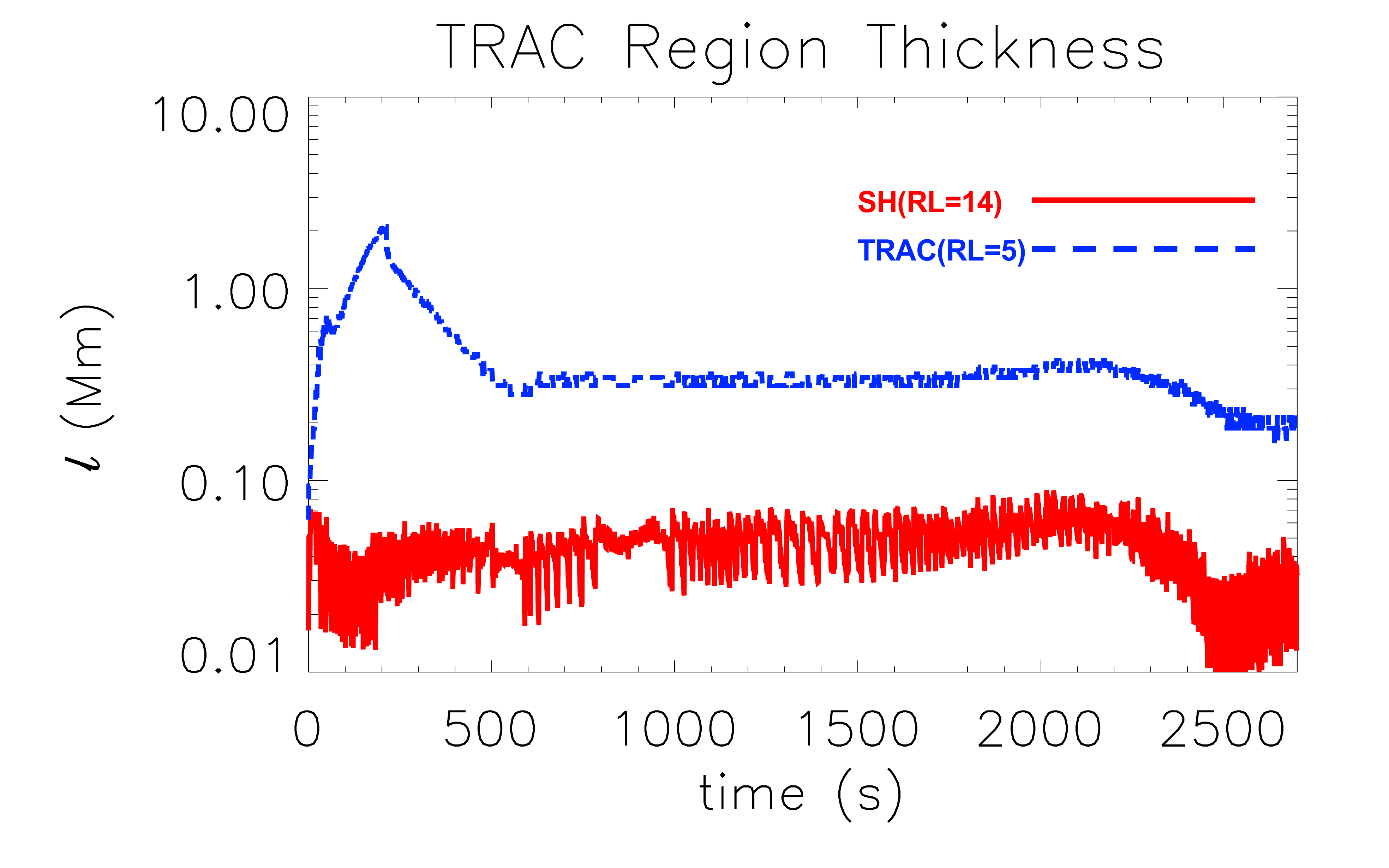}}
  \\[-4mm]
  \hspace*{0.075\linewidth}
  \subfigure{\includegraphics[width=0.4\linewidth]
  {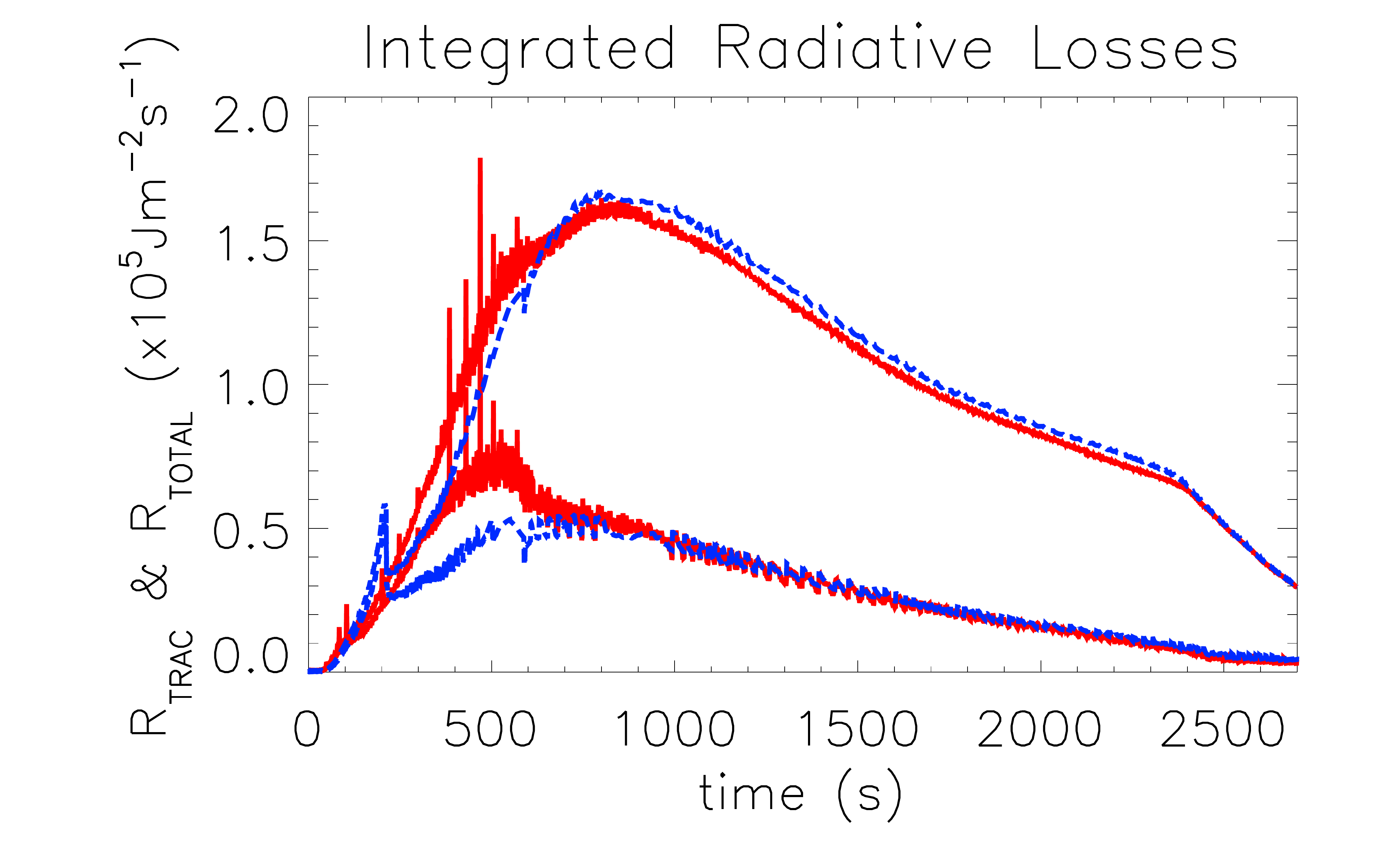}}
  \subfigure{\includegraphics[width=0.4\linewidth]
  {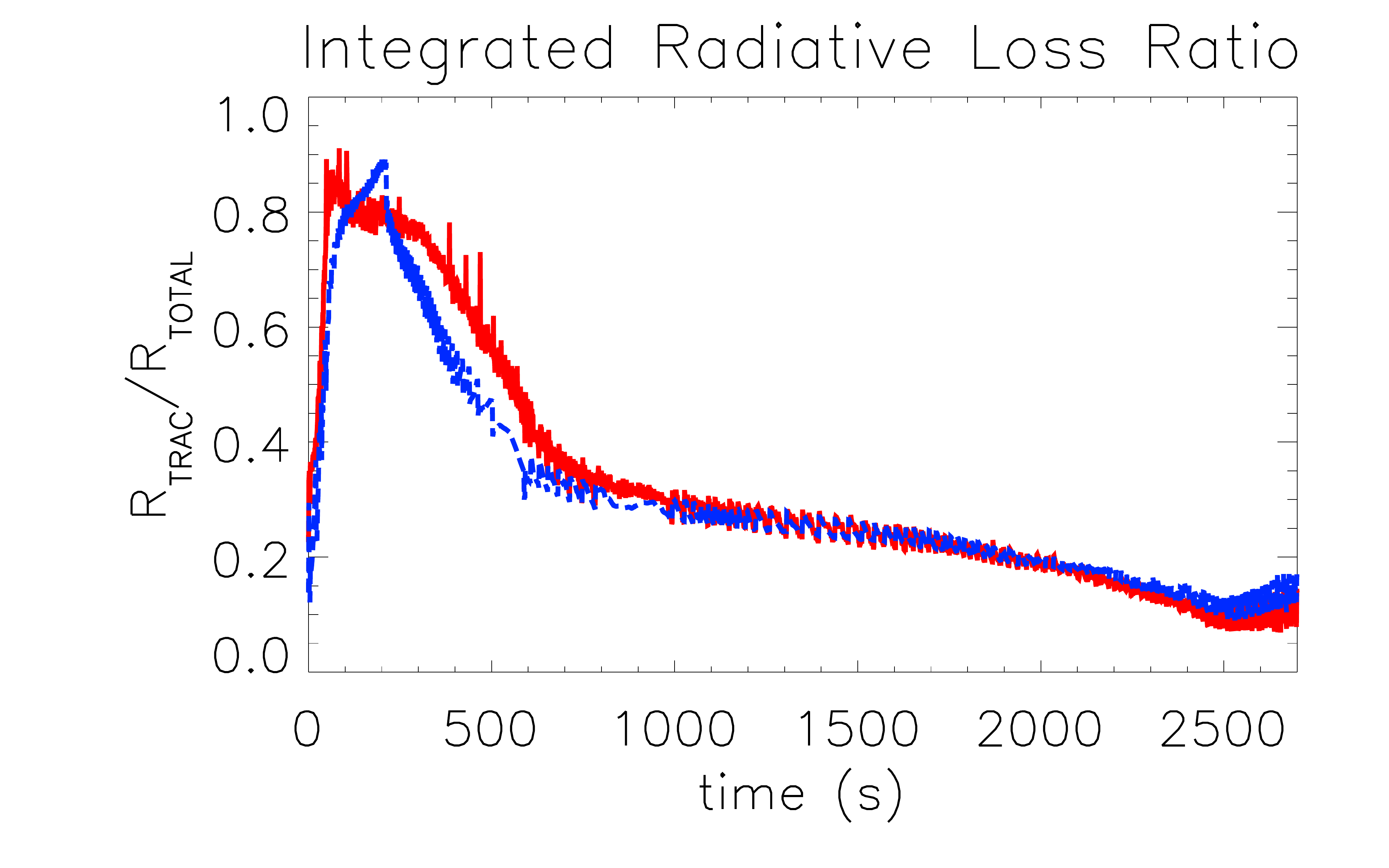}}
  \vspace*{-4mm}
  \caption{Results for the 600 s heating pulse simulations
    (Section \ref{Sect:Long_pulse}).
    The upper four panels show
    the coronal averaged temperature and density as functions
    of time,
    and their 
    respective normalised differences.
    The lines are colour coded in a way that reflects
    the conduction method used 
    and different values of RL are separated by different line
    styles as shown in the right-hand panels.
    The central two panels present a 
    comparison of the energetically dominant quantities
    that are associated with the
    TRAC region,
    namely, the  
    heat ($F_{c,c}$, blue line)
    and enthalpy ($F_{e,c}$, orange line) 
    fluxes at the top of the 
    TRAC region, 
    and the 
    radiative losses ($\mathcal{R}_{\textsc{trac}}$, red line)
    and
    heating 
    ($Q_{\textsc{trac}}$, green line)
    integrated across
    the TRAC region.
    Solid (dashed) lines indicate where 
    the enthalpy flux is upflowing (downflowing).
    The lower four panels show the time evolution of the
    cutoff temperature ($T_c$), TRAC region 
    thickness~($\ell$),
    radiative losses integrated across the TRAC region 
    ($\mathcal{R}_{\textsc{trac}}$) and total over half of the
    loop ($\mathcal{R}_{\textsc{total}}$), and
    the ratio between these losses 
   ($\mathcal{R}_{\textsc{trac}}/\mathcal{R}_{\textsc{total}}$).
   We note that the temperature  of the TRAC(RL=5) and 
   SH(RL=14) 
   solutions overlay~in~the~top~left-hand~panel.
   \label{Fig:600s_pulse_RL_5_time_evolution}
  }
\end{figure*}
  %
  %
  %
  %
  %%%%%%%%%%%%%%%%%%%%%%%%%%%%%%%%%%%%%%%%%%%%%%%%%%%%%%%%%%%%%    
  %
  % Fig:60s_pulse_RL_5_time_evolution
  %
\begin{figure*}
  \hspace*{0.075\linewidth}
  \subfigure{\includegraphics[width=0.4\linewidth]
  {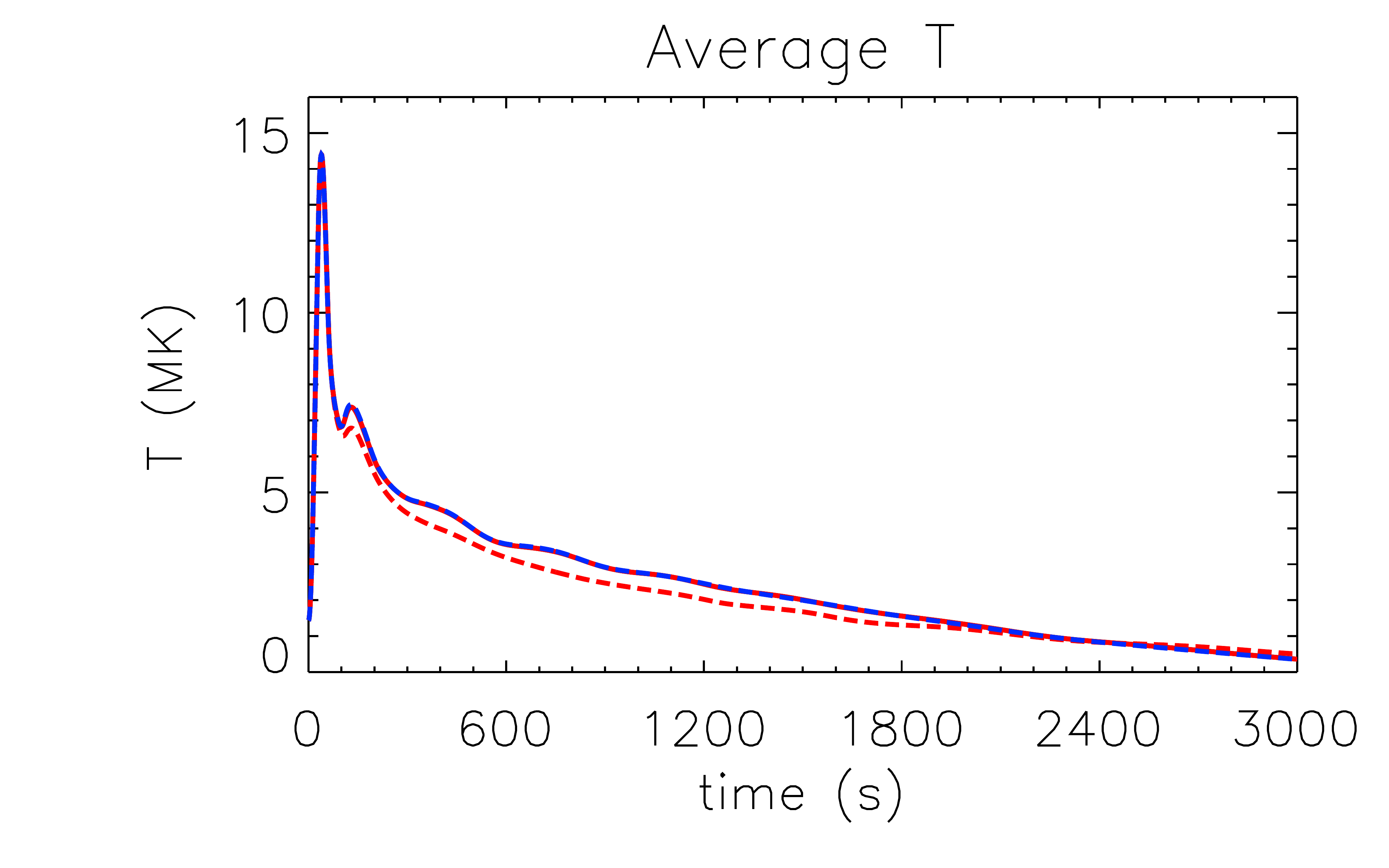}}
  \subfigure{\includegraphics[width=0.4\linewidth]
  {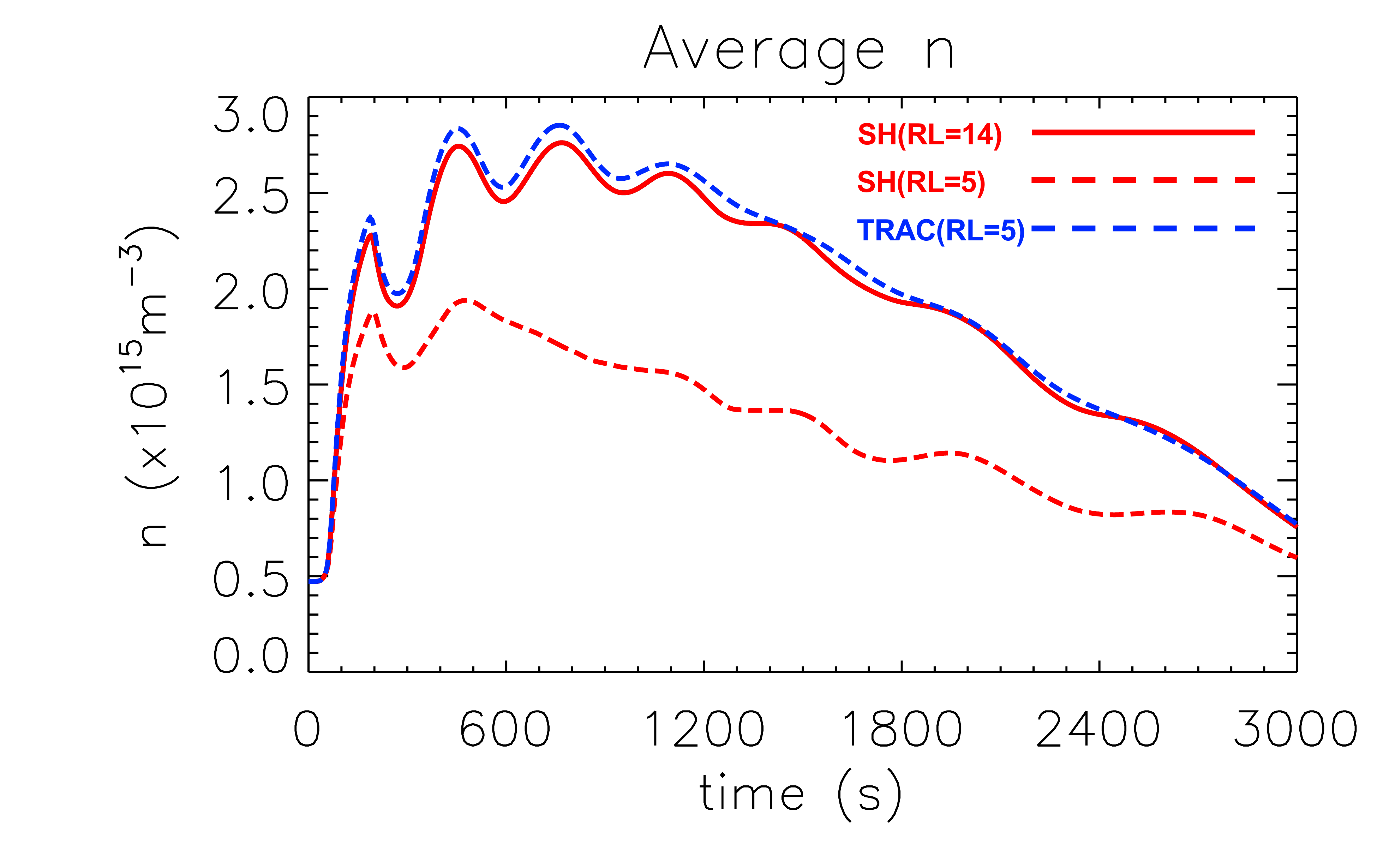}}
  \\[-4mm]
  \hspace*{0.075\linewidth}
  \subfigure{\includegraphics[width=0.4\linewidth]
  {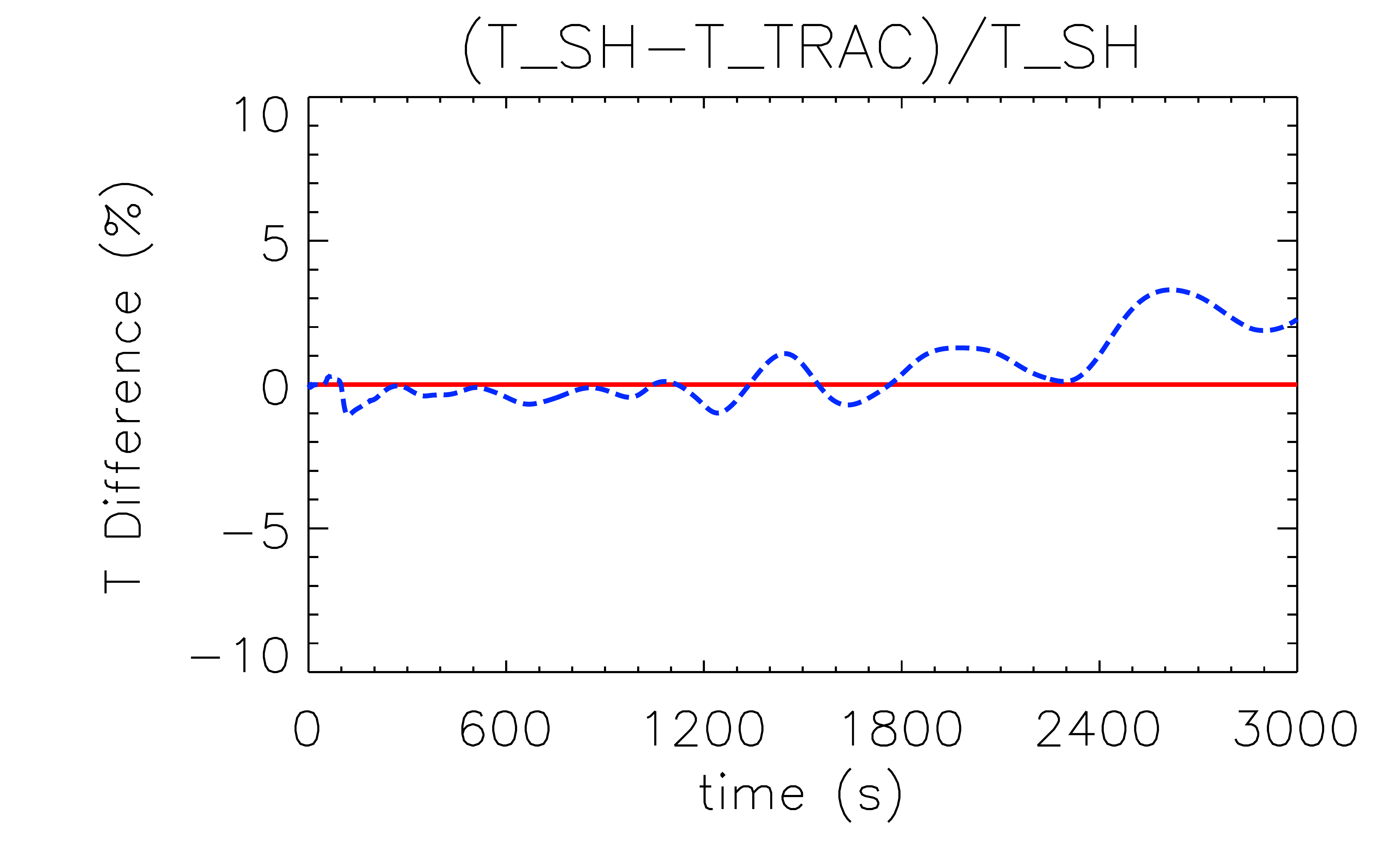}}
  \subfigure{\includegraphics[width=0.4\linewidth]
  {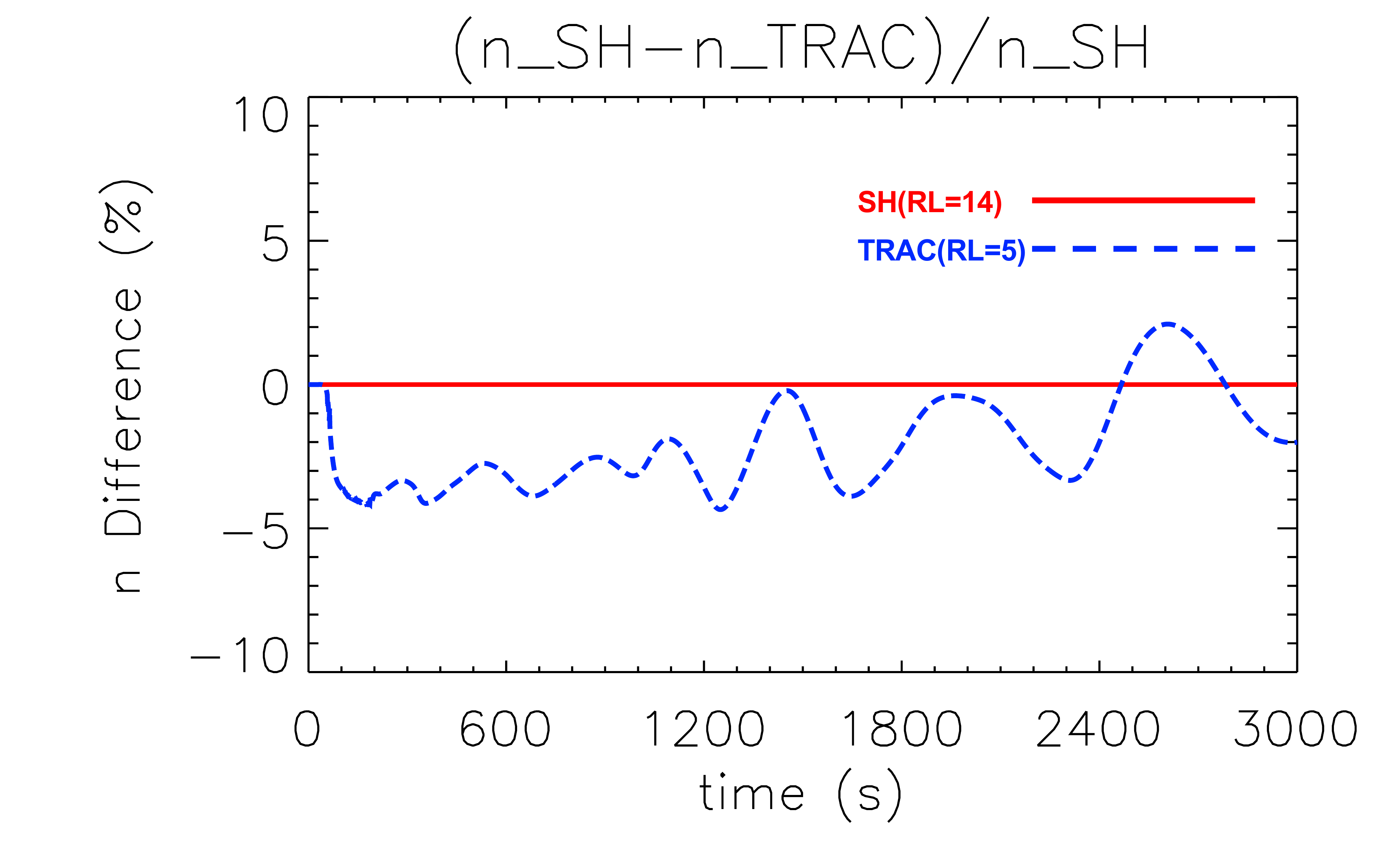}}
  \\[-4mm]
  \hspace*{0.075\linewidth}
  \subfigure{\includegraphics[width=0.4\linewidth]
  {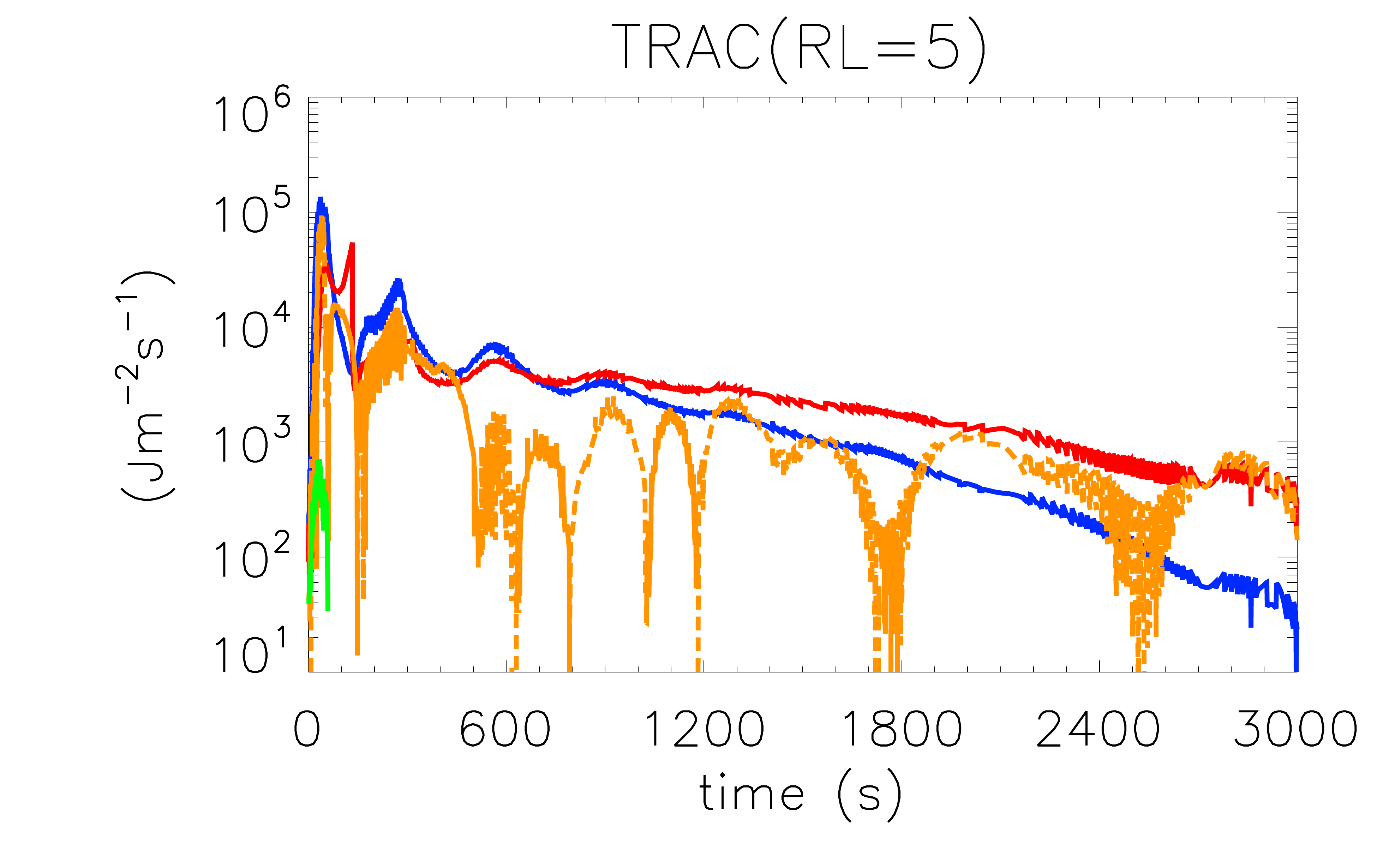}}
  \subfigure{\includegraphics[width=0.4\linewidth]
  {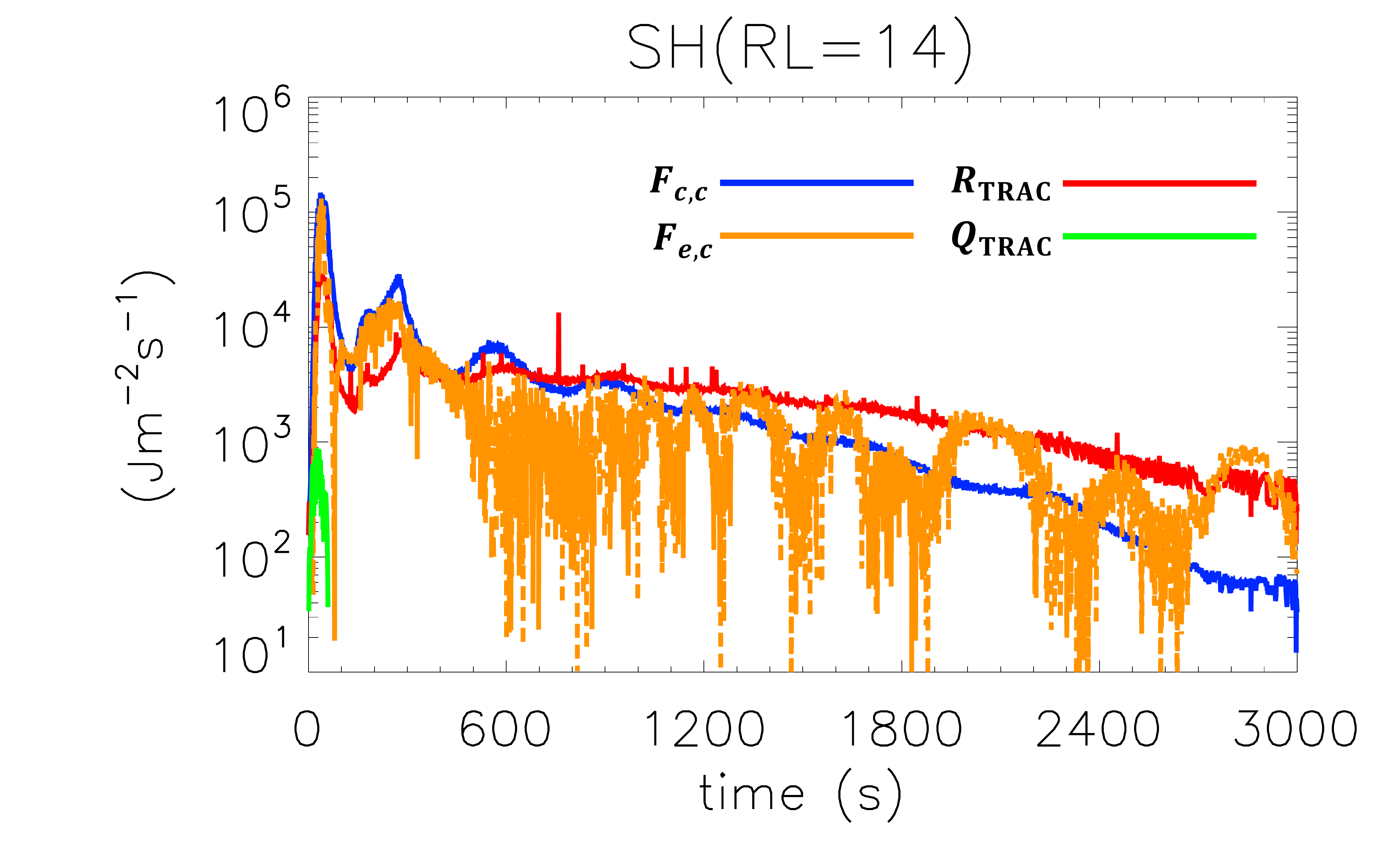}}
  \\[-4mm]
  \hspace*{0.075\linewidth}
  \subfigure{\includegraphics[width=0.4\linewidth]
  {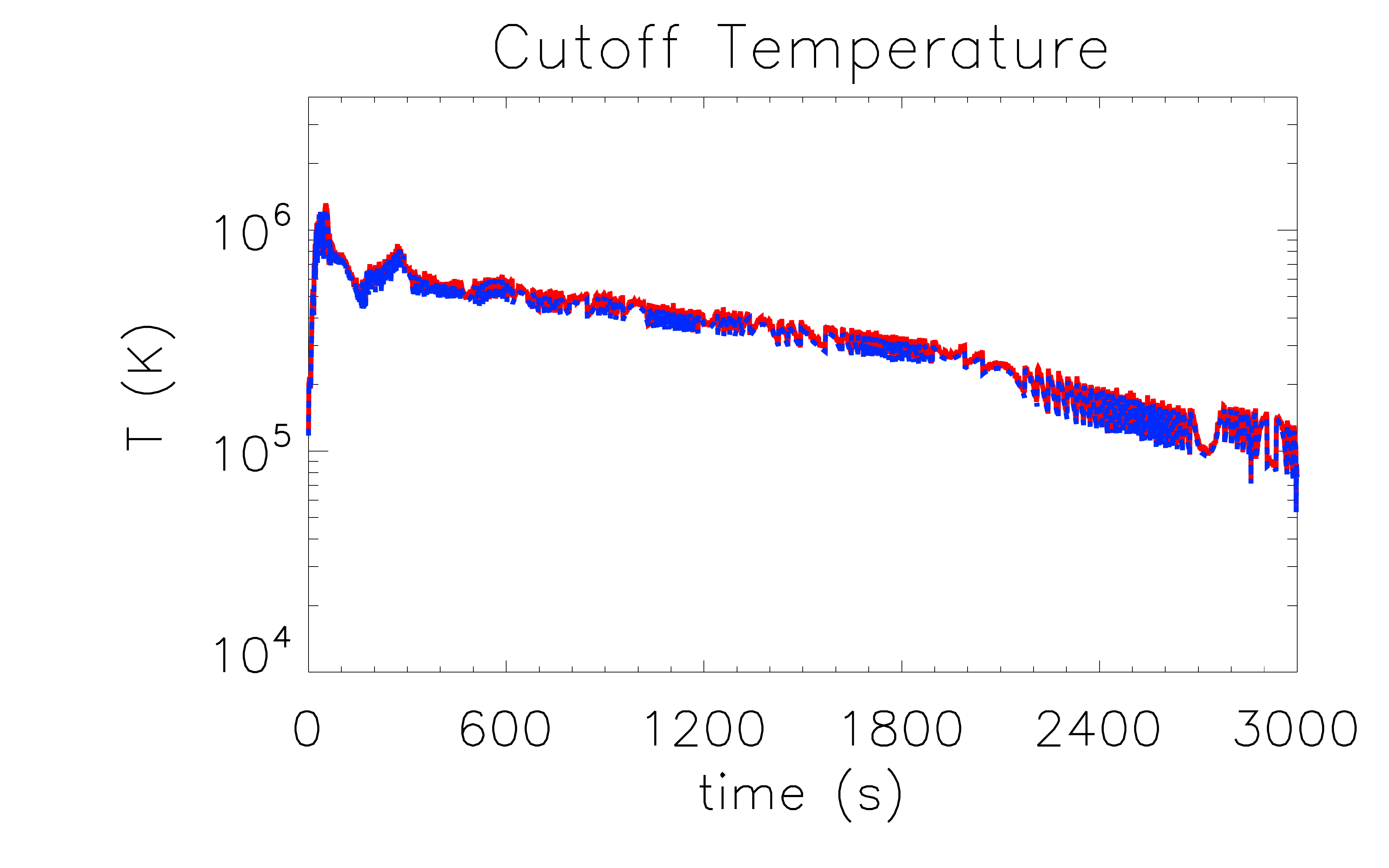}}
  \subfigure{\includegraphics[width=0.4\linewidth]
  {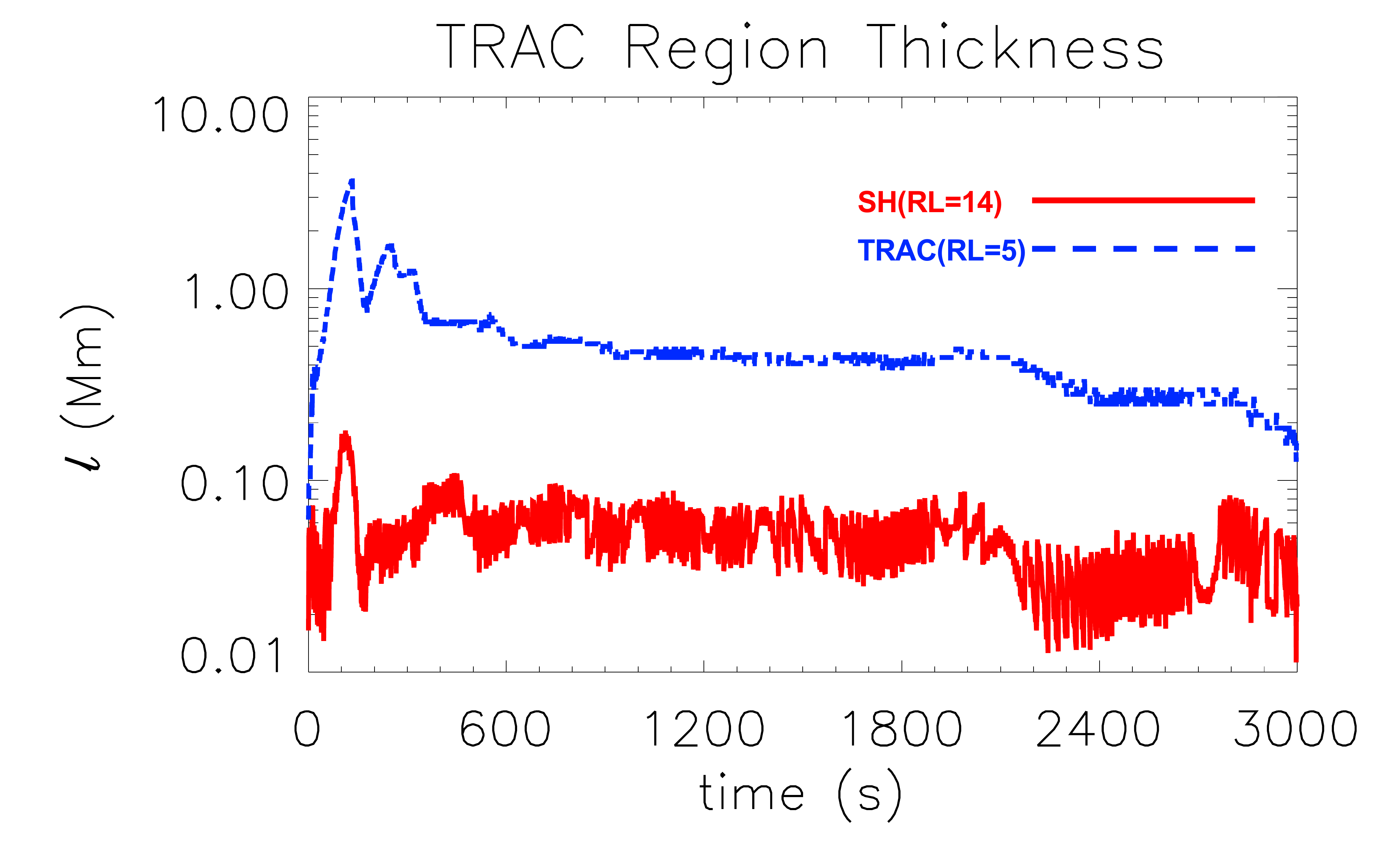}}
  \\[-4mm]
  \hspace*{0.075\linewidth}
  \subfigure{\includegraphics[width=0.4\linewidth]
  {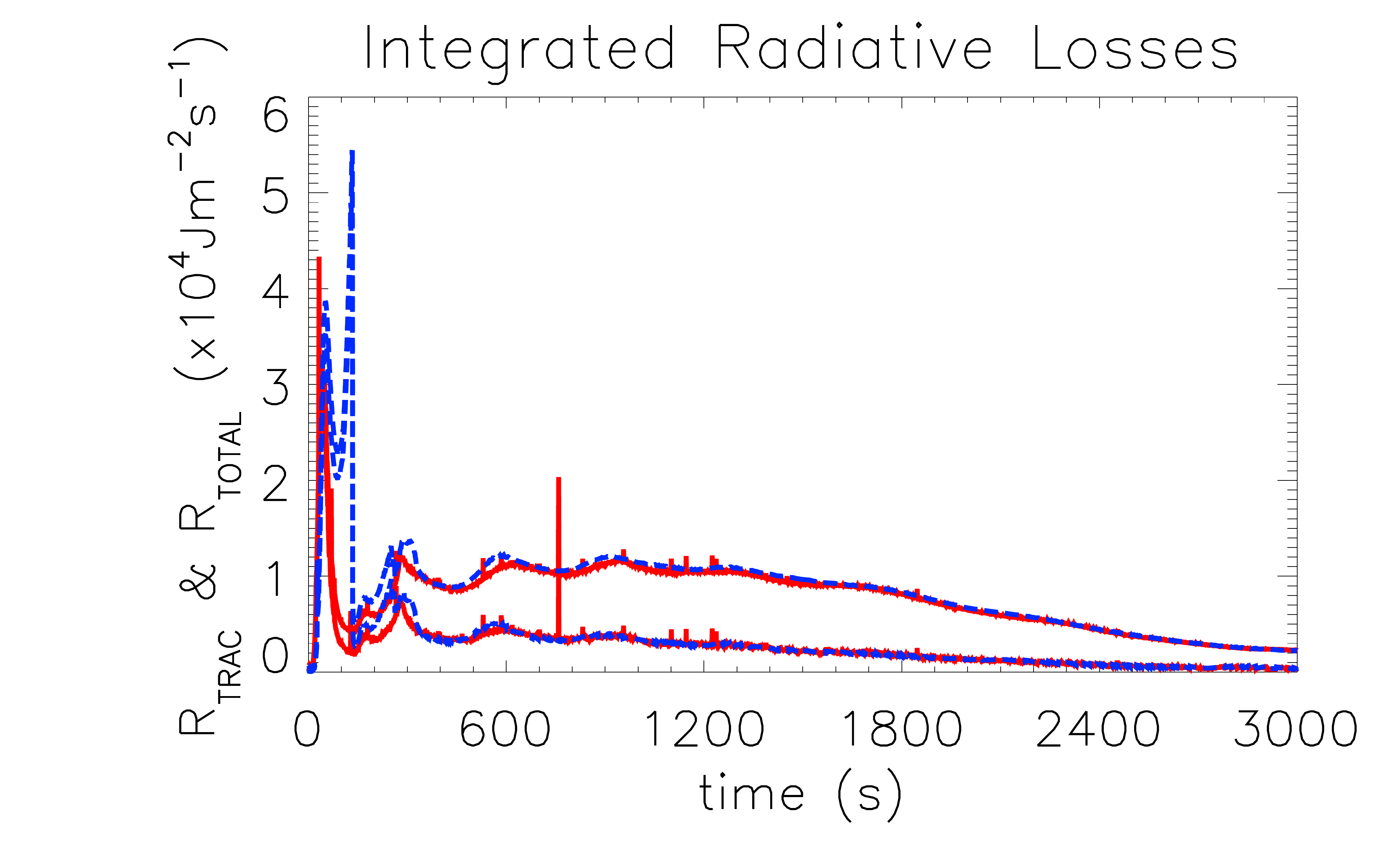}}
  \subfigure{\includegraphics[width=0.4\linewidth]
  {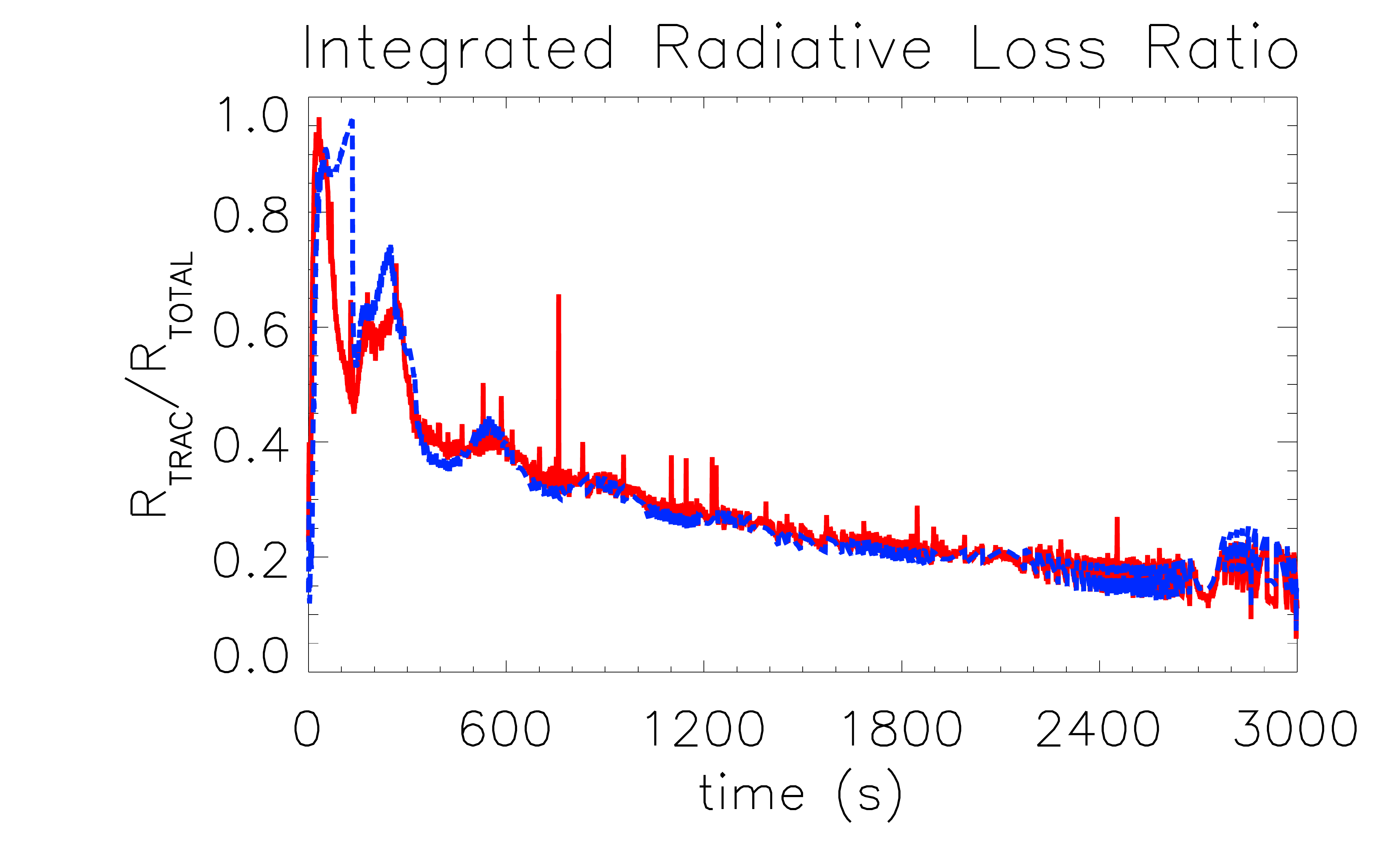}}
  \vspace*{-4mm}
  \caption{Results for the 60~s heating pulse simulations
  (Section \ref{Sect:Short_pulse}).
  Notation is the same as that in Fig.
  \ref{Fig:600s_pulse_RL_5_time_evolution}.
  \label{Fig:60s_pulse_RL_5_time_evolution}
  }
\end{figure*}
  \clearpage
  %
  %
  %%%%%%%%%%%%%%%%%%%%%%%%%%%%%%%%%%%%%%%%%%%%%%%%%%%%%%%%%%%%%    
  %
  % Table:uniform_simulations
  %
  \begin{sidewaystable*}
    \caption{
    \label{Table:uniform_simulations}
    A summary of the parameter space used and results from
    the uniform heating study presented in
    Section \ref{Sect:Uniform_heating}.} 
    \centering
    \resizebox{\hsize}{!}
    {
    \begin{tabular}{lcccccccccc}
    \hline\hline
    \\
    Case & 2L & $Q_H$ & $\tau_H$ &Time&$T$(SH(RL=14))&
    $T$(TRAC(RL=5,[3]))&
    $\Delta T$ &
    $n$(SH)& $n$(TRAC(RL=5,[3]))&
    $\Delta n$
    \\
    (Stage of Evolution) & (Mm) & ($10^{-4}$Jm$^{-3}$s$^{-1}$) 
    &(s)&(s)&(MK)&(MK)&  
    ($\%$)& {($10^{15}$m$^{-3}$)}& {($10^{15}$m$^{-3}$)} & 
    ($\%$)
    \\
    \hline
    \\[-0.8em]
    1 (peak heating) & 60 & 8 & 60 & 30 & 1.77 & 1.77 [1.77]
    &0.00 [0.00] & 0.35	& 0.35 [0.35] & 0.00 [0.00]
    \\
    1 (peak density) & 60 & 8 & 60 & 550 & 1.35 & 1.34 [1.32]
    &	0.74 [2.22] &	0.58&	0.59 [0.60]&	 -1.72 [-3.45]
    \\
    1 (decay) & 60 & 8 & 60 & 900 & 1.14& 1.13 [1.11] & 0.88
    [2.63]&0.58&	0.59 [0.59] &	-1.72 [-1.72]
    \\
    \hline
    \\[-0.8em]
    2 (peak heating) & 60  & 80  &  60 & 30&	6.48&	6.47
    [6.46] &	0.15 [0.31] &	0.35&	0.35 [0.35] &	0.00 
    [0.00] 
    \\
     2 (peak density) & 60  & 80  &  60 & 600&	2.11&	2.10
     [2.01] &	0.47 [4.74] &	1.85&	1.93 [1.88] & -4.32
     [-1.62]
    \\
     2 (decay) & 60  & 80  &  60 & 800&	1.73&	1.70 [1.63]
      &	1.73 [5.78]	&1.71&	1.75 [1.71] &	-2.34 [0.00]
    \\
     \hline
    \\[-0.8em]
   3 (peak heating) & 60  & 800  &  60 & 30&	13.81&	13.77 [13.71] &	0.29 [0.72]&	0.65	&0.65 [0.65] &	0.00 [0.00]
    \\
    3 (peak density) & 60  & 800  &  60 & 450	&4.45&	4.45 [4.35] &0.00 [2.25] &	8.24&	8.80 [8.70] &	-6.80 [-5.58]
    \\
    3 (decay) & 60  & 800  &  60 & 800&	3.06&	3.02 [2.95] &	1.31 [3.59]&	7.26&	7.63 [7.52] &	-5.10 [-3.58] 
    \\
     \hline
    \\[-0.8em]
   4 (peak heating) & 60 & 8 & 600 & 300	&3.79 &3.76 [3.73] &	0.79 [1.58]&	0.79&	0.80 [0.80]&	-1.27 [-1.27]
    \\
    4 (peak density) & 60 & 8 & 600 & 900&2.17&	2.16	 [2.12]&0.46  [2.30]	&1.87&	1.89 [1.90] &	-1.07 [-1.60]
    \\
    4 (decay) & 60 & 8 & 600 & 1700&	1.05	&1.04 [1.01]&	0.95 [3.81]	&1.42&	1.44 [1.42]  &	-1.41 [0.00]
    \\
     \hline
    \\[-0.8em]
    5 (peak heating) & 60  & 80  &  600 & 300&	7.65&	7.57 [7.46] &	1.05 [2.48] &	3.45&	3.56 [3.57]  &	-3.19 [-3.48]
    \\
    5 (peak density) & 60  & 80  &  600 & 800	&4.41&	4.40 [4.32] &	0.23 [2.04] &	8.22&	8.48 [8.51]  &	-3.16 [-3.53]
    \\
    5 (decay) & 60  & 80  &  600 & 1400&	2.29&	2.27 [2.17] &	0.87 [5.24] &	6.26&	6.36 [6.19] &	-1.60 [1.12]
    \\
     \hline
    \\[-0.8em]
   6 (peak heating) & 60  & 800  &  600 & 300 & 15.04&	14.88 
   [14.65] & 1.06 [2.59]  &	16.95&	17.78 [17.78]&	-4.90 [-4.90]
    \\
    6 (peak density) & 60  & 800  &  600 & 700&	9.19	&9.12 [8.97] &0.76 [2.39]  &	36.97	&38.77 [38.82] &	-4.87 [-5.00]
    \\
    6 (decay) & 60  & 800  &  600 & 1300&3.43	&3.28 [3.20] 	&4.37 [6.71] 	&25.69&	25.91 [25.55]	&-0.86 [0.54]
    \\
    \hline
    \\[-0.8em]
    7 (peak heating) & 180 & 0.5   &  60  &  30&	1.41 &	1.41	 [1.41] &0.00 [0.00]	&0.068 	&0.068 [0.068] &0.00 [0.00]
    \\
    7 (peak density) & 180 & 0.5   &  60  &  750  &  1.34  &  1.34 [1.34] & 0.00 [0.00]   &  0.084  &  0.084 [0.085]   & 0.00 [-1.19] 
    \\
     7 (decay) & 180 & 0.5   &  60  &  2000	&  1.19	&  1.19
     [1.19]	& 0.00 [0.00] &  0.083	&  0.083 [0.084] 	& 0.00 
     [-1.20]
    \\
     \hline
    \\[-0.8em]
   8 (peak heating) & 180 & 5   &  60  &  30	&  3.60	&  3.60
   [3.60]& 0.00 [0.00]	&  0.068	&  0.068 [0.068]	& 0.00 [0.00]
    \\
    8 (peak density) & 180 & 5   &  60  &  525	&  2.31	&  2.31 [2.28]& 0.00 [1.30] &  0.20	&  0.20 [0.20] 	& 0.00 [0.00]
    \\
    8 (decay) & 180 & 5   &  60  &  2500	&  1.33	&  1.32
    [1.30]	& 0.75 [2.26] 	&  0.17	&  0.17 [0.17] 	& 0.00 [0.00]
    \\
     \hline
    \\[-0.8em]
   9 (peak heating) & 180 & 50   &  60  &  30	&  8.06	&  8.06	 [8.06]& 0.00 [0.00] &  0.26	&  0.26 [0.26]& 0.00 [0.00] 
     \\
     9 (peak density & 180 & 50   &  60  &  1000	&  3.54	&  3.54 [3.50]	& 0.00 [1.13] 	&  0.96	&  0.98 [0.98]	& -2.08 [-2.08]
     \\
     9 (decay) & 180 & 50   &  60  &  3000	& 2.02	& 2.01
     [1.97]	&0.50 [2.48] 	& 0.68& 0.69 [0.67] 	&-1.47 [1.47]
     \\
       \hline
    \\[-0.8em]
   10 (peak heating) & 180 & 0.5   &  600  &  300	& 2.87	& 2.87 [2.87]&0.00 [0.00] 	& 0.071	& 0.071 [0.071] 	&0.00 [0.00]
    \\
    10 (peak density) & 180 & 0.5   &  600  &  900	& 2.22	& 2.22	 [2.21]&0.00 [0.45]& 0.19	& 0.19 [0.19] &0.00 [0.00]
    \\
    10 (decay) & 180 & 0.5   &  600  &  3500	& 1.18	& 1.18 
    [1.17]	&0.00 [0.85] 	& 0.14& 	0.14 [0.14]  	&0.00 [0.00] 
    \\
      \hline
    \\[-0.8em]
   11 (peak heating) & 180 & 5   &  600  &  300	&  6.42	&  6.40 [6.38] 	& 0.31 [0.62] 	&  0.15	&  0.15 [0.15] 	& 0.00 [0.00] 
    \\
    11 (peak density) & 180 & 5   &  600  &  1350 & 3.33	& 3.33
    [3.31] & 0.00 [0.60]	& 0.82	& 0.83 [0.83]	&-1.22 [-1.22]
    \\
    11 (decay) & 180 & 5   &  600  &  4000	& 1.37	& 1.37 
    [1.35]	&0.00 [1.46] & 	0.51& 	0.52 [0.51]&  -1.96	[0.00] 
    \\
      \hline
    \\[-0.8em]
   12 (peak heating) & 180 & 50   &  600  &  300	&  13.59	&  13.56 [13.51]&  	0.52 [0.59] &  	0.94 &  	0.96 [0.95]  & 	-2.13
   [-1.06]
    \\
    12 (peak density) & 180 & 50   &  600  &  1500 & 5.69 &	5.71 [5.67]&	-0.35 [0.35] &4.00 &	4.09 [4.08] &	-2.25 [-2.00]
    \\
    12 (decay) & 180 & 50   &  600  &  4000	& 2.01	& 2.01
    [1.98] &0.00 [1.49]	& 2.07& 	2.11 [2.07]&-1.93 [0.00]
    \\
    \hline
    \end{tabular}
    }
    \tablefoot{
    From left to right
    the columns show 
    the case number and stage of evolution,
    the total length of the loop, 
    the maximum heating rate,
    the duration of the heating pulse, 
    the sample time,
    the coronal averaged temperature and density
    attained by the properly resolved SH solution
    obtained with 61~m grid cells (RL=14)
    and the TRAC simulation computed with 31.25~km grid cells 
    (RL=5) 
    at the sample time
    and their 
    respective normalised differences.
    We note that the corresponding results 
    when the TRAC simulations are computed with 125~km grid 
    cells (RL=3)  are shown in square brackets. 
    }
  \end{sidewaystable*}
  %
  %
  %
  %
  %%%%%%%%%%%%%%%%%%%%%%%%%%%%%%%%%%%%%%%%%%%%%%%%%%%%%%%%%%%%%    
  %
  % Fig:fp2_Case_600s_RL_5_time_evolution
  %
\begin{figure*}
  \hspace*{0.075\linewidth}
  \subfigure{\includegraphics[width=0.4\linewidth]
  {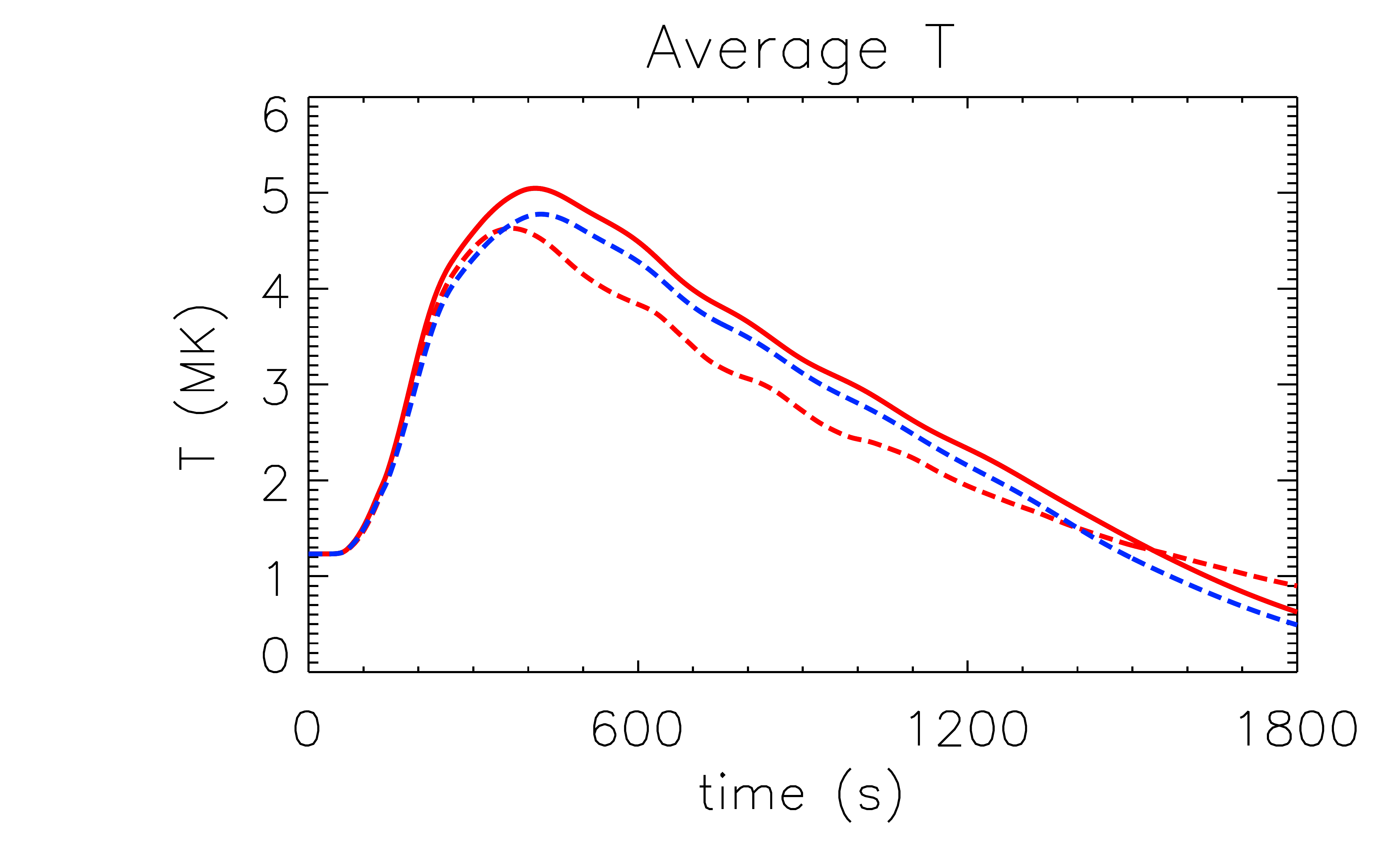}}
  \subfigure{\includegraphics[width=0.4\linewidth]
  {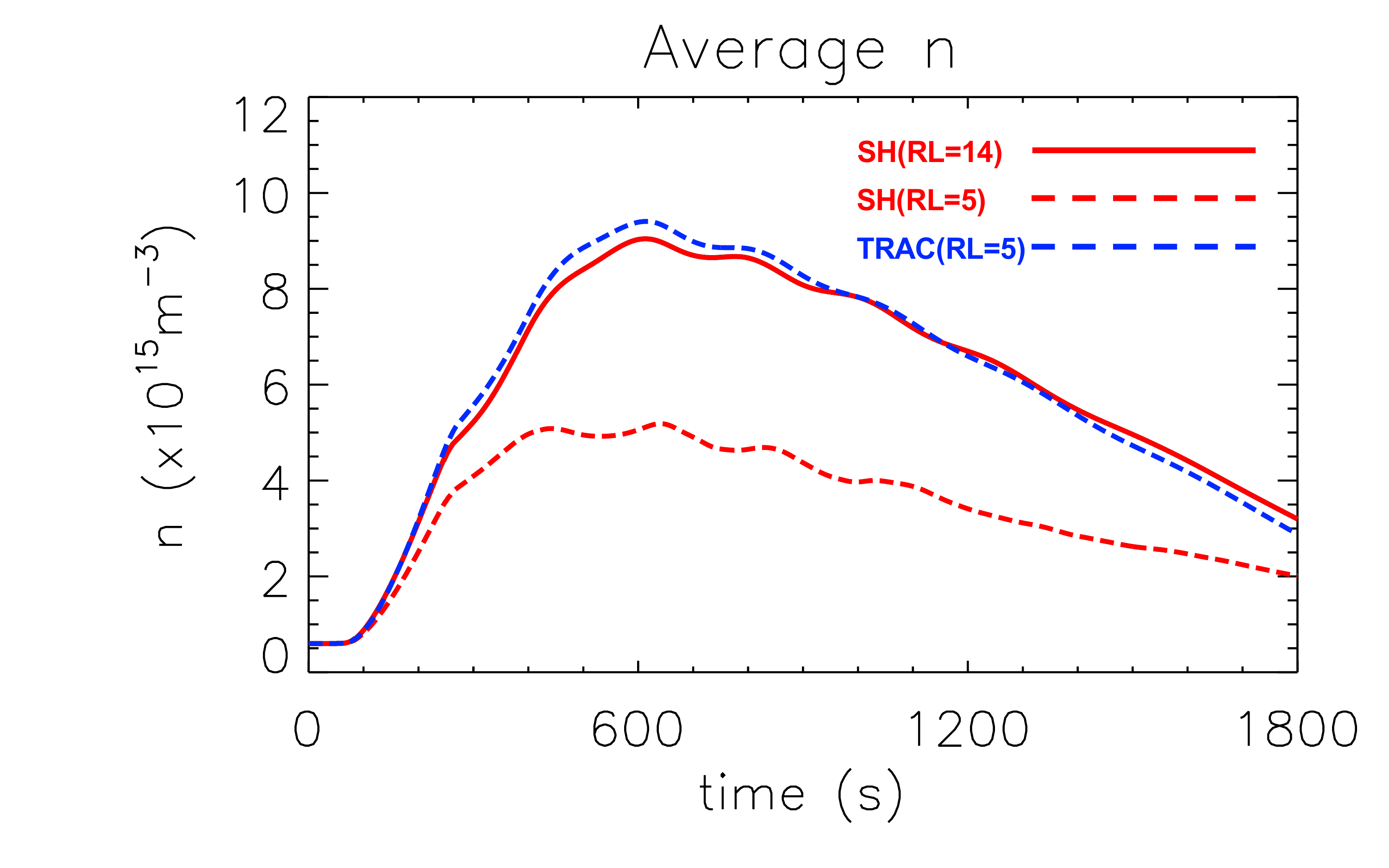}}
  \\[-4mm]
  \hspace*{0.075\linewidth}
  \subfigure{\includegraphics[width=0.4\linewidth]
  {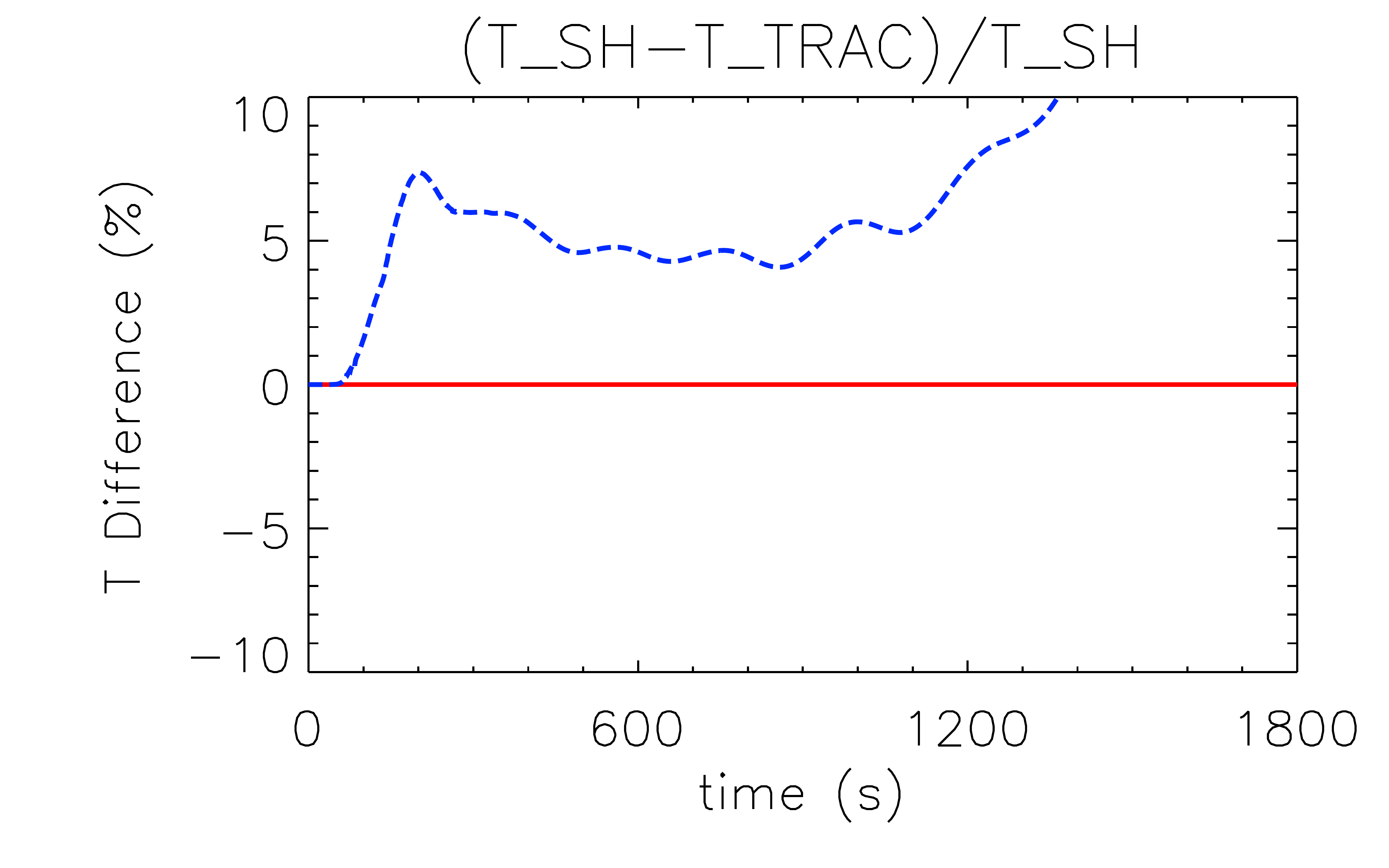}}
  \subfigure{\includegraphics[width=0.4\linewidth]
  {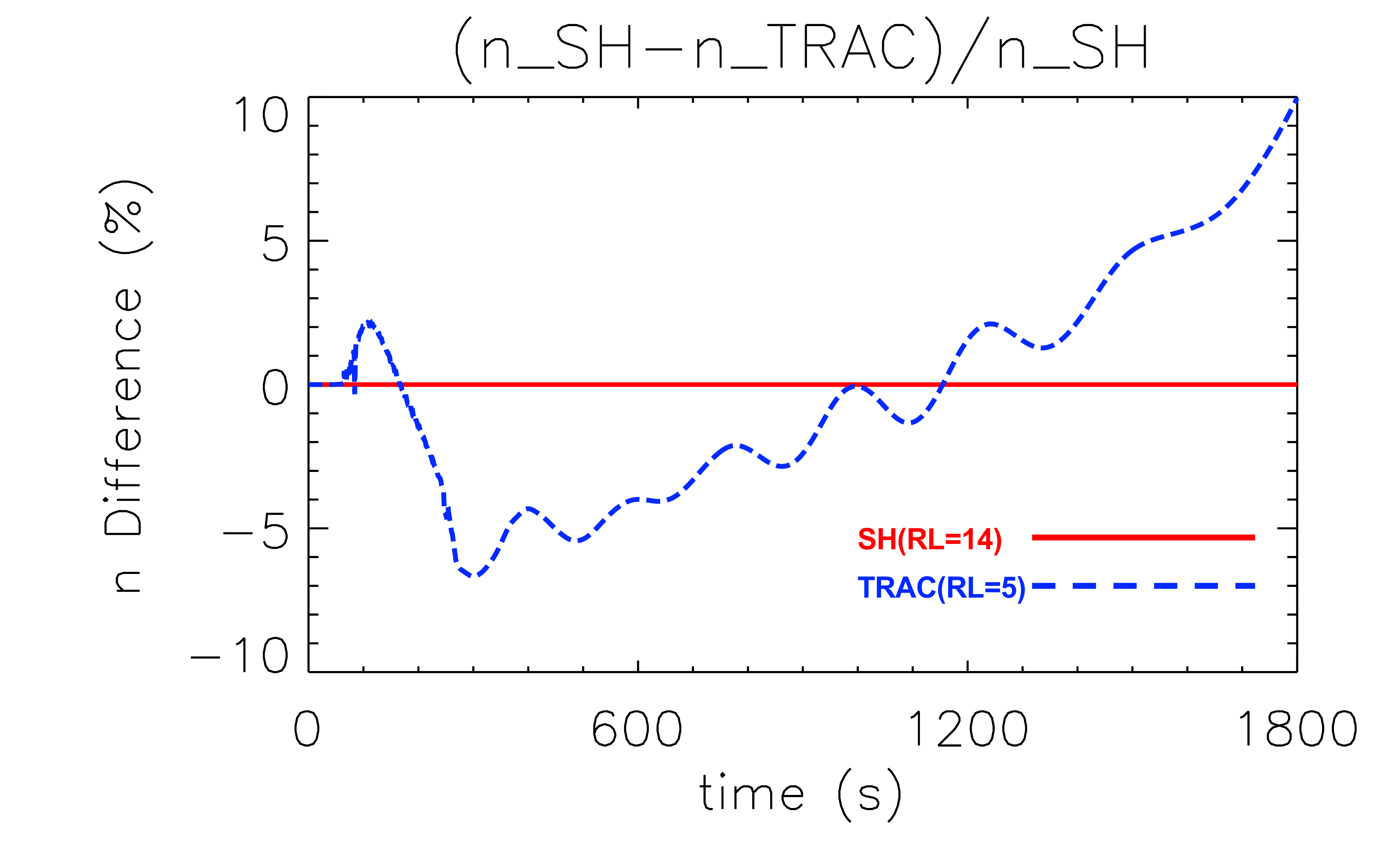}}
  \\[-4mm]
  \hspace*{0.075\linewidth}
  \subfigure{\includegraphics[width=0.4\linewidth]
  {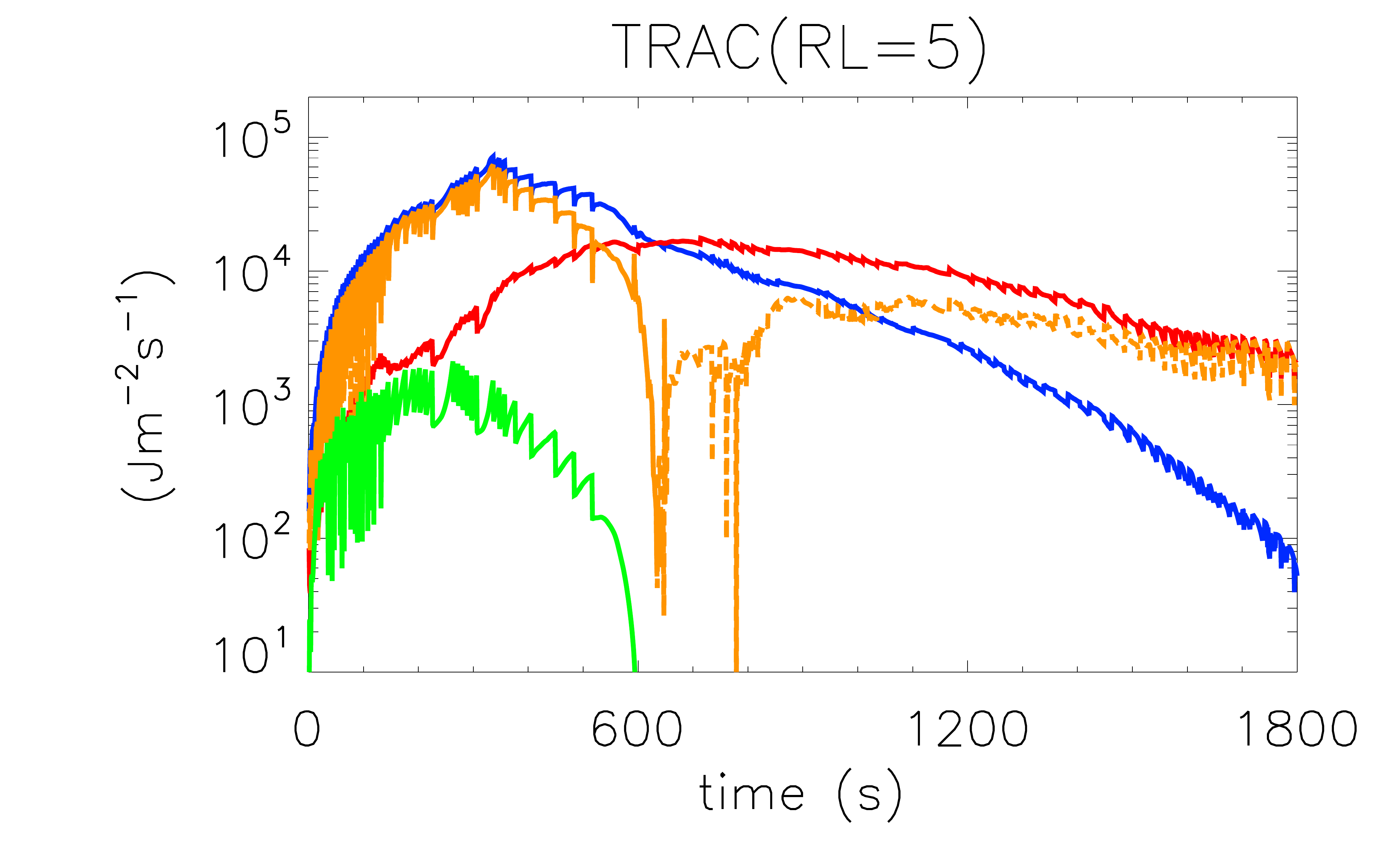}}
  \subfigure{\includegraphics[width=0.4\linewidth]
  {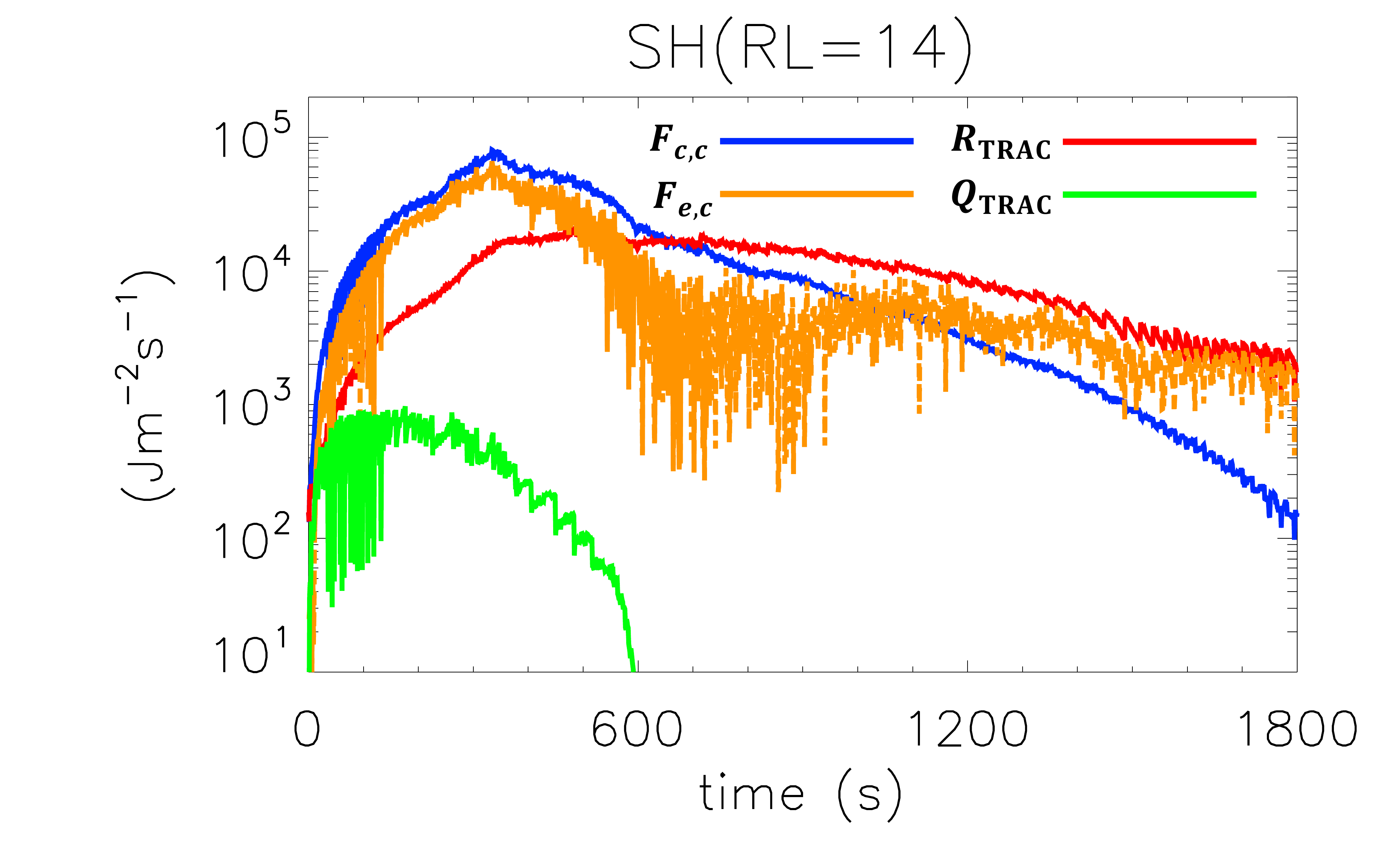}}
  \\[-4mm]
  \hspace*{0.075\linewidth}
  \subfigure{\includegraphics[width=0.4\linewidth]
  {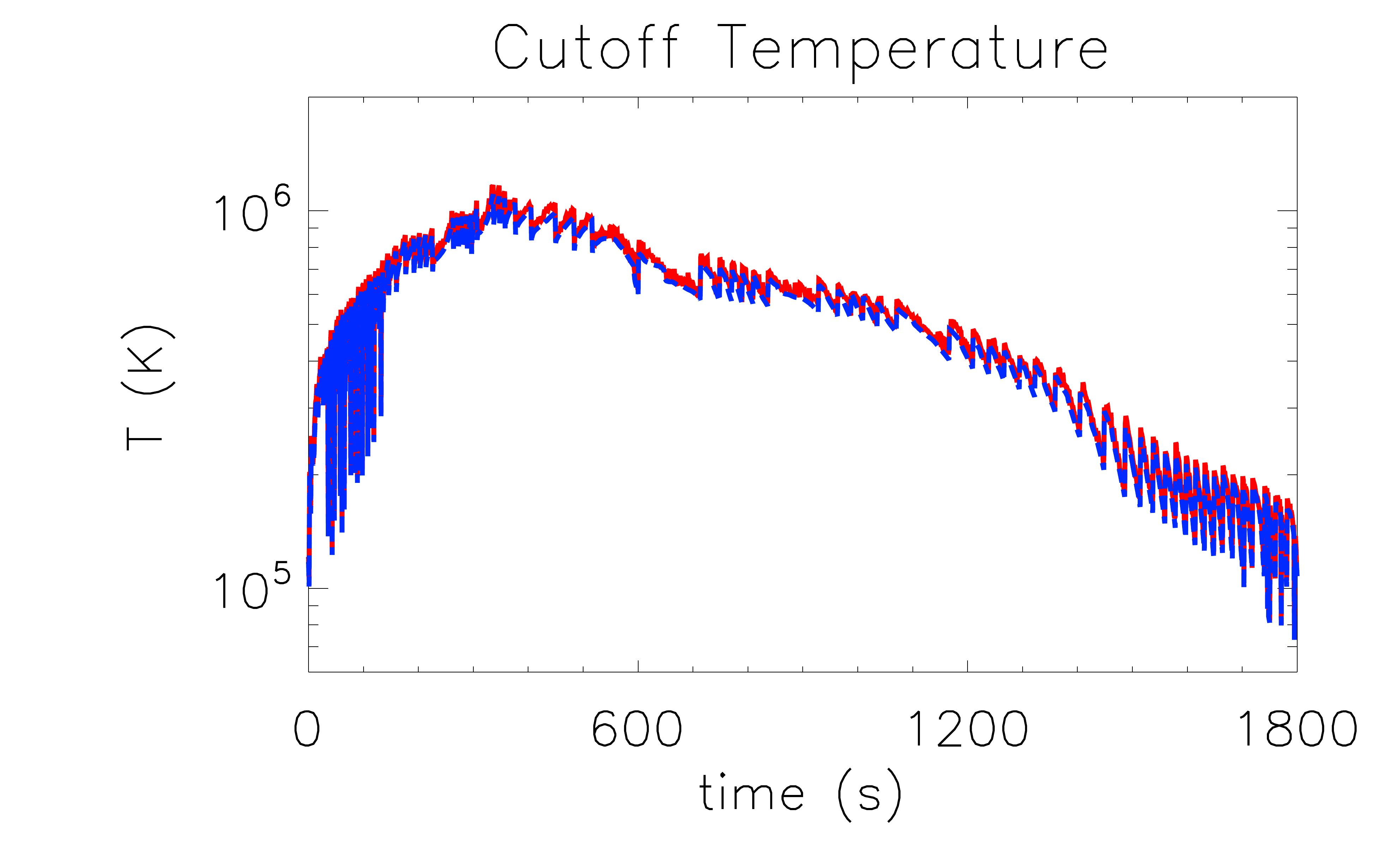}}
  \subfigure{\includegraphics[width=0.4\linewidth]
  {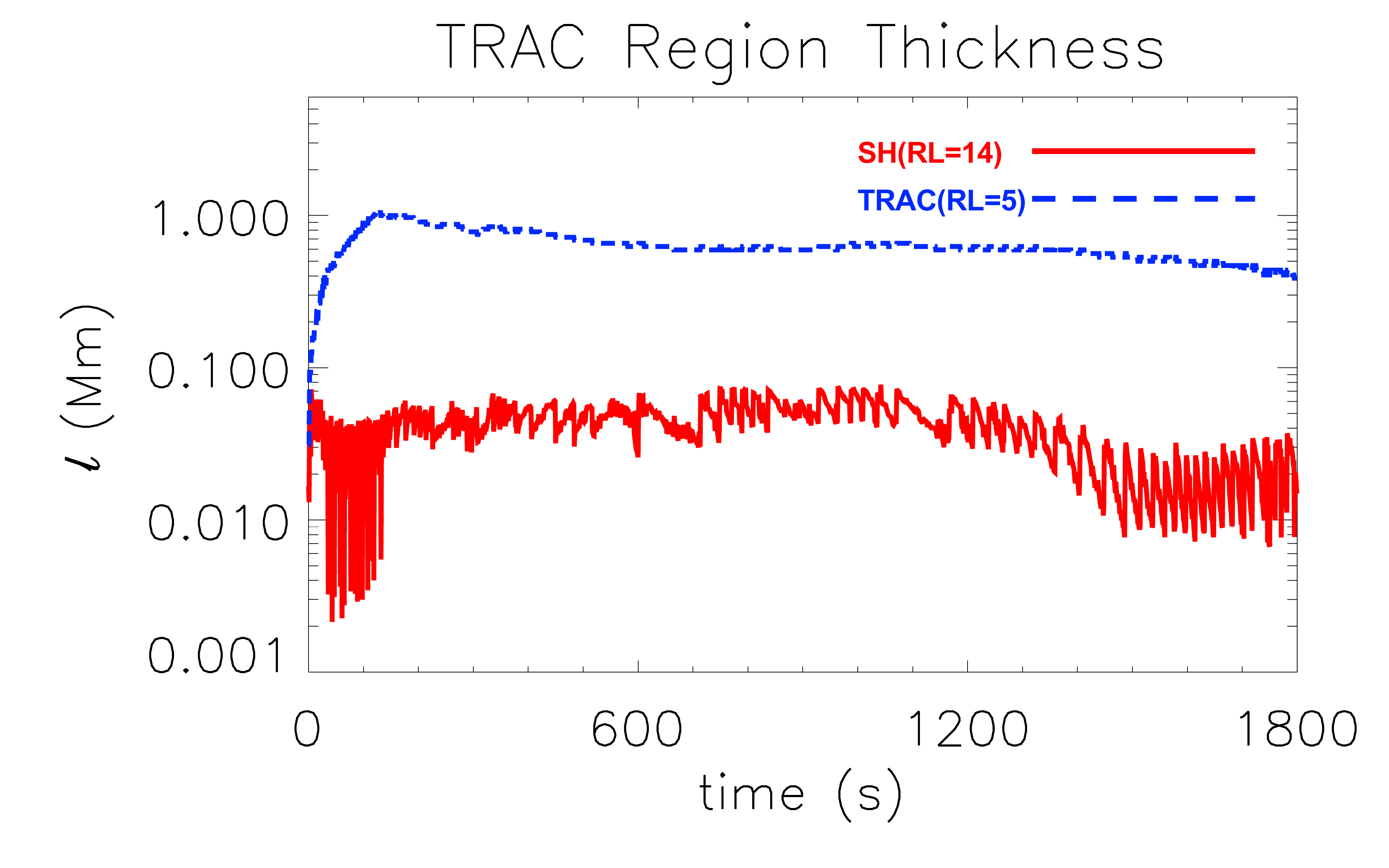}}
  \\[-4mm]
  \hspace*{0.075\linewidth}
  \subfigure{\includegraphics[width=0.4\linewidth]
  {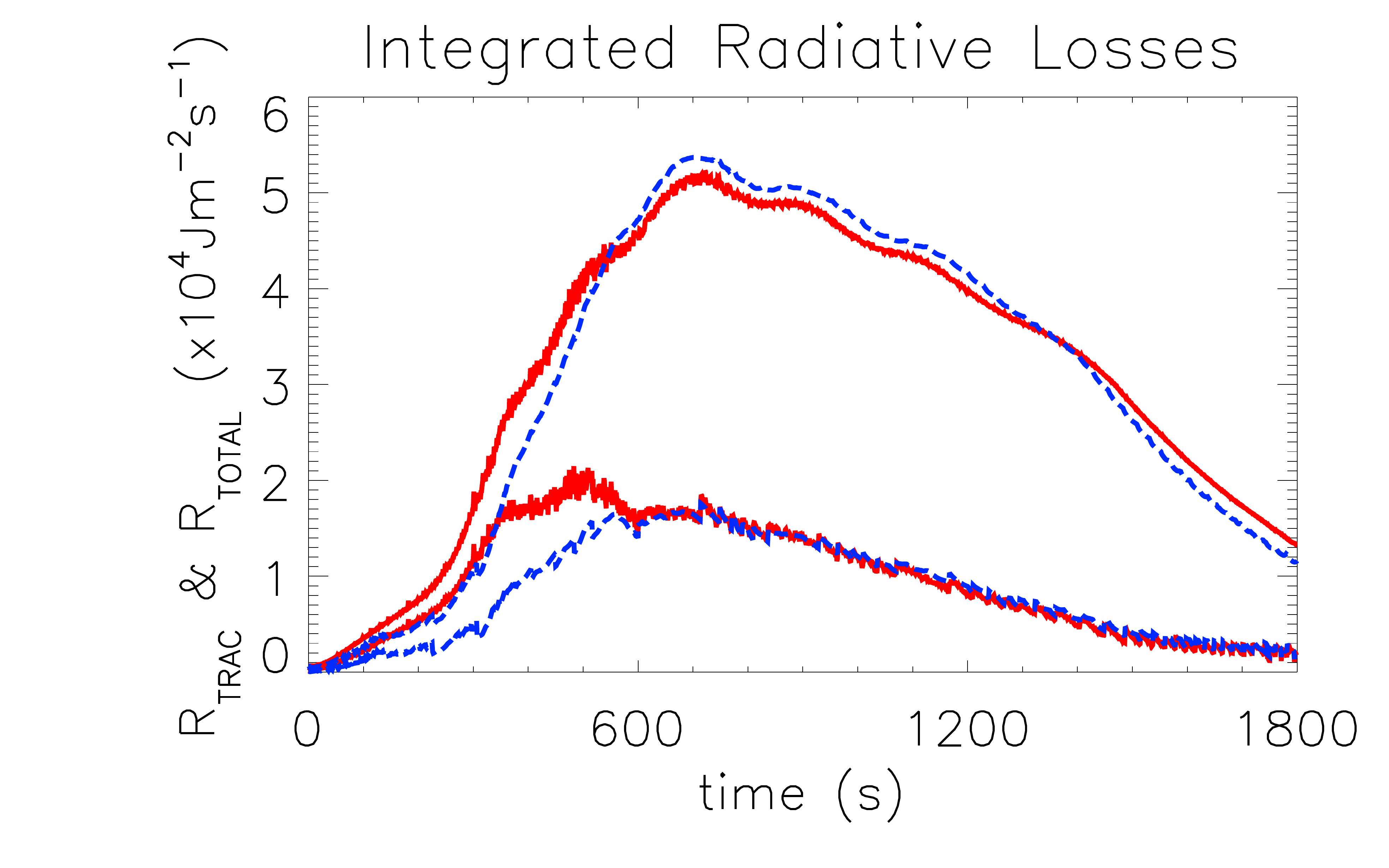}}
  \subfigure{\includegraphics[width=0.4\linewidth]
  {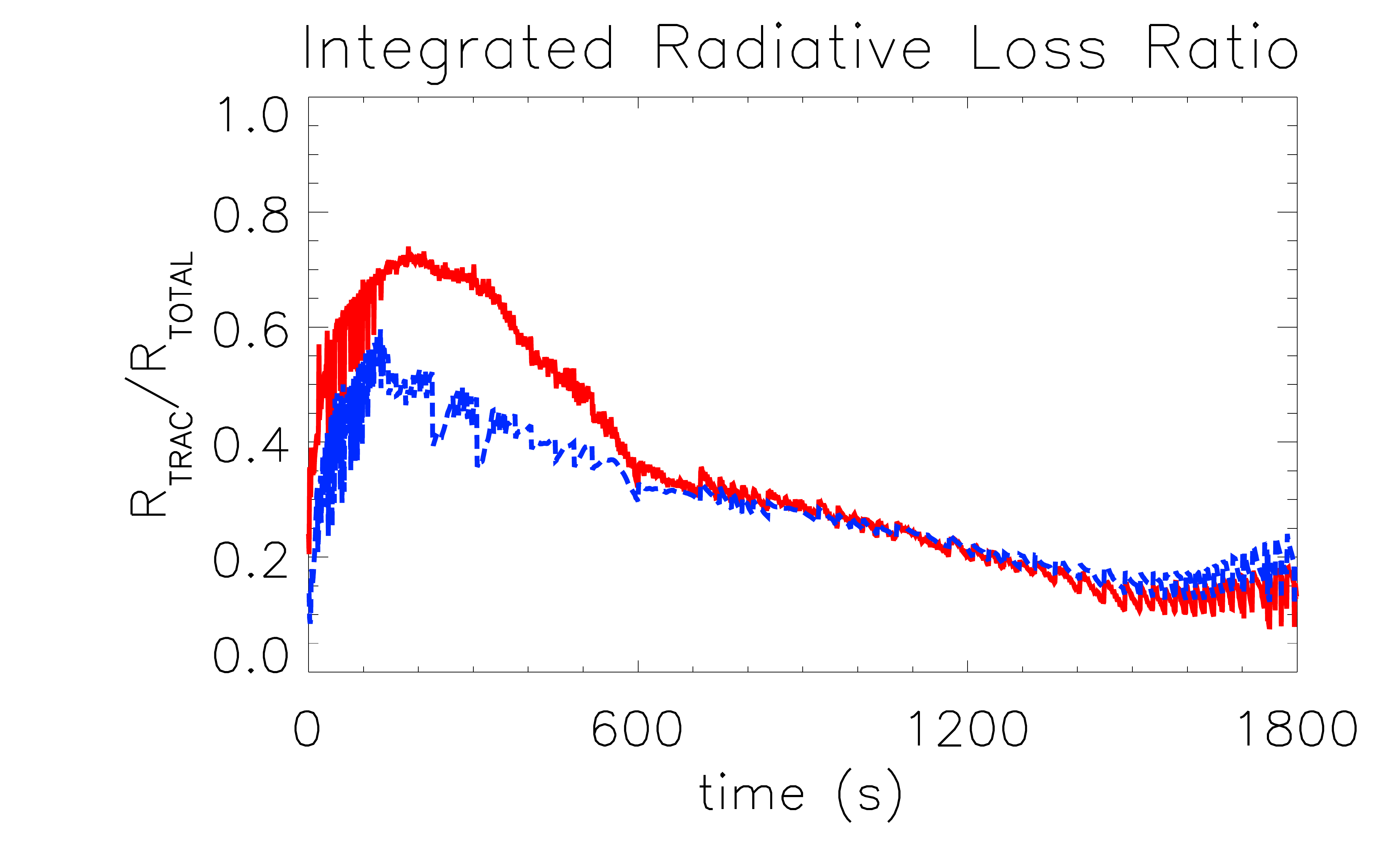}}
  \vspace*{-4mm}
  \caption{Results for the 600~s footpoint heating
  simulations (Section 
  \ref{Sect:Impulsive_footpoint_heating}).
  Notation is the same as that in 
  Fig.~\ref{Fig:600s_pulse_RL_5_time_evolution}.
  \label{Fig:fp2_Case_600s_RL_5_time_evolution}
  }
\end{figure*}
  %
  %
  %
  %
  %%%%%%%%%%%%%%%%%%%%%%%%%%%%%%%%%%%%%%%%%%%%%%%%%%%%%%%%%%%%%    
  %
  % Fig:fp2_Case_60s_RL_5_time_evolution
  %
\begin{figure*}
  \hspace*{0.075\linewidth}
  \subfigure{\includegraphics[width=0.4\linewidth]
  {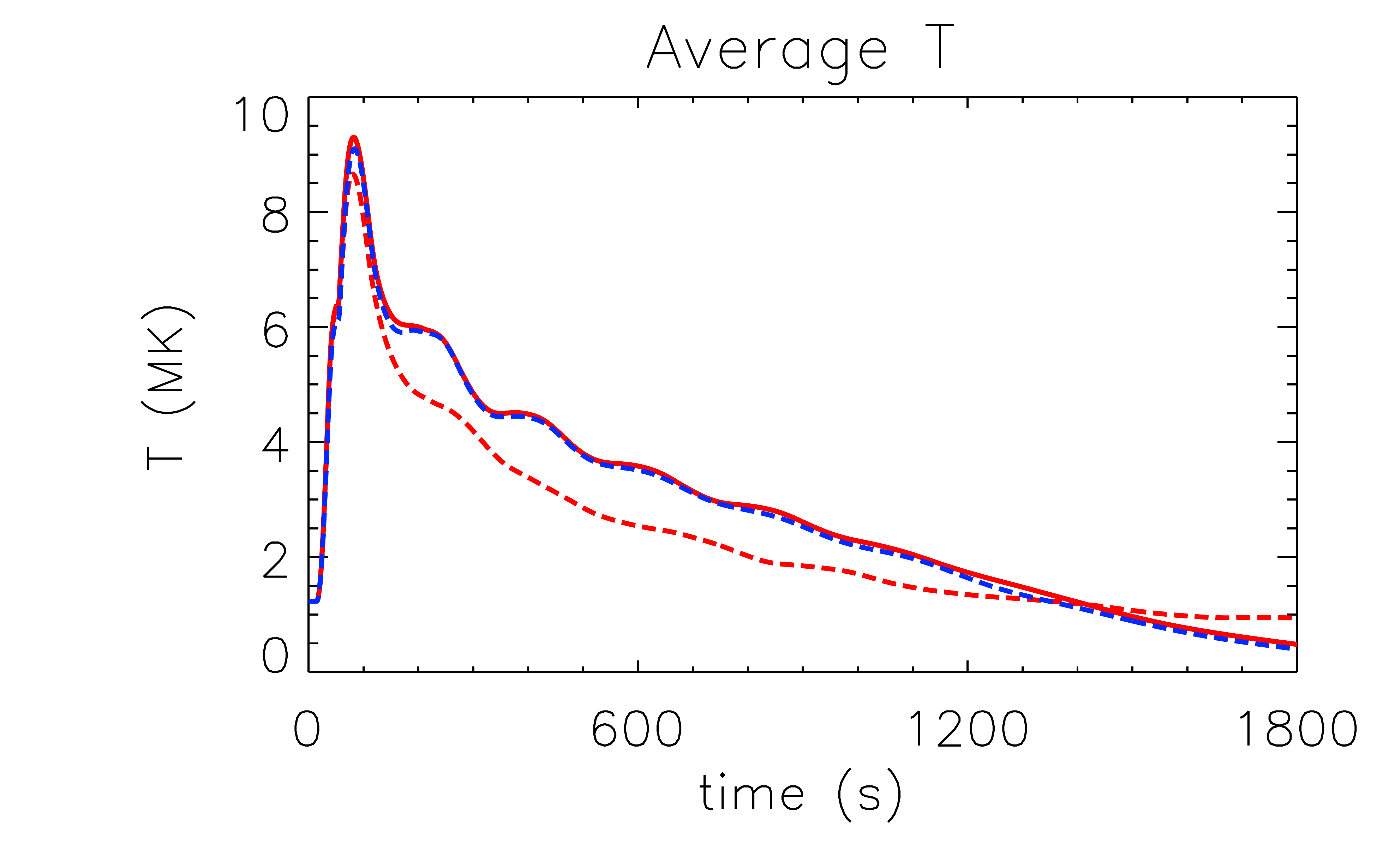}}
  \subfigure{\includegraphics[width=0.4\linewidth]
  {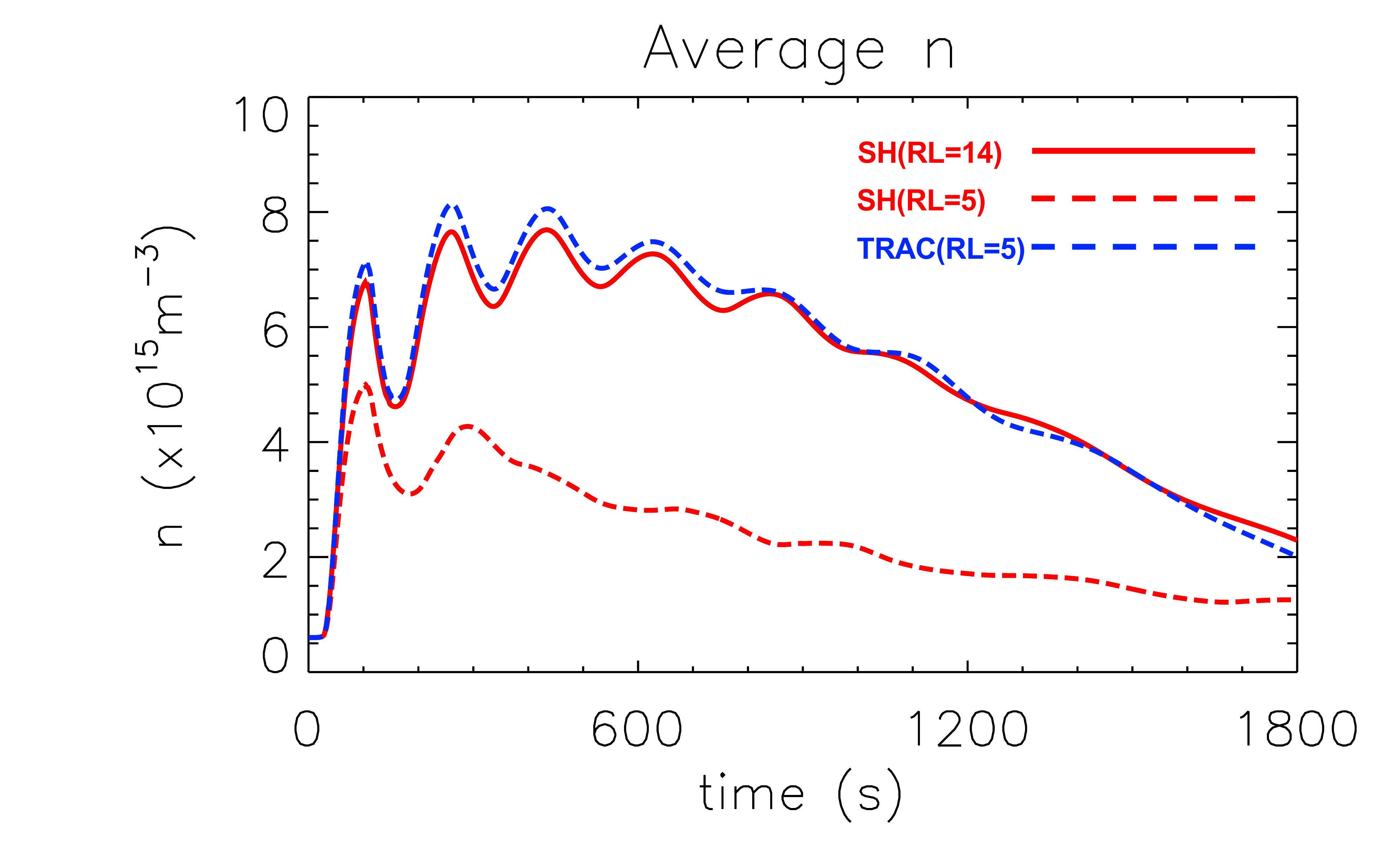}}
  \\[-4mm]
  \hspace*{0.075\linewidth}
  \subfigure{\includegraphics[width=0.4\linewidth]
  {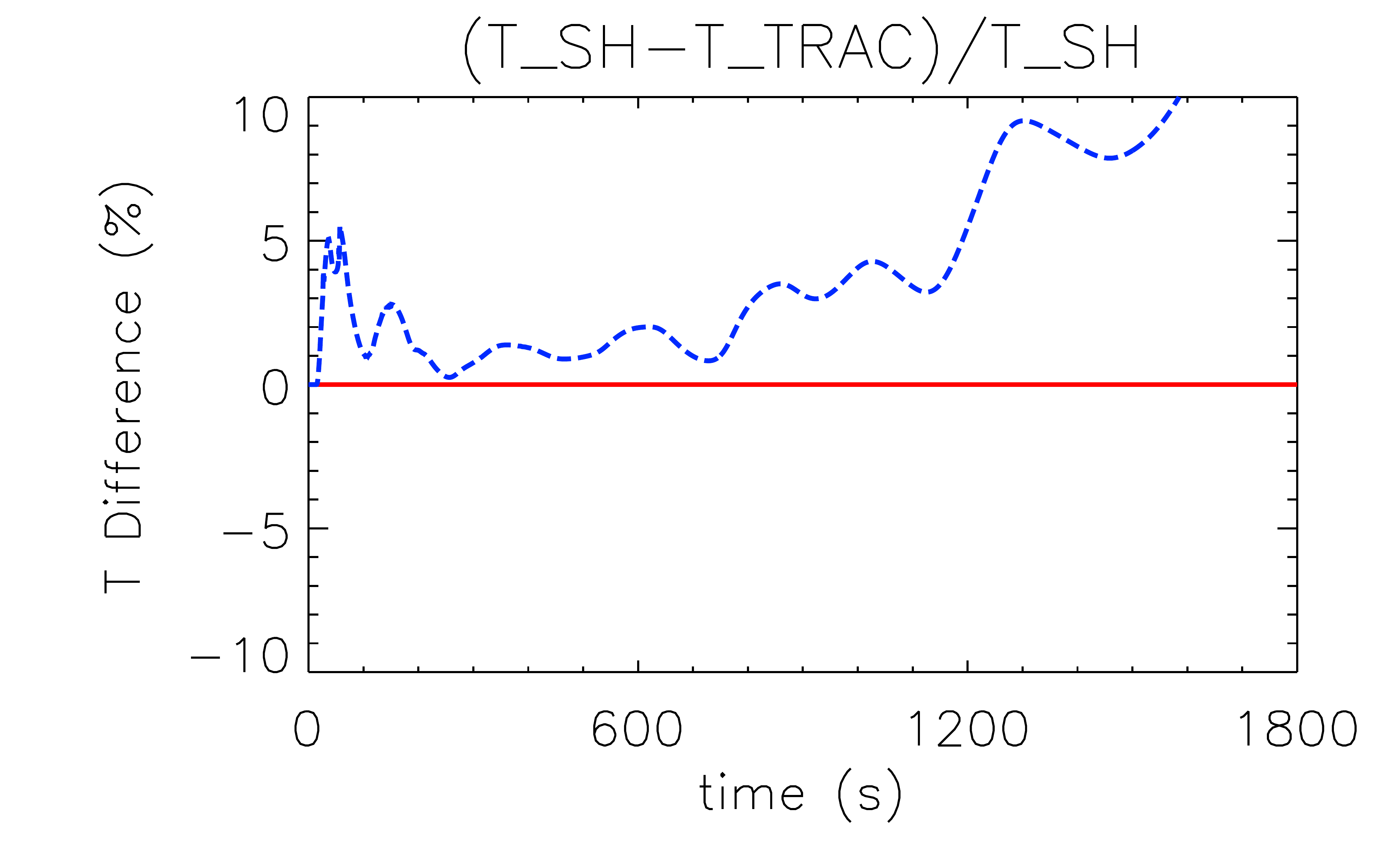}}
  \subfigure{\includegraphics[width=0.4\linewidth]
  {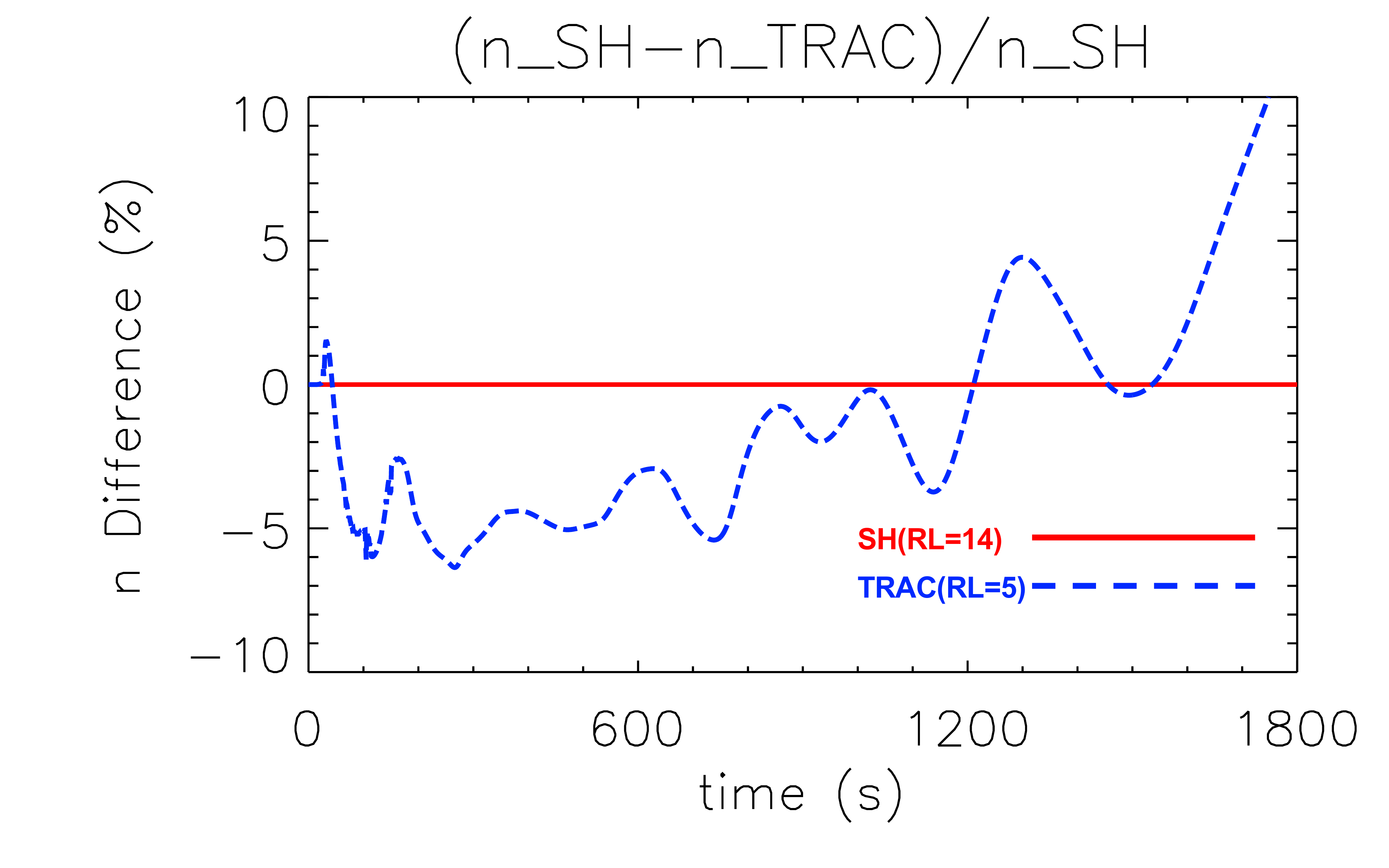}}
  \\[-4mm]
  \hspace*{0.075\linewidth}
  \subfigure{\includegraphics[width=0.4\linewidth]
  {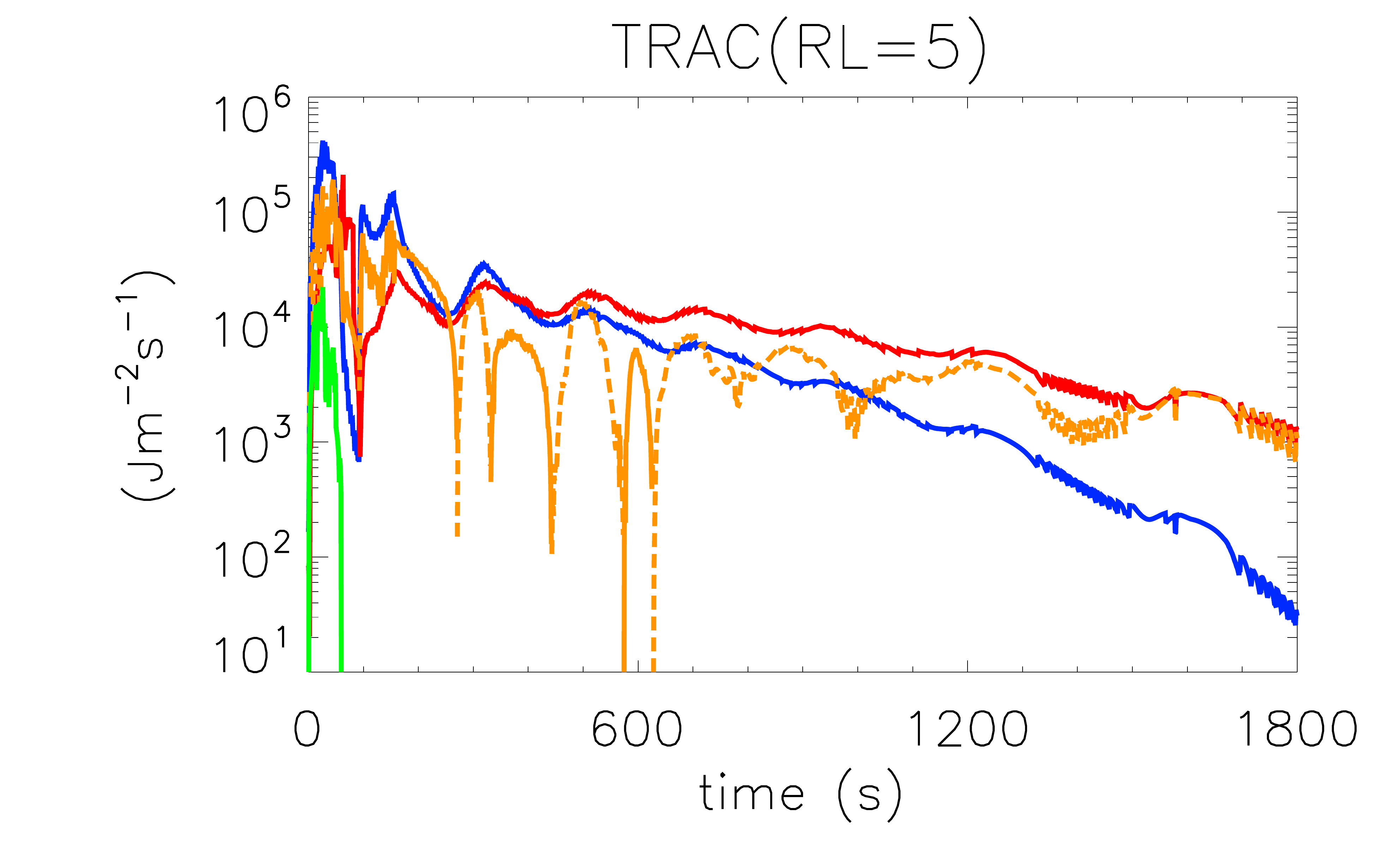}}
  \subfigure{\includegraphics[width=0.4\linewidth]
  {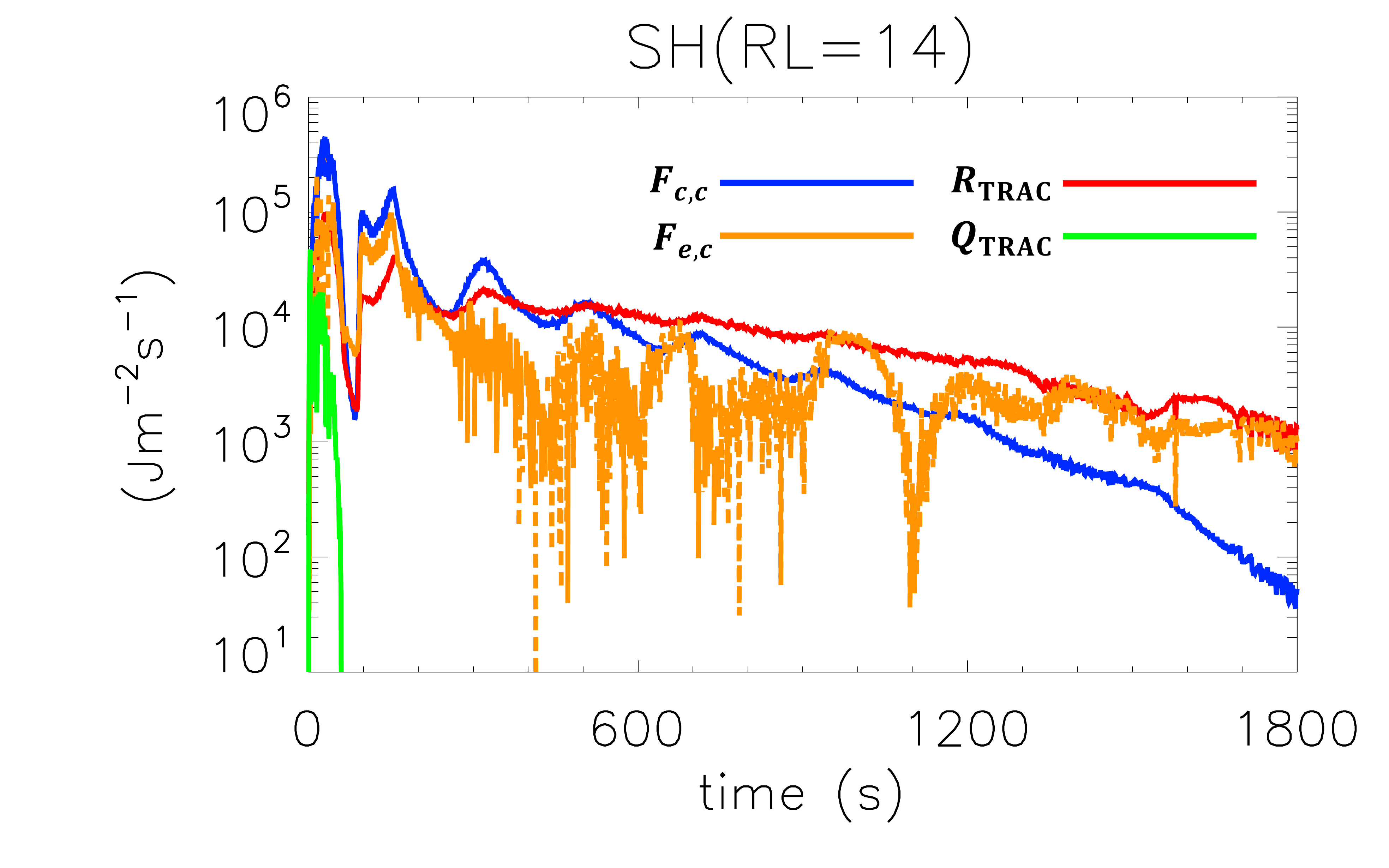}}
  \\[-4mm]
  \hspace*{0.075\linewidth}
  \subfigure{\includegraphics[width=0.4\linewidth]
  {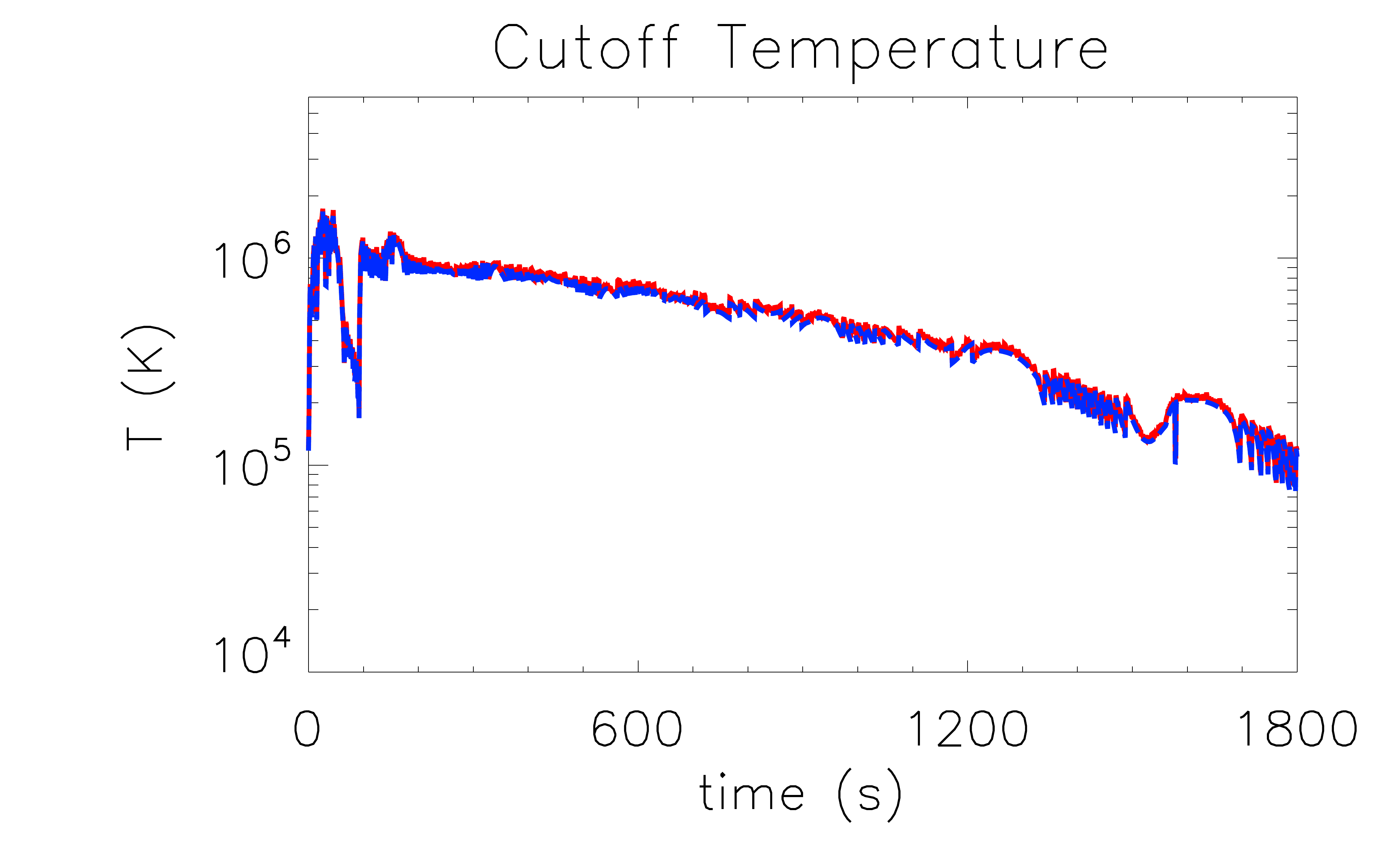}}
  \subfigure{\includegraphics[width=0.4\linewidth]
  {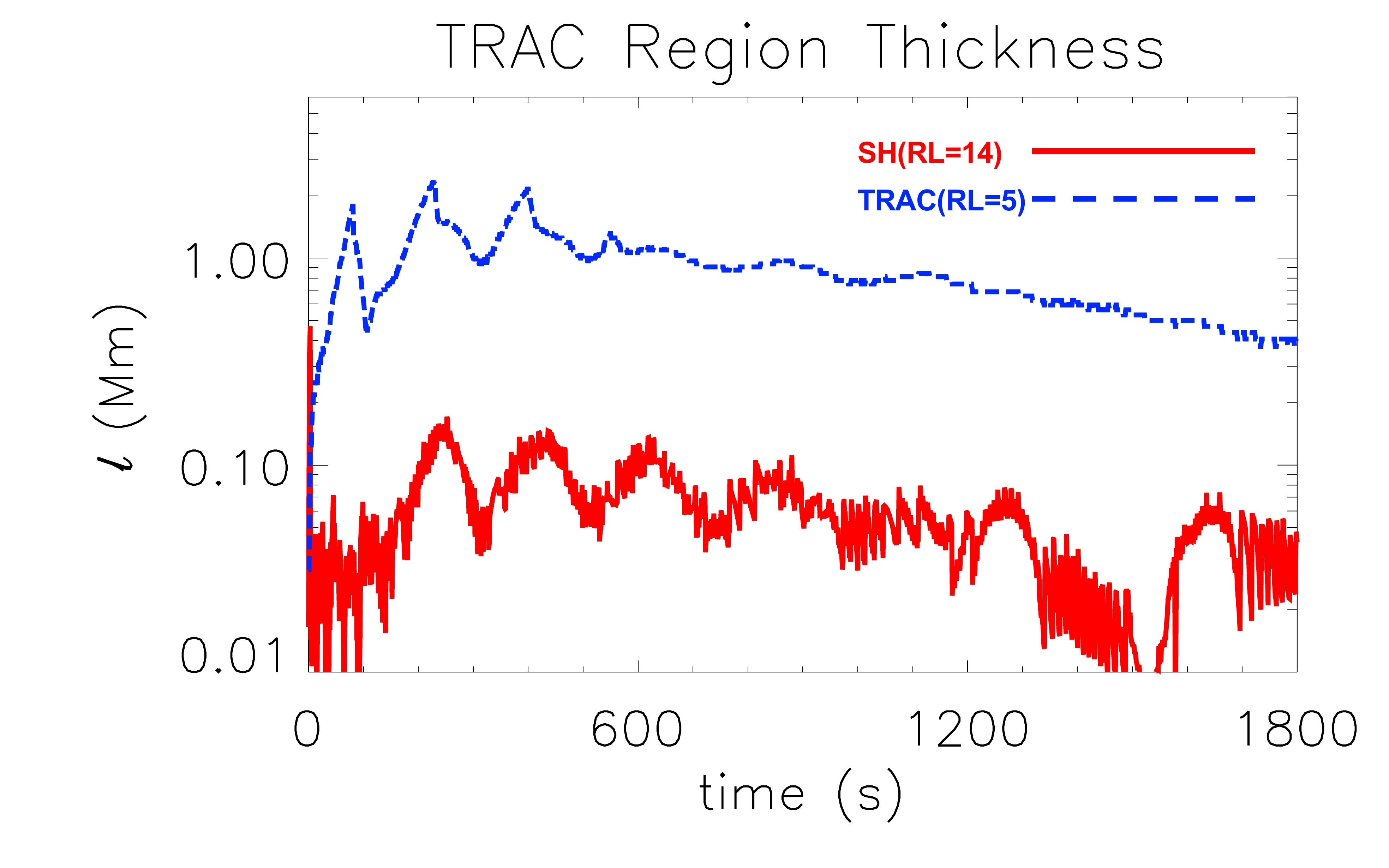}}
  \\[-4mm]
  \hspace*{0.075\linewidth}
  \subfigure{\includegraphics[width=0.4\linewidth]
  {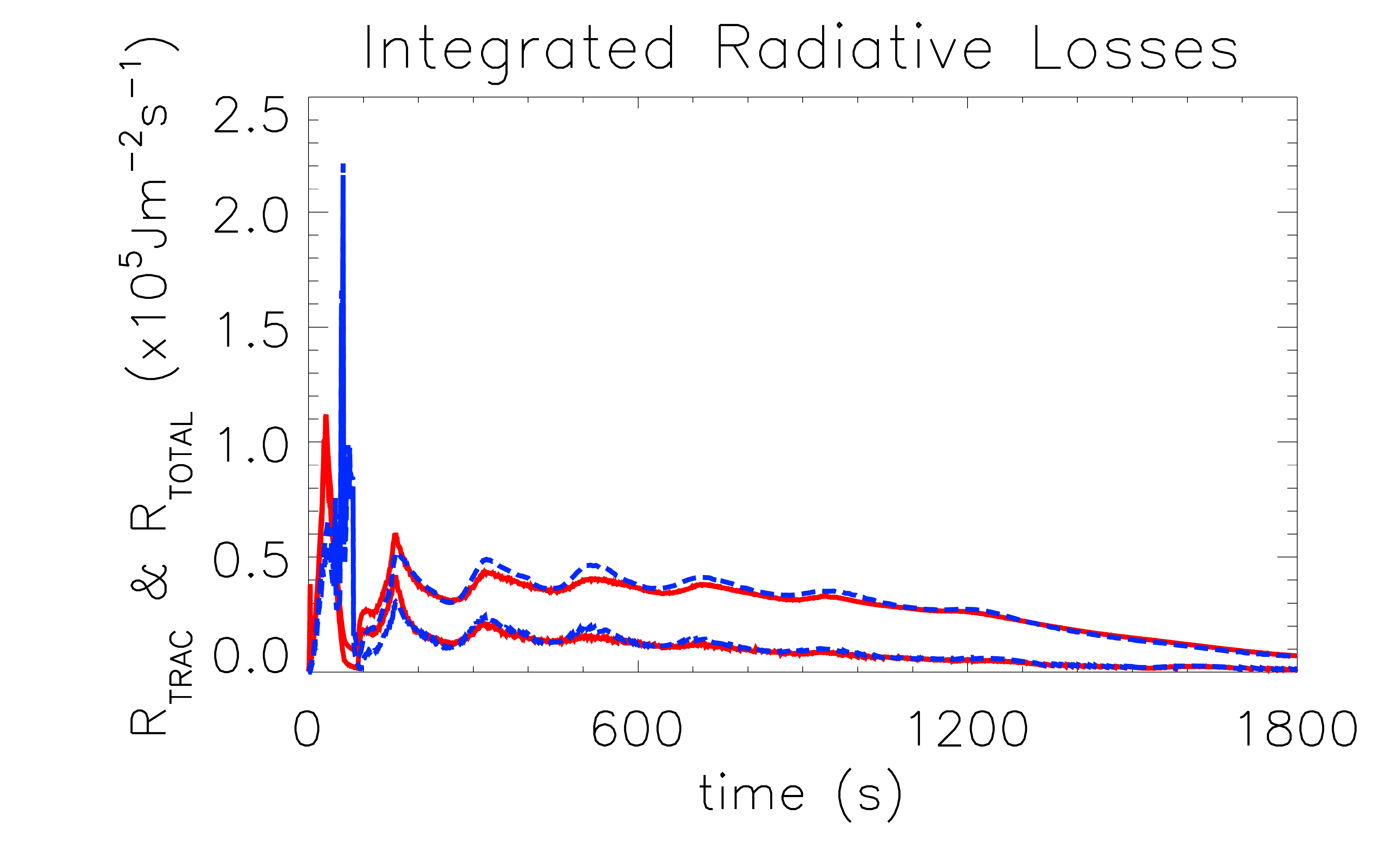}}
  \subfigure{\includegraphics[width=0.4\linewidth]
  {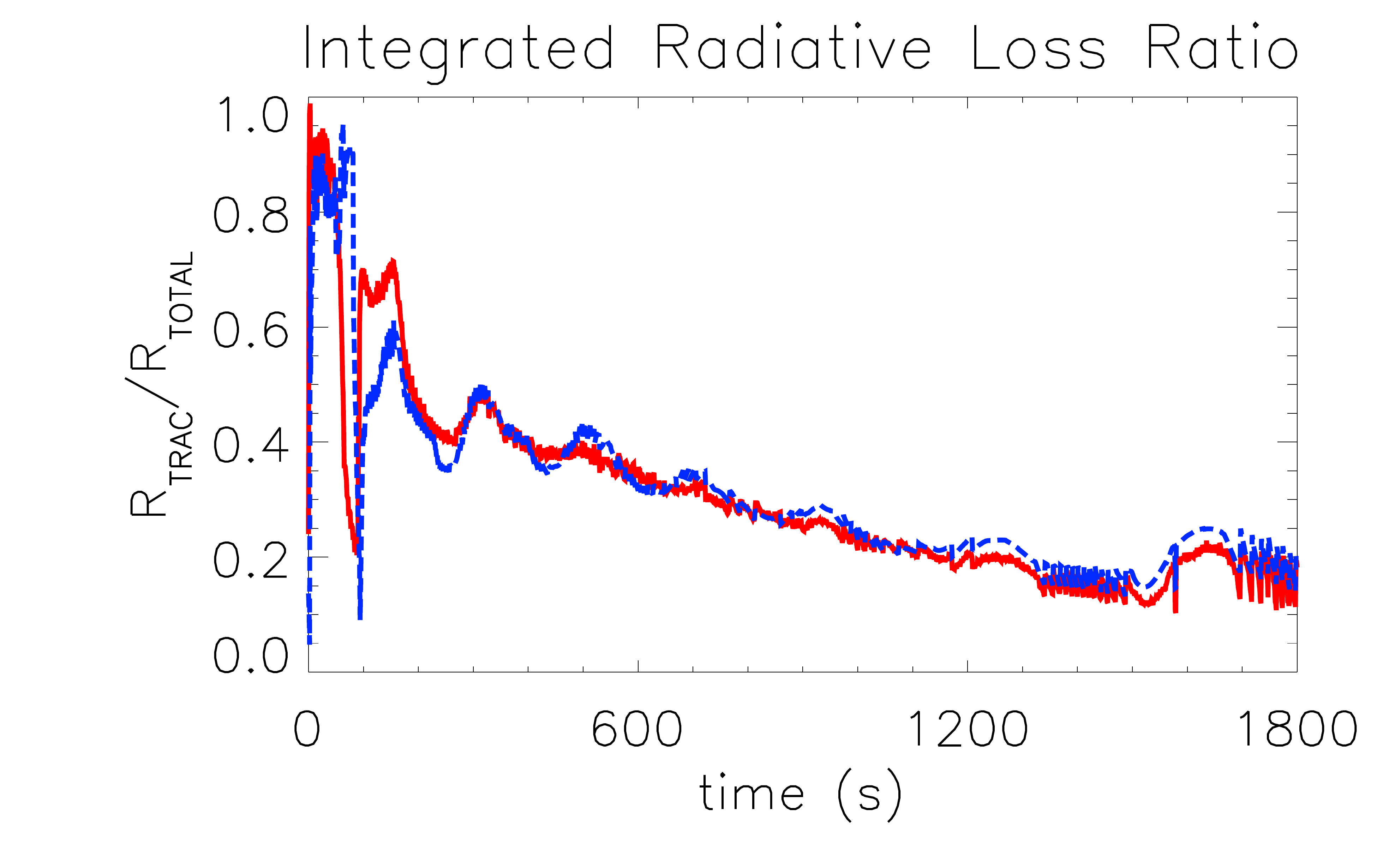}}
  \vspace*{-4mm}
  \caption{Results for the 60~s
  footpoint heating
  simulations (Section 
  \ref{Sect:Impulsive_footpoint_heating}).
  Notation is the same as that in Fig.
  \ref{Fig:600s_pulse_RL_5_time_evolution}.
  \label{Fig:fp2_Case_60s_RL_5_time_evolution}
  }
\end{figure*}
  %
  %
  %
  %
  %%%%%%%%%%%%%%%%%%%%%%%%%%%%%%%%%%%%%%%%%%%%%%%%%%%%%%%%%%%%%    
  %
  % Fig:TNE_f_vs_resolution
  %
\begin{figure*}
  \vspace*{-0.3cm}
  \hspace*{1cm}
  \subfigure{\includegraphics[width=0.85\linewidth]
  {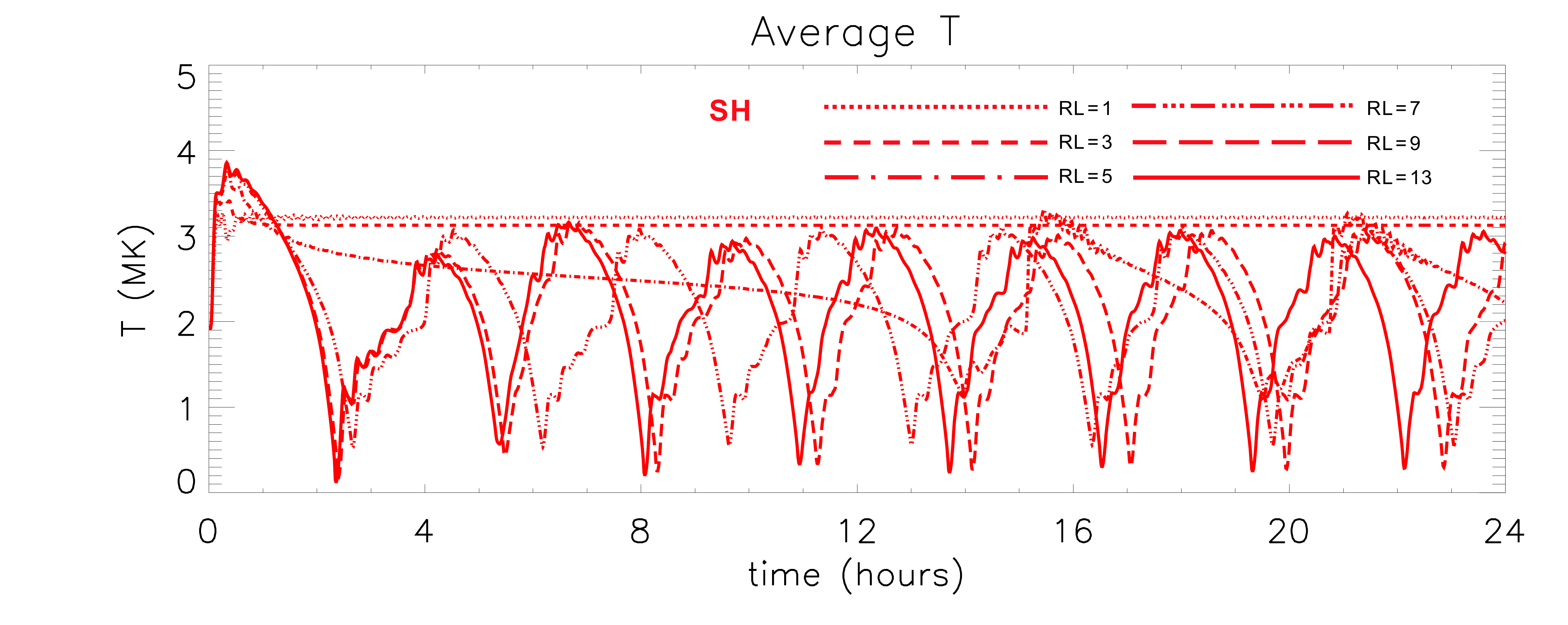}}
  \\[-0.5cm]
  \hspace*{1cm}
  \subfigure{\includegraphics[width=0.85\linewidth]
  {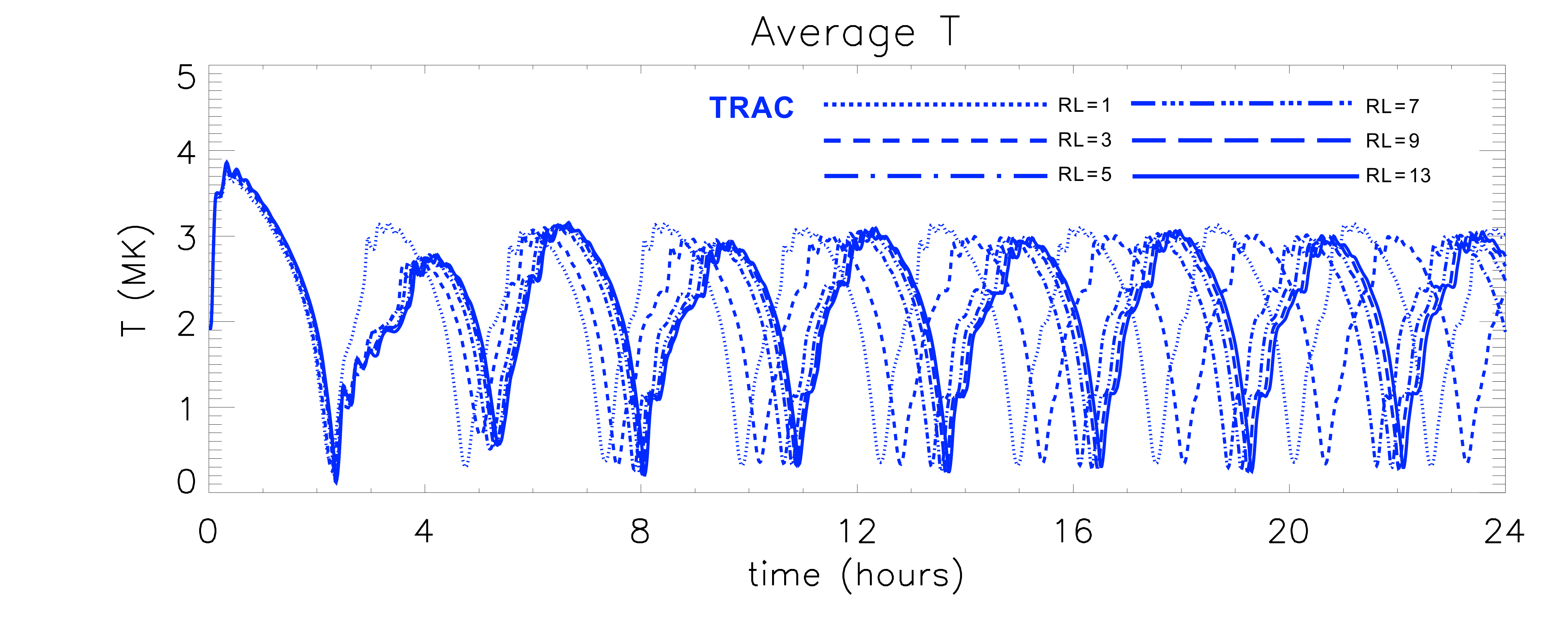}}
  \\[-0.5cm]
  \hspace*{1cm}
  \subfigure{\includegraphics[width=0.85\linewidth]
  {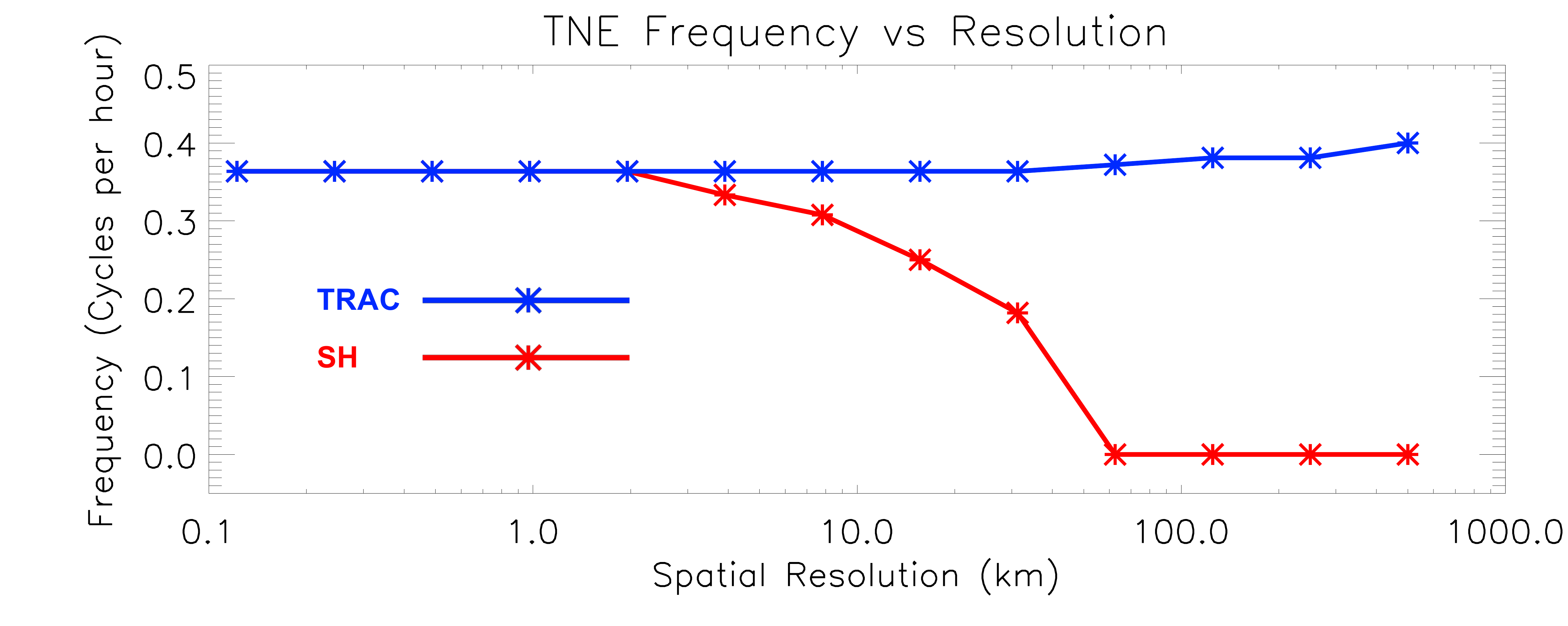}}
  \\[-0.5cm]
  \hspace*{1cm}
  \subfigure{\includegraphics[width=0.85\linewidth]
  {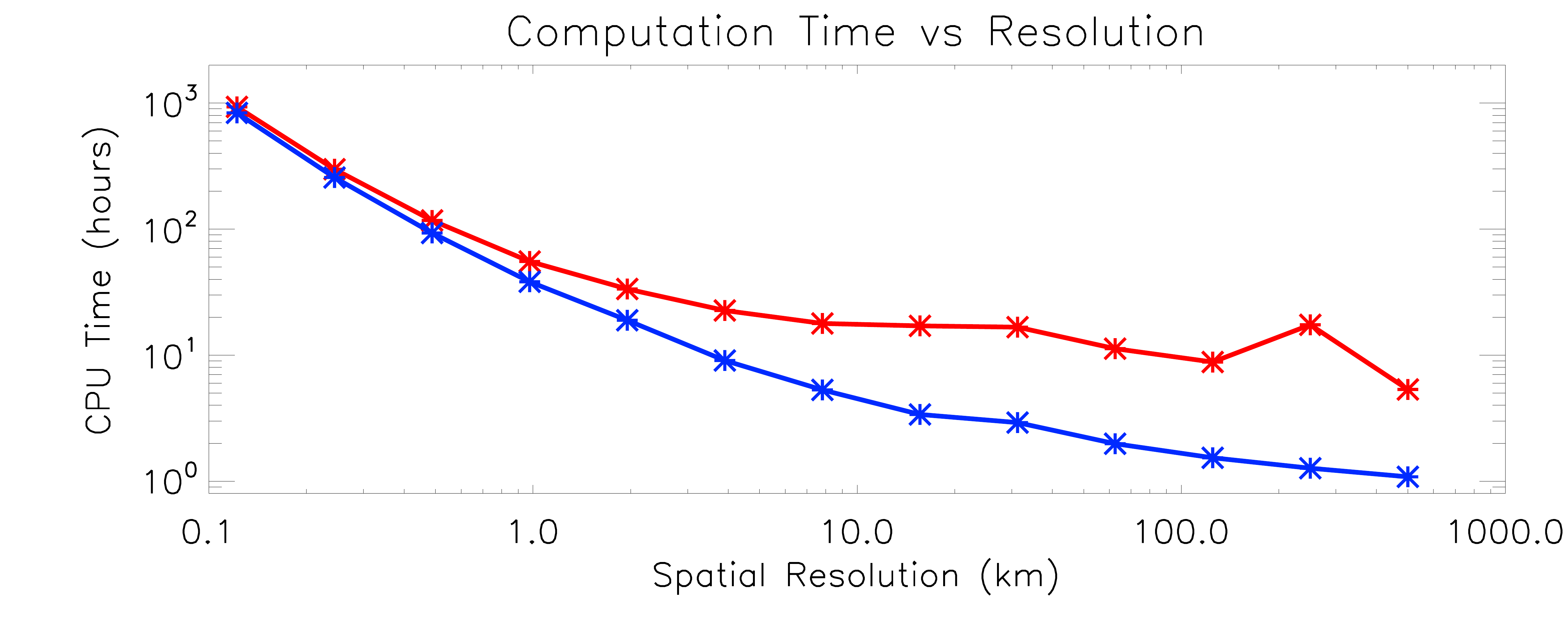}}
  \vspace*{-0.5cm}
  \caption{Results for the steady footpoint heating simulations
    run with the
    Spitzer-H{\"a}rm (SH)
    and TRAC conduction 
    methods
    (Section \ref{Sect:Steady_footpoint_heating}).
    The upper two panels show 
    the coronal averaged temperature as a
    function of time, for six 
    values of RL. 
    The lower two panels 
    show how the
    TNE cycle frequency and computation time
    depend on the minimum permitted 
    spatial resolution (coarser resolution is associated with 
    smaller RL). 
    The lines are 
    colour-coded in a way that reflects the 
    conduction method used.
    \label{Fig:TNE_f_vs_resolution}
  }
\end{figure*}
  \clearpage
  %
  %
%%%%%%%%%%%%%%%%%%%%%%%%%%%%%%%%%%%%%%%%%%%%%%%%%%%%%%%%%%%%%%
%
% Appendices
%
%%%%%%%%%%%%%%%%%%%%%%%%%%%%%%%%%%%%%%%%%%%%%%%%%%%%%%%%%%%%%%
\begin{appendix}
  %
  %
%%%%%%%%%%%%%%%%%%%%%%%%%%%%%%%%%%%%%%%%%%%%%%%%%%%%%%%%%%%%%%
%
% Alternative TRAC method options
%
%%%%%%%%%%%%%%%%%%%%%%%%%%%%%%%%%%%%%%%%%%%%%%%%%%%%%%%%%%%%%%
%
\section{Alternative TRAC method options
  \label{App:A}}
  \indent
  %
  %
%%%%%%%%%%%%%%%%%%%%%%%%%%%%%%%%%%%%%%%%%%%%%%%%%%%%%%%%%%%%%%
%
% JB19 formulation
%
%%%%%%%%%%%%%%%%%%%%%%%%%%%%%%%%%%%%%%%%%%%%%%%%%%%%%%%%%%%%%%
\subsection{JB19 formulation of the TRAC method}
  \indent
  As noted in Section \ref{Sect:TRAC_identification},
  the algorithm described in JB19 to identify
  the adaptive cutoff temperature ($T_c$)
  can give rise to
  oscillations in the $T_c$ value prescribed.
  However, these oscillations
  do not affect the ability of the TRAC method
  to accurately capture the coronal response to heating
  but they 
  can introduce errors in the 
  TRAC region
  integrated radiative losses (and other 
  TRAC region integrated 
  quantities too). 
  \\
  \indent
  In particular, the 
  algorithm for moving from time step $j$ to $j+1$ allows 
  the cutoff temperature to drop on occasions.
  For example, to calculate $T^{j+1}$ etc, we first scan 
  through
  the solution at time step $j$  to calculate $T_c$.
  Once this is obtained, we then modify the 
  parallel thermal conductivity,
  radiative loss rate and heating rate 
  at 
  time step $j$ as specified in Section 
  \ref{Sect:TRAC_broadening}, and then calculate 
  $T^{j+1}$ in
  the usual way. 
  This then gives the TRAC region 
  diagnostics at $j+1$. 
  Therefore, the solution and diagnostics at time step 
  $j+1$ ($j$) are based
  on the cutoff temperature calculated at time step $j$ 
  ($j-1$).  
  Thus, it is possible that at time step $j$ the TR has become 
  fully (or partially) resolved as a result of the 
  broadening below $T_c(j-1)$
  (i.e. no (or significantly fewer) 
  grid points violate the resolution criteria
  Eq. \eqref{Eqn:T_c}) 
  and so the cutoff temperature
  drops to the minimum value  
  $T_c(j)= T_{\textrm{chrom}}$ (or
  to a lower unresolved temperature in the TR) 
  for the update to
  time step $j+1$. 
  \\
  \indent
  The response of the integrated radiative losses 
  to the drop in $T_c$
  is to significantly
  increase in magnitude. This happens because 
  $\Lambda^{\prime}(T)$ scales with $(T/T_c)^{5/2}$
  and the density stratification at time step $j+1$ remains
  identical to that at time step $j$  
  since the lower atmosphere has not had time to respond 
  to the changing TR conditions.
  \\
  \indent
  The upper six panels of Fig. \ref{Fig:Coronal_Averages_3000s_LI_PI_Q_mod_fixed_percentage_Tc} 
  show the results of the 
  600~s heating pulse simulations considered 
  in Section \ref{Sect:Long_pulse}, when run with the JB19
  formulation of the TRAC method.
  The format is similar to
  Fig. \ref{Fig:600s_pulse_RL_5_time_evolution}.
  \\
  \indent
  It is clear that
  oscillations in the cutoff temperature and spikes in the 
  integrated radiative losses occur 
  throughout the evolution.
  The characteristics of these oscillations and spikes
  are similar
  during periods of
  (i)~strong evaporation ($0-300$~s) and
  (ii) peak density ($600-1200$~s), 
  with the drops in the 
  $T_c$ 
  triggered by the TR becoming either fully or
  partially resolved.
  We note that these oscillations are short lived (< 1~s)
  because the TR re-steepens when the radiative losses spike.
  On the other hand, the oscillations seen in the 
  decay phase 
  ($1800-2400$~s)
  are driven by increases in the $T_c$ when the TR becomes
  temporarily under-resolved as the loop cools. 
  \\
  \indent
  The main aim of 
  the TRAC method is to provide coronal diagnostics
  and the TRAC solutions 
  converge to the coronal response of  
  the properly resolved SH solution. 
  However, 
  if one is interested in improved TRAC region
  diagnostics, then
  the subsequent section provides a solution.
  %
  %
%%%%%%%%%%%%%%%%%%%%%%%%%%%%%%%%%%%%%%%%%%%%%%%%%%%%%%%%%%%%%%
%
% Cutoff temperature limiter
%
%%%%%%%%%%%%%%%%%%%%%%%%%%%%%%%%%%%%%%%%%%%%%%%%%%%%%%%%%%%%%%
\subsection{Cutoff temperature limiter
  \label{App:T_c_limiter}
  }
  \indent
  It is possible to remove 
  the jumps in the cutoff temperature and thus the radiation
  spikes, described in the previous section,
  by
  limiting the decreases in the 
  $T_c$ to a small percentage per time step. 
  This can be achieved by imposing the following
  limiter on the cutoff temperature.
  \\
  \indent
  We define the time interval over which we 
  limit decreases in $T_c$ as,
  \begin{align}
    \tau=n \delta t,
  \end{align}
  where $\delta t$ is 
  the current time step in the simulation,
  given by the minimum of the advection and conduction 
  time steps,
  \begin{align}
    \delta t = 
    \textrm{min} 
    (\delta t_{\textrm{adv}},\, \delta t_{\textrm{cond}}).
  \end{align}
  \indent
  Confining the compound reduction in $T_c$ over the $n$ time 
  steps
  in $\tau$ to a
  maximum percentage ($d_\textrm{max}$) 
  then takes the form,
  \begin{align}
    (1-\alpha)^n=d_\textrm{max}.
  \end{align}
  \indent
  In this paper, we have taken the time interval as 
  $\tau=1$~s
  and the maximum percentage change in
  $T_c$ over this interval
  is prescribed as
  $d_\textrm{max}=0.1$ (i.e. 10\%).
  \\
  \indent
  Using these definitions, 
  when $T_{c_{j+1}}<T_{c_{j}}$,
  decreases in the cutoff temperature
  are then limited as follows,
  \begin{align}
    T_{c_{j+1}}=(1-\alpha)T_{c_{j}},
  \end{align}
  where $T_{c_{j+1}}$ $(T_{c_{j}})$ is the cutoff temperature
  at time step $j+1$ $(j)$
  and $\alpha=1-d^{1/n}_\textrm{max}$. 
  This corresponds to limiting
  the decreases in $T_c$ to 10\% over any 1~s interval.
  However, we note that
  $T_{c_{j+1}}$ is also required to satisfy the criteria,
  \begin{align}
    T_{\textrm{chrom}} \leq T_{c_{j+1}} \leq  0.2 \,
    \textrm{max}(T(s)).
    \label{Eqn:T_c_criteria_app} 
  \end{align}	
  which we outlined previously in Eq. 
  \eqref{Eqn:T_c_criteria}.
  \\
  \indent
  Fig. 1
  shows
  the results of the 600~s heating pulse simulations
  when TRAC is run with the cutoff temperature 
  limiter described above.
  The figure clearly demonstrates
  the benefit 
  of imposing such a limiter: 
  the removal of the radiation spikes enables the
  TRAC method to provide accurate diagnostics
  from both the TRAC region and TR (e.g.
  the total radiative losses integrated across the TR). 
  Furthermore, 
  limiting the decreases in the cutoff temperature
  does not influence the accuracy of the coronal
  evolution
  and so
  the method also retains
  the ability to provide accurate coronal diagnostics.
  %
  %
%%%%%%%%%%%%%%%%%%%%%%%%%%%%%%%%%%%%%%%%%%%%%%%%%%%%%%%%%%%%%%
%
% Fixed percentage cutoff temperature
%
%%%%%%%%%%%%%%%%%%%%%%%%%%%%%%%%%%%%%%%%%%%%%%%%%%%%%%%%%%%%%%
\subsection{Fixed percentage cutoff temperature}
\begin{figure*}
  \hspace*{-0.05\linewidth}
  \subfigure{\includegraphics[width=0.375\linewidth]
  {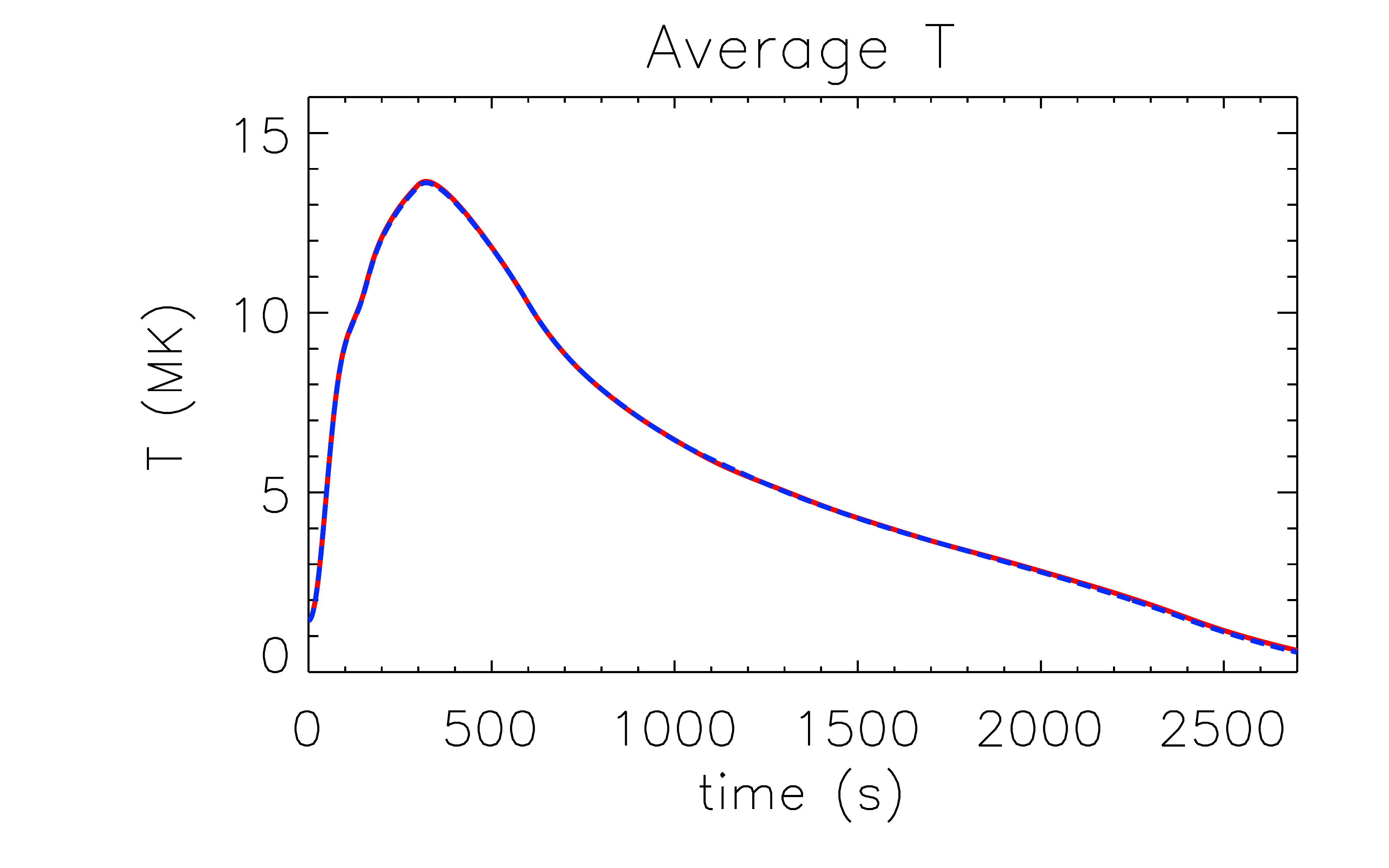}}
  \hspace*{-0.03\linewidth}
  \subfigure{\includegraphics[width=0.375\linewidth]
  {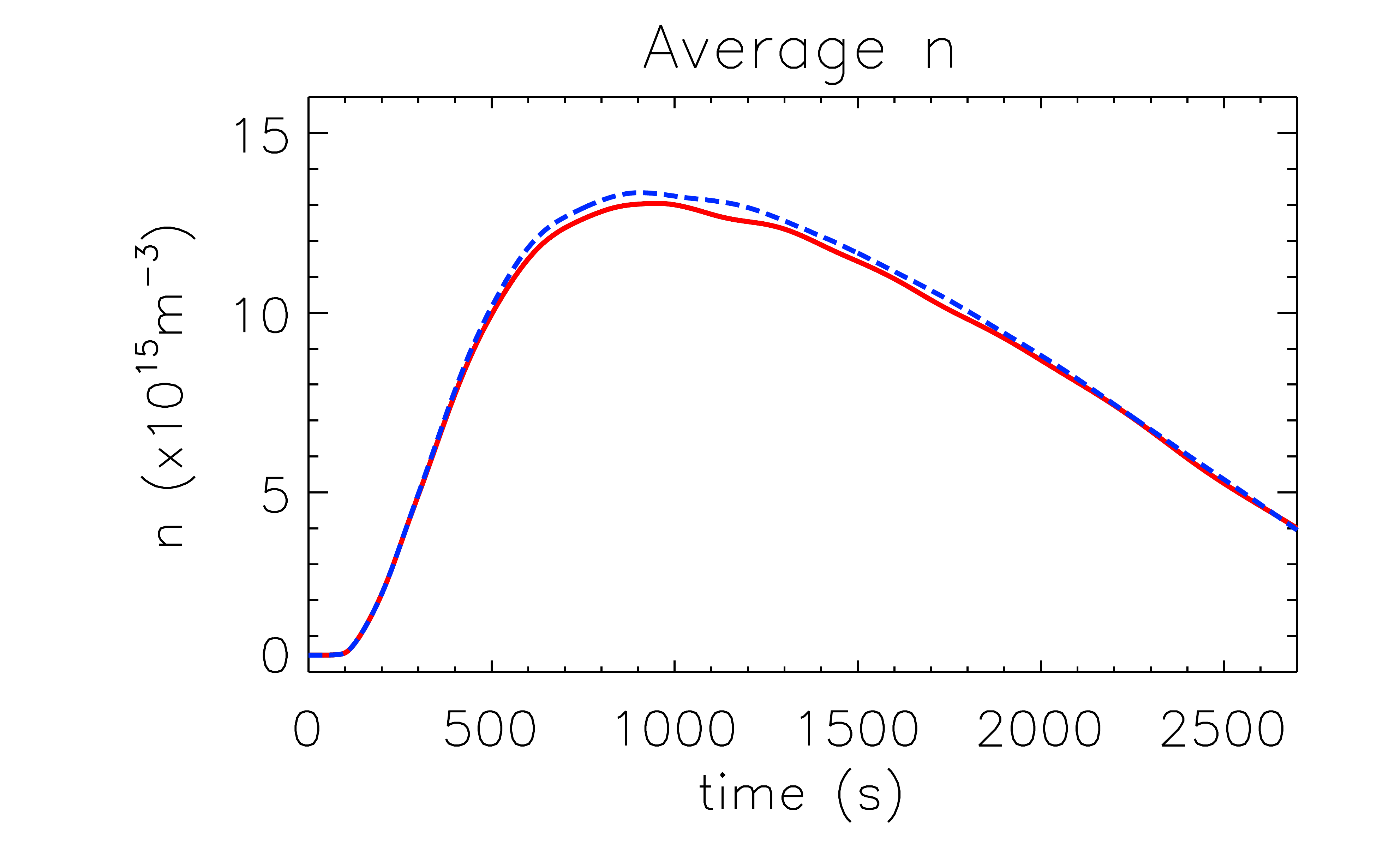}}
  \hspace*{-0.03\linewidth}
  \subfigure{\includegraphics[width=0.375\linewidth]
  {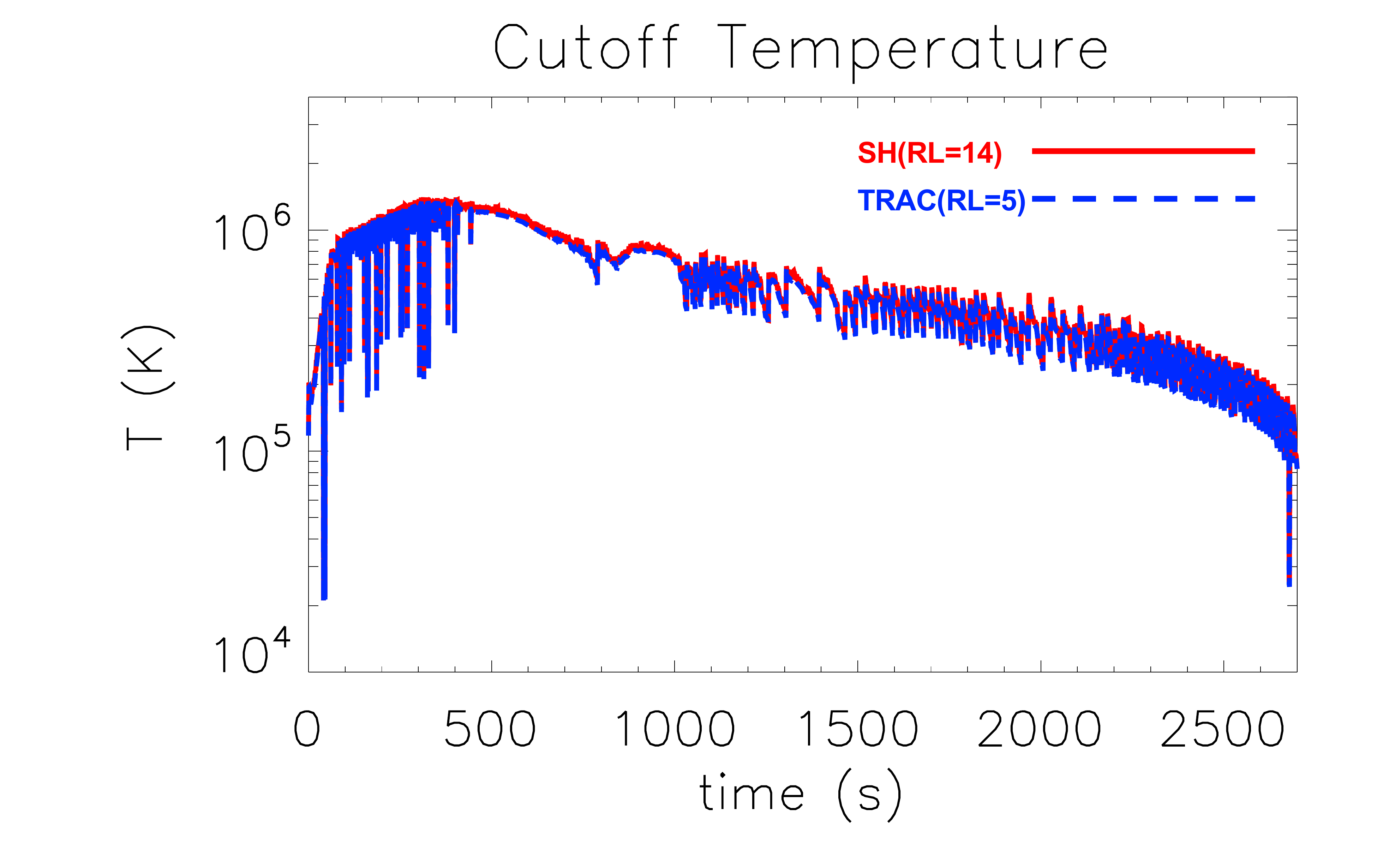}}
  \\[-4mm]
  \hspace*{-0.05\linewidth}
  \subfigure{\includegraphics[width=0.375\linewidth]
  {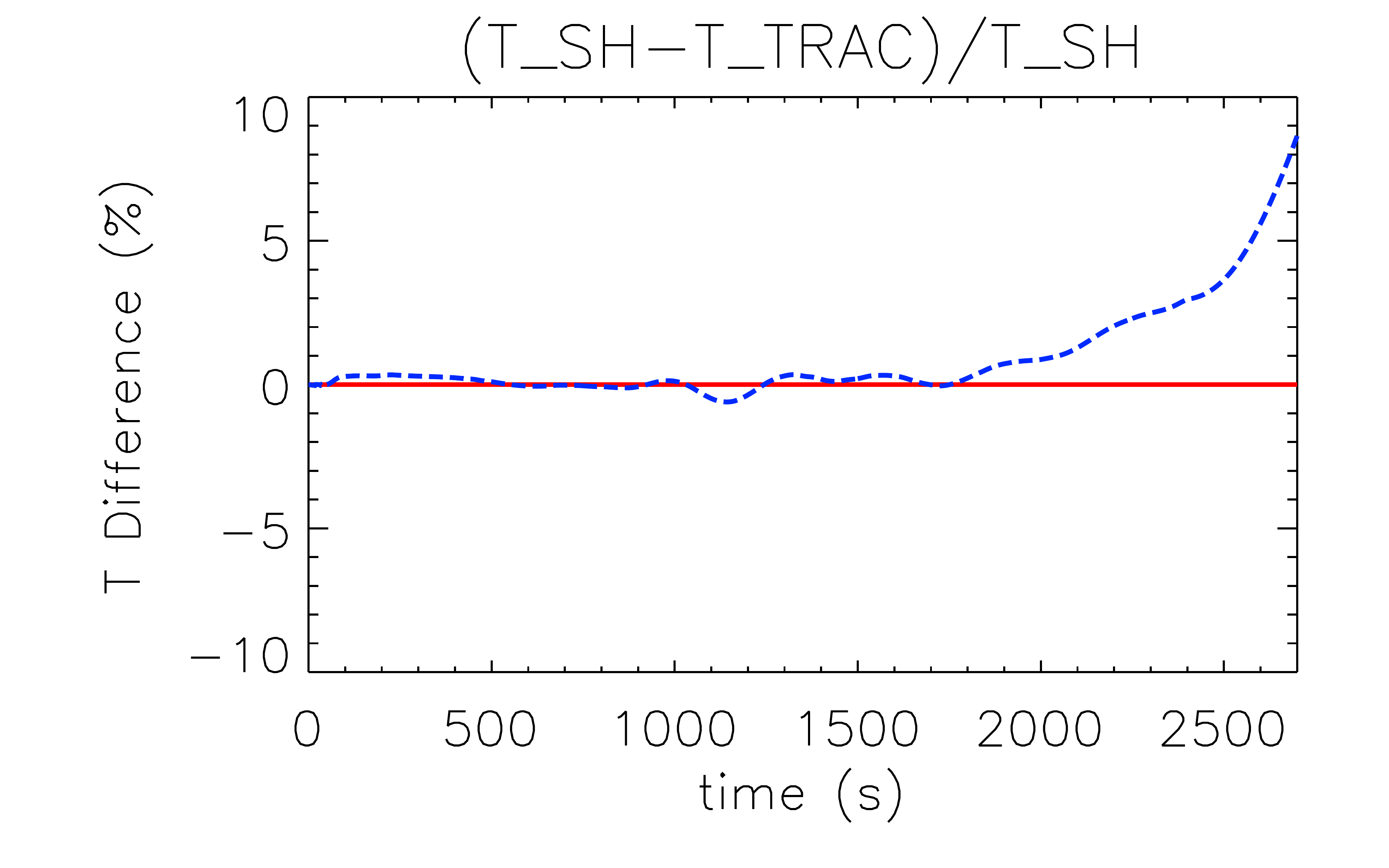}}
  \hspace*{-0.03\linewidth}
  \subfigure{\includegraphics[width=0.375\linewidth]
  {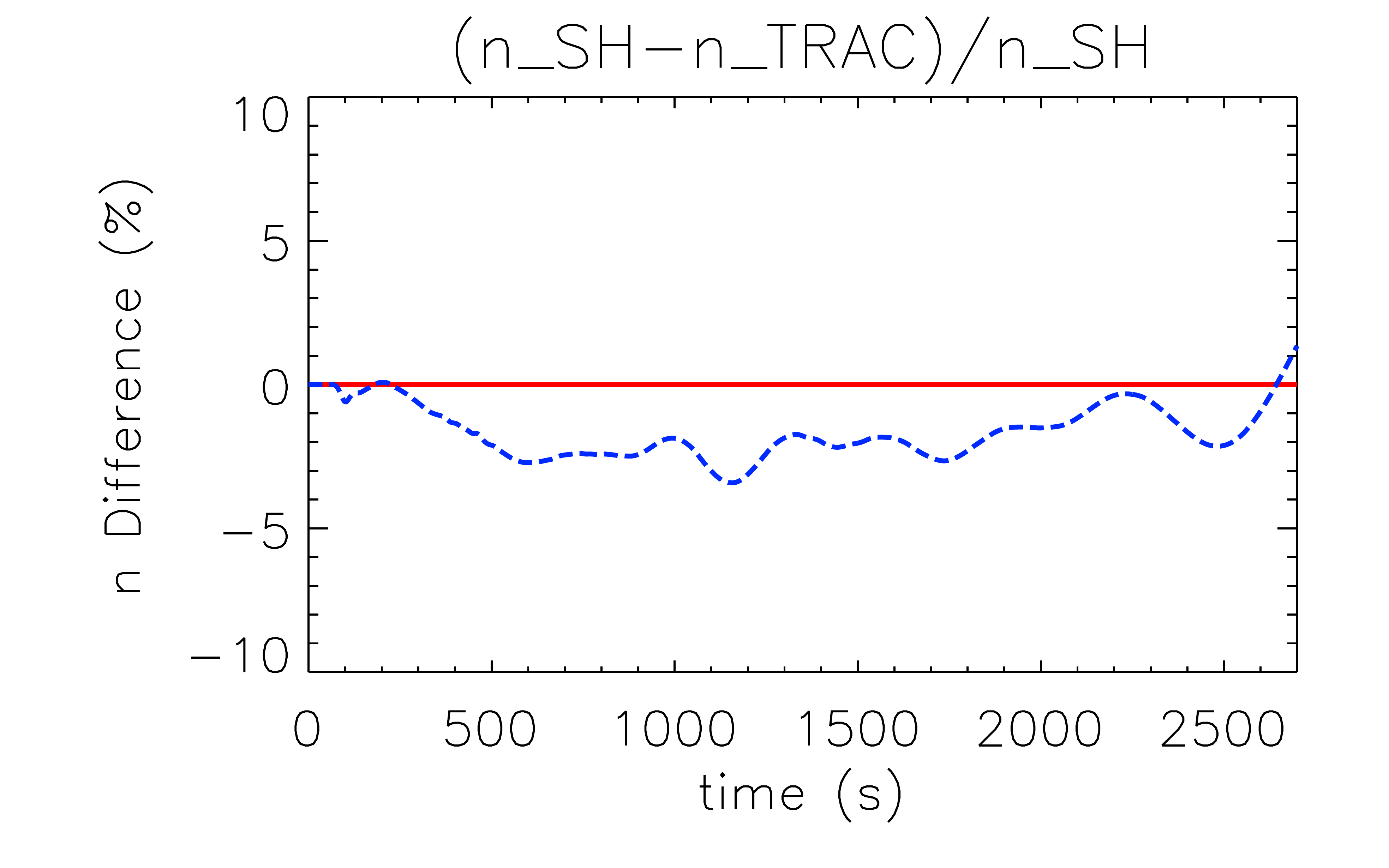}}
  \hspace*{-0.03\linewidth}
  \subfigure{\includegraphics[width=0.375\linewidth]
  {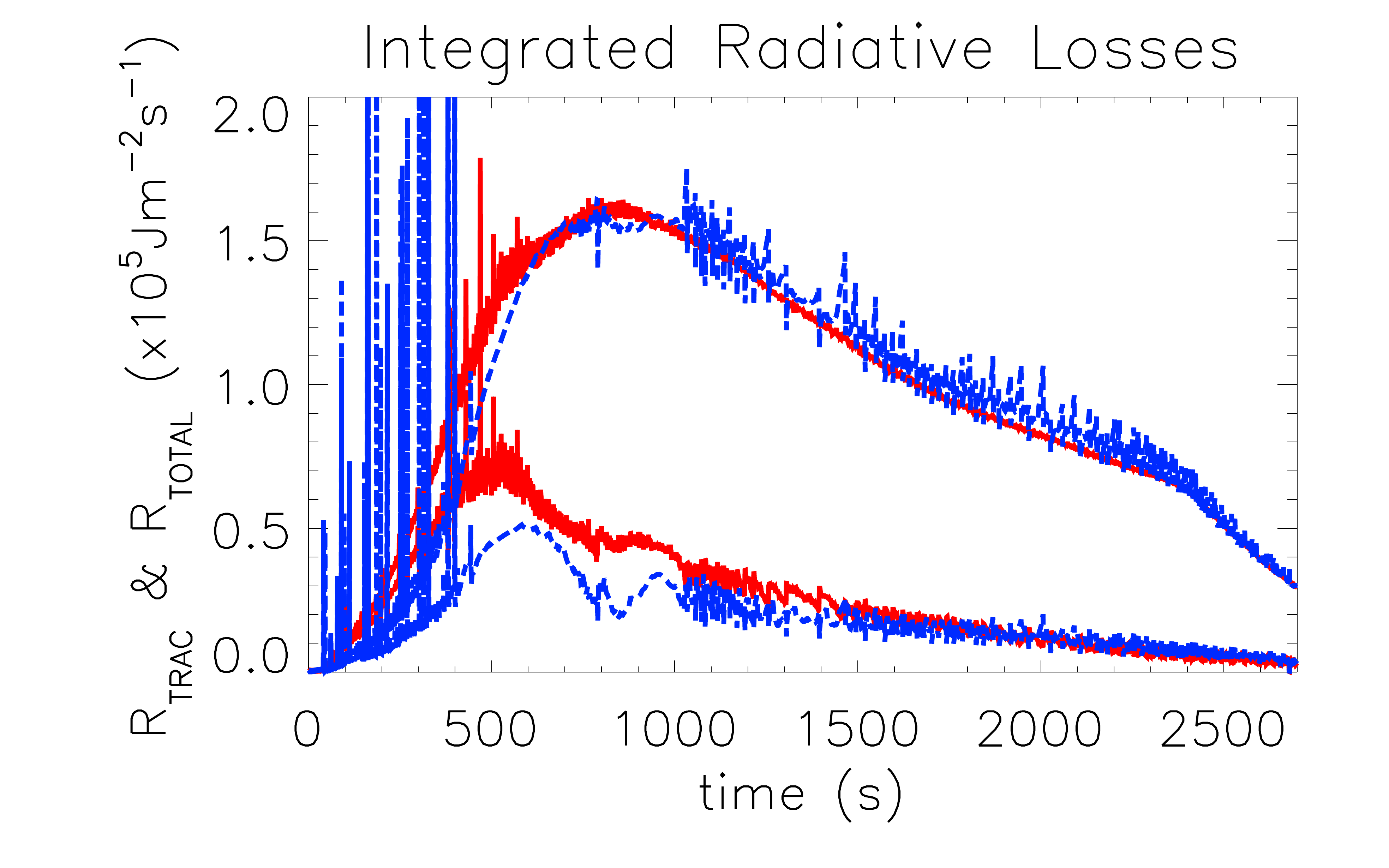}}
  \\[-2mm]
  \hspace*{-0.05\linewidth}
  \subfigure{\includegraphics[width=0.375\linewidth]
  {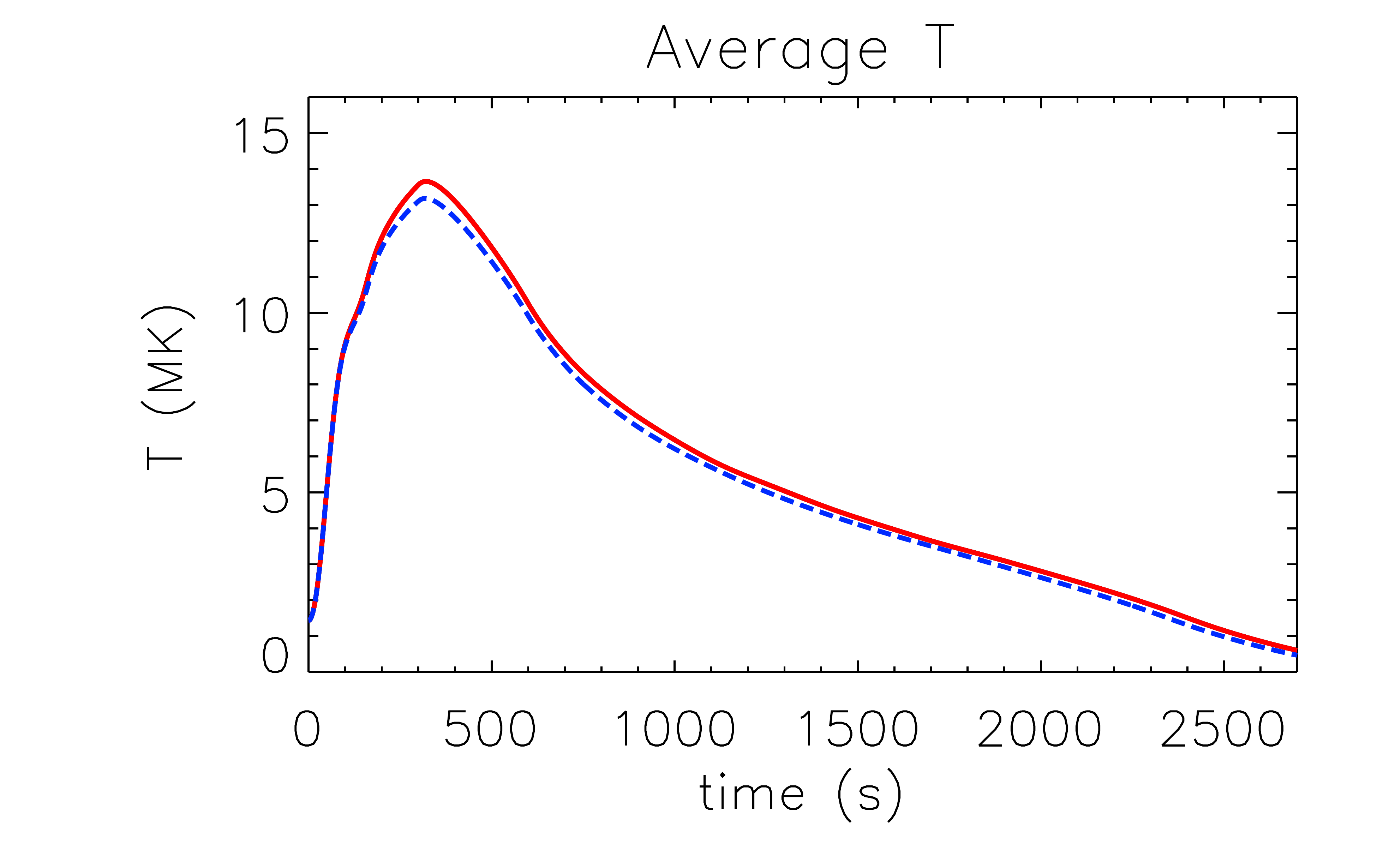}}
  \hspace*{-0.03\linewidth}
  \subfigure{\includegraphics[width=0.375\linewidth]
  {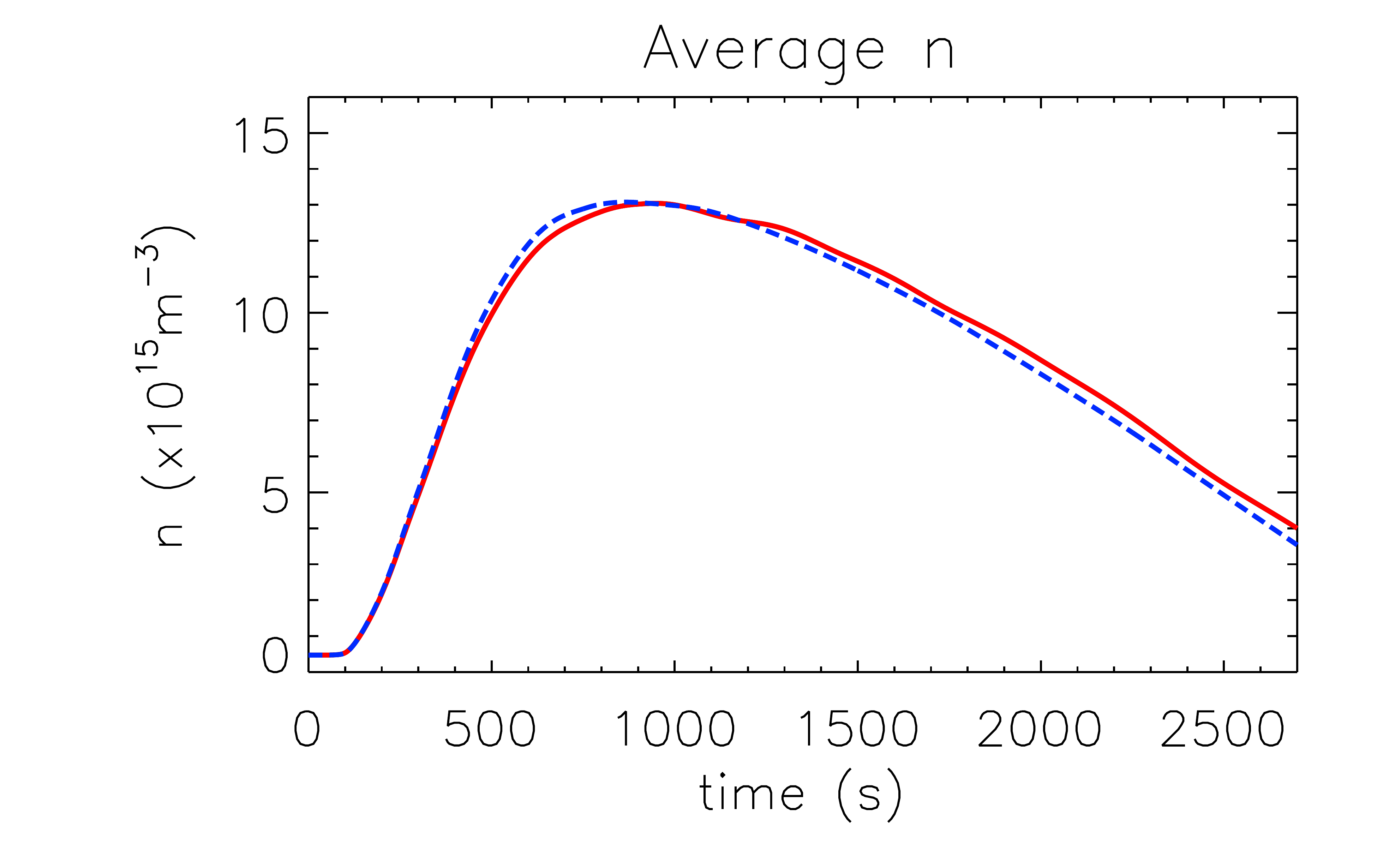}}
  \hspace*{-0.03\linewidth}
  \subfigure{\includegraphics[width=0.375\linewidth]
  {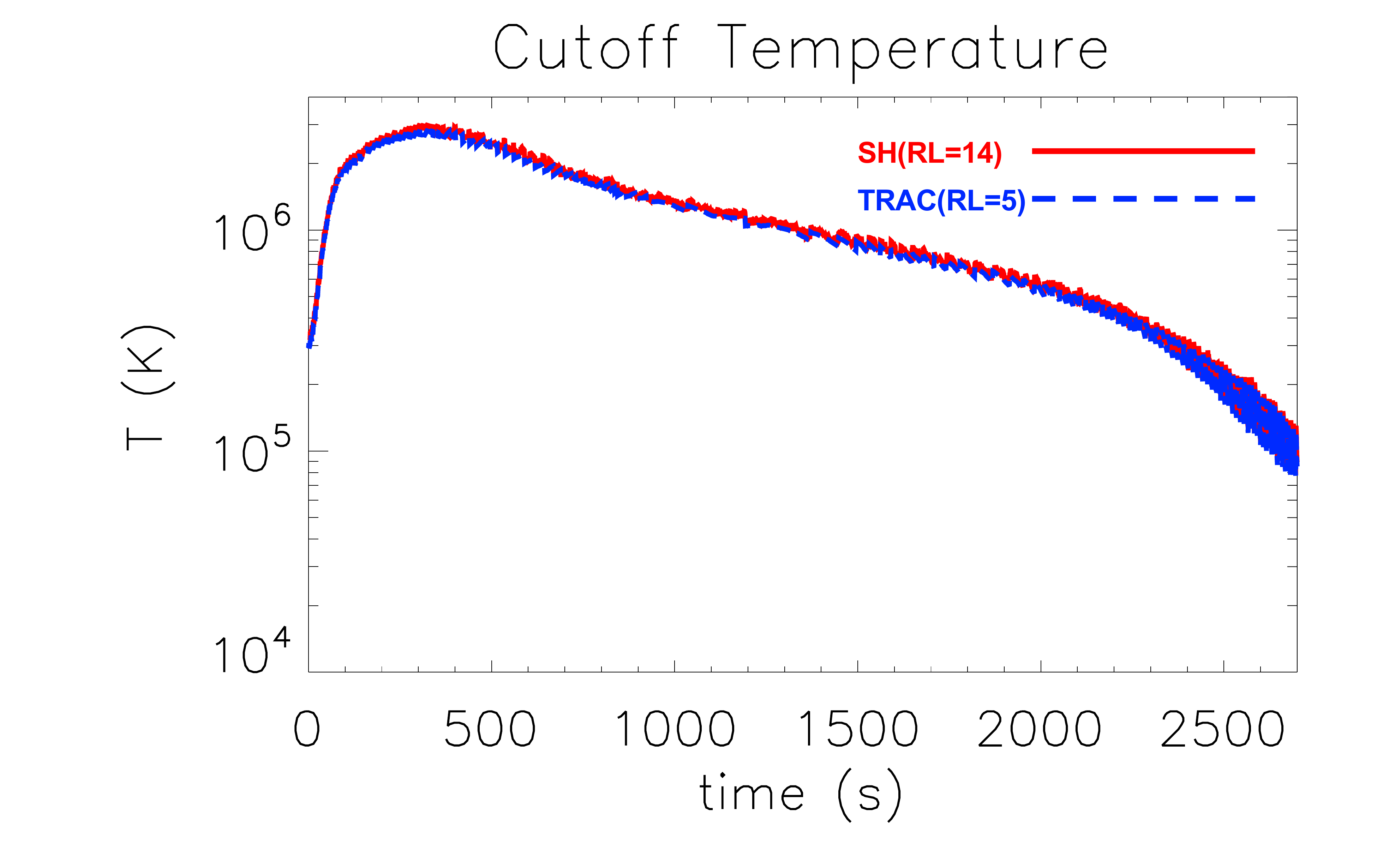}}
  \\[-4mm]
  \hspace*{-0.05\linewidth}
  \subfigure{\includegraphics[width=0.375\linewidth]
  {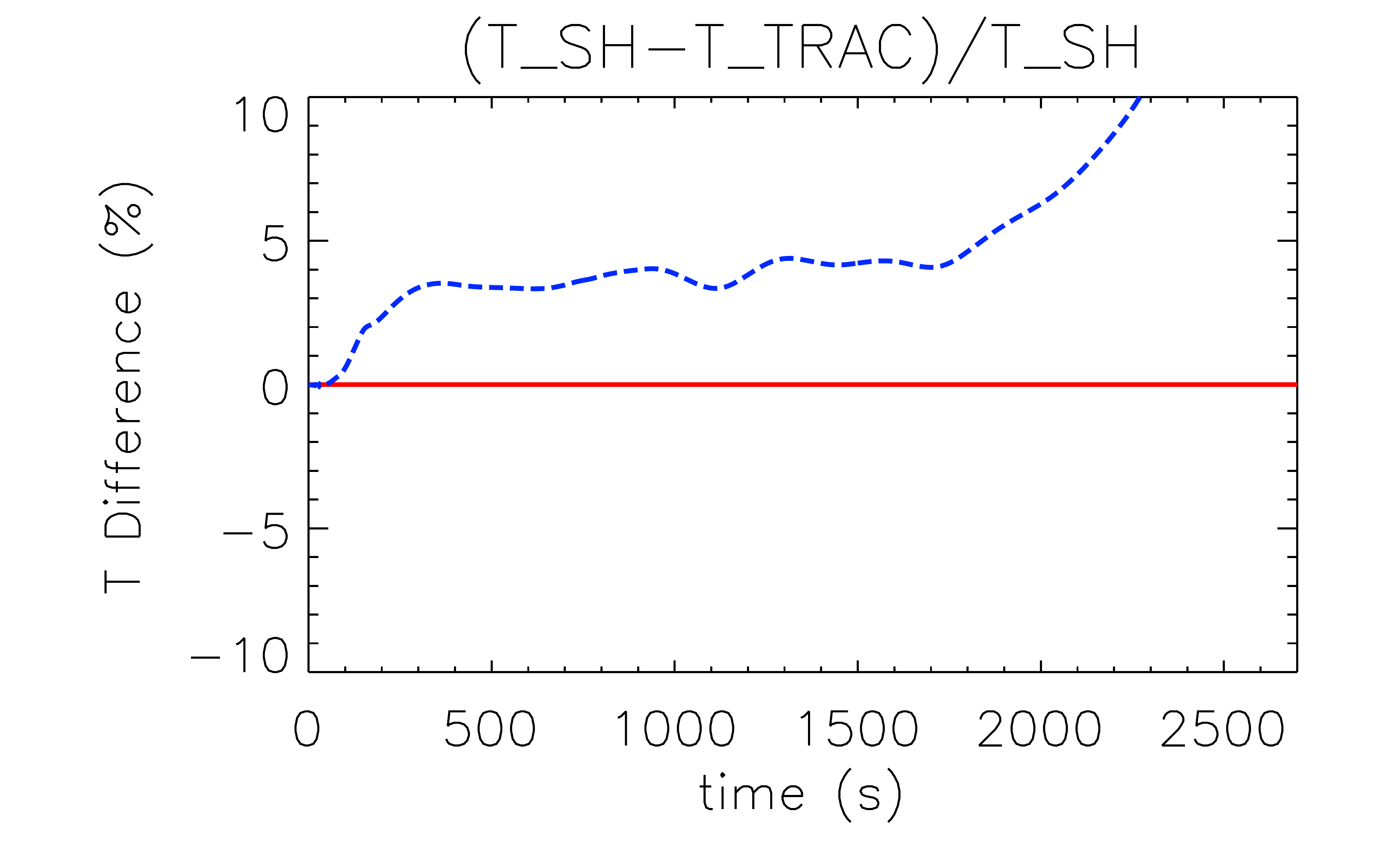}}
  \hspace*{-0.03\linewidth}
  \subfigure{\includegraphics[width=0.375\linewidth]
  {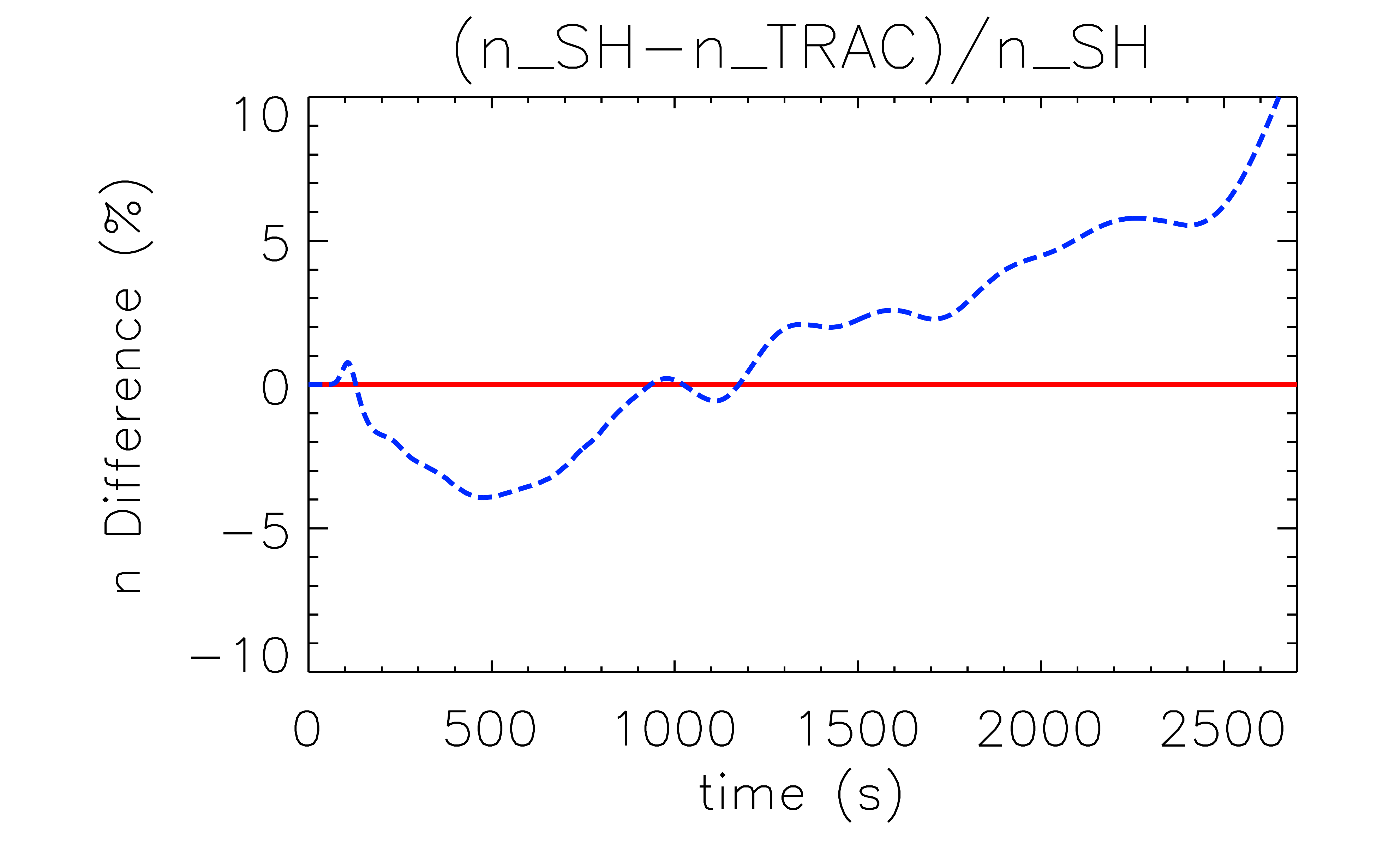}}
  \hspace*{-0.03\linewidth}
  \subfigure{\includegraphics[width=0.375\linewidth]
  {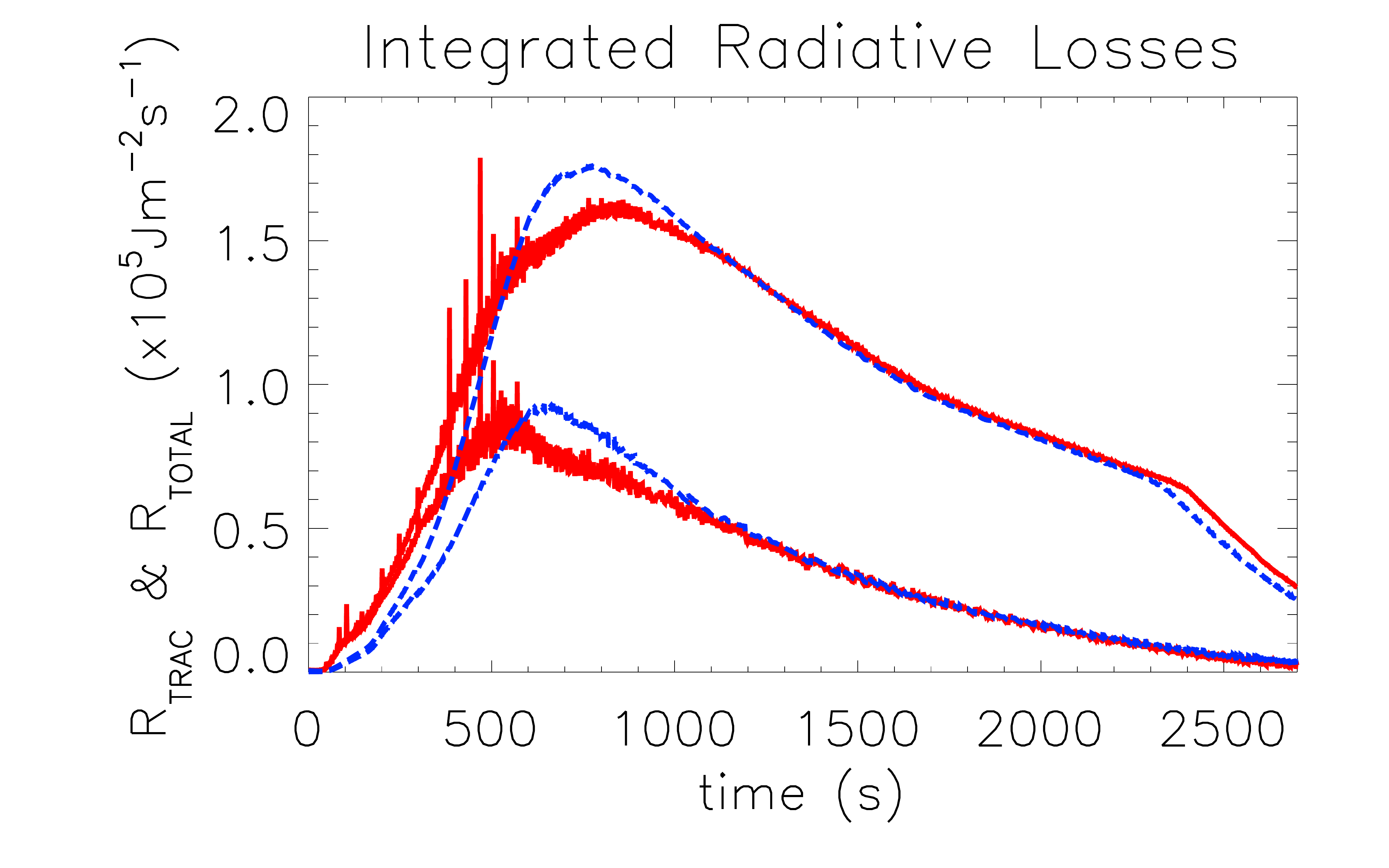}}
  \vspace*{-4mm}
  \caption{Results for the 600 s heating pulse simulations
    when run with the JB19 formulation of the TRAC method
    (upper six panels)
    and a 
    fixed percentage cutoff temperature 
    (lower six panels).
    The panels show
    the coronal averaged temperature and density,
    as functions
    of time,
    the time evolution of the
    cutoff temperature ($T_c$), 
    the normalised differences between the coronal quantities,
    and the
    radiative losses integrated across the TRAC region 
    ($\mathcal{R}_{\textsc{trac}}$) and total over half of the
    loop ($\mathcal{R}_{\textsc{total}}$).
    The lines are colour coded in a way that reflects
    the conduction method used 
    with dashed blue (solid red) representing the TRAC (SH)
    solution. 
    Each set of six panels (upper and lower)
    can be compared with Fig. 
    \ref{Fig:600s_pulse_RL_5_time_evolution},
    which shows the results when TRAC is run
    with the cutoff temperature limiter. 
    Starting from the top left, the corresponding panels in
    Fig. \ref{Fig:600s_pulse_RL_5_time_evolution} are
    a, b, g, c, d and i, respectively.
    \label{Fig:Coronal_Averages_3000s_LI_PI_Q_mod_fixed_percentage_Tc}
  }  
\end{figure*}

  \indent
  An alternative option 
  for the TRAC method is 
  just to
  set the cutoff temperature as a fixed percentage 
  of the peak coronal temperature.
  For example, one can prescribe the adaptive $T_c$
  to be the upper bound value,
  \begin{align}
    T_c=0.2 \,\textrm{max}(T(s)).
    \label{Eqn:T_c_max} 
  \end{align}
  \indent
  The lower six panels of
  Fig.
  \ref{Fig:Coronal_Averages_3000s_LI_PI_Q_mod_fixed_percentage_Tc}
  show the results when using this approach of a
  fixed $20\%$ cutoff temperature.
  The main compromise made for simplicity 
  in the implementation is that 
  the method no longer
  reduces to the SH formulation when the TR is 
  fully resolved (i.e. when running with high
  spatial resolution). 
  The outcome is that 
  the error in the coronal response does not 
  scale with the resolution (i.e. the RL value).
  On the other hand,
  (i)
  the error in the coronal response at the particular times of
  interest (peak heating, peak density and
  during the decay phase) 
  remains bounded by around 5$\%$,
  (ii) the temporal evolution of the cutoff temperature  is  
  smooth and so 
  (iii)
  the TRAC region diagnostics show good agreement 
  with the fully resolved SH solution.
  %
  %
  %
  %
  %%%%%%%%%%%%%%%%%%%%%%%%%%%%%%%%%%%%%%%%%%%%%%%%%%%%%%%%%%%%%    
  %
  % Fig:Coronal_Averages_3000s_LI_PI_Q_mod_fixed_percentage_Tc
  %
  %
  %
%%%%%%%%%%%%%%%%%%%%%%%%%%%%%%%%%%%%%%%%%%%%%%%%%%%%%%%%%%%%%%
%
% Long pulse - global evolution
%
%%%%%%%%%%%%%%%%%%%%%%%%%%%%%%%%%%%%%%%%%%%%%%%%%%%%%%%%%%%%%%
%
\section{Details of the long heating pulse results - 
  global evolution
  \label{App:B}}
\begin{figure*}
  \vspace*{-4.5mm}
  \hspace*{-0.05\linewidth}
  \subfigure{\includegraphics[width=0.53\linewidth]
  {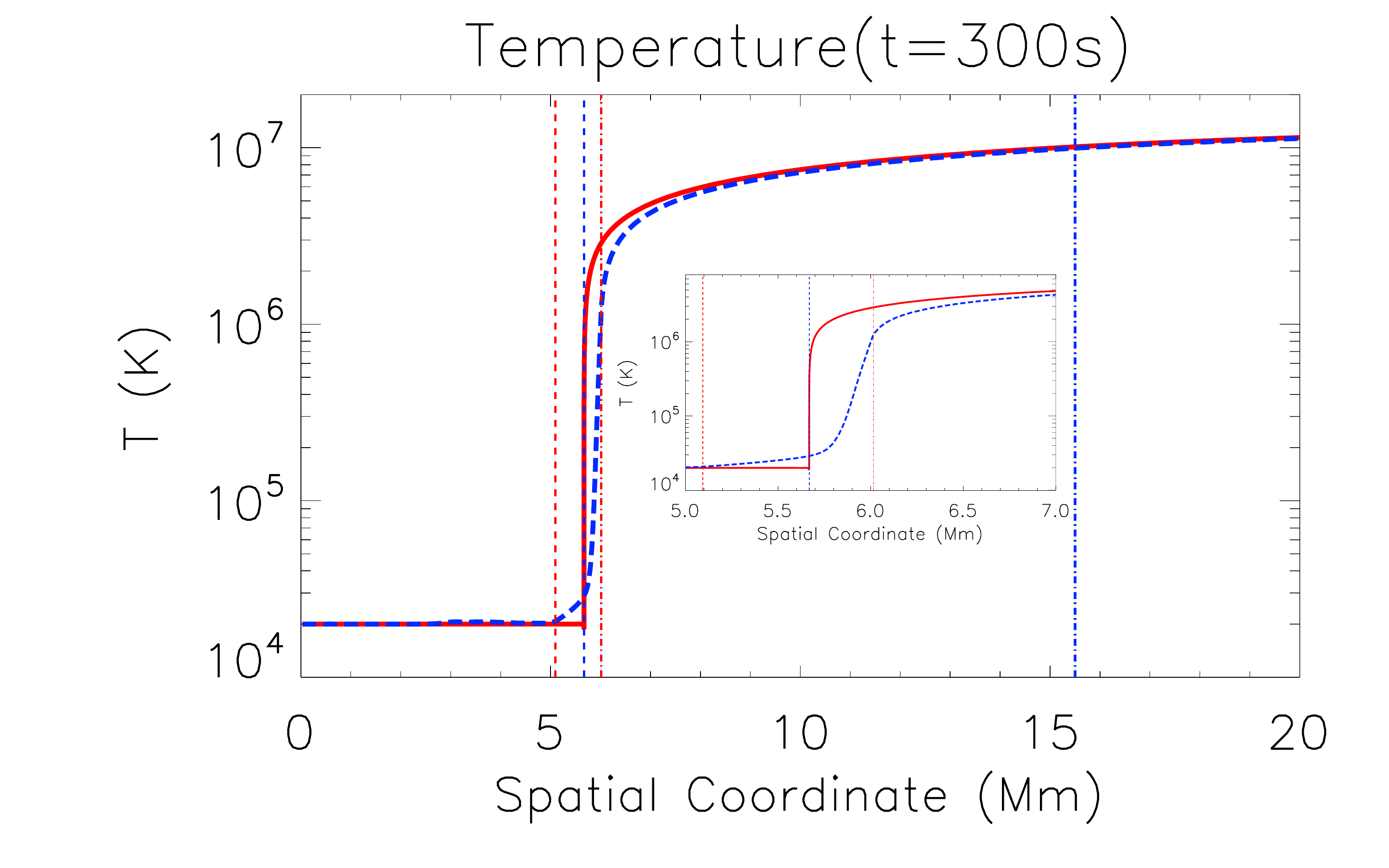}}
  \hspace*{-0.03\linewidth}
  \subfigure{\includegraphics[width=0.53\linewidth]
  {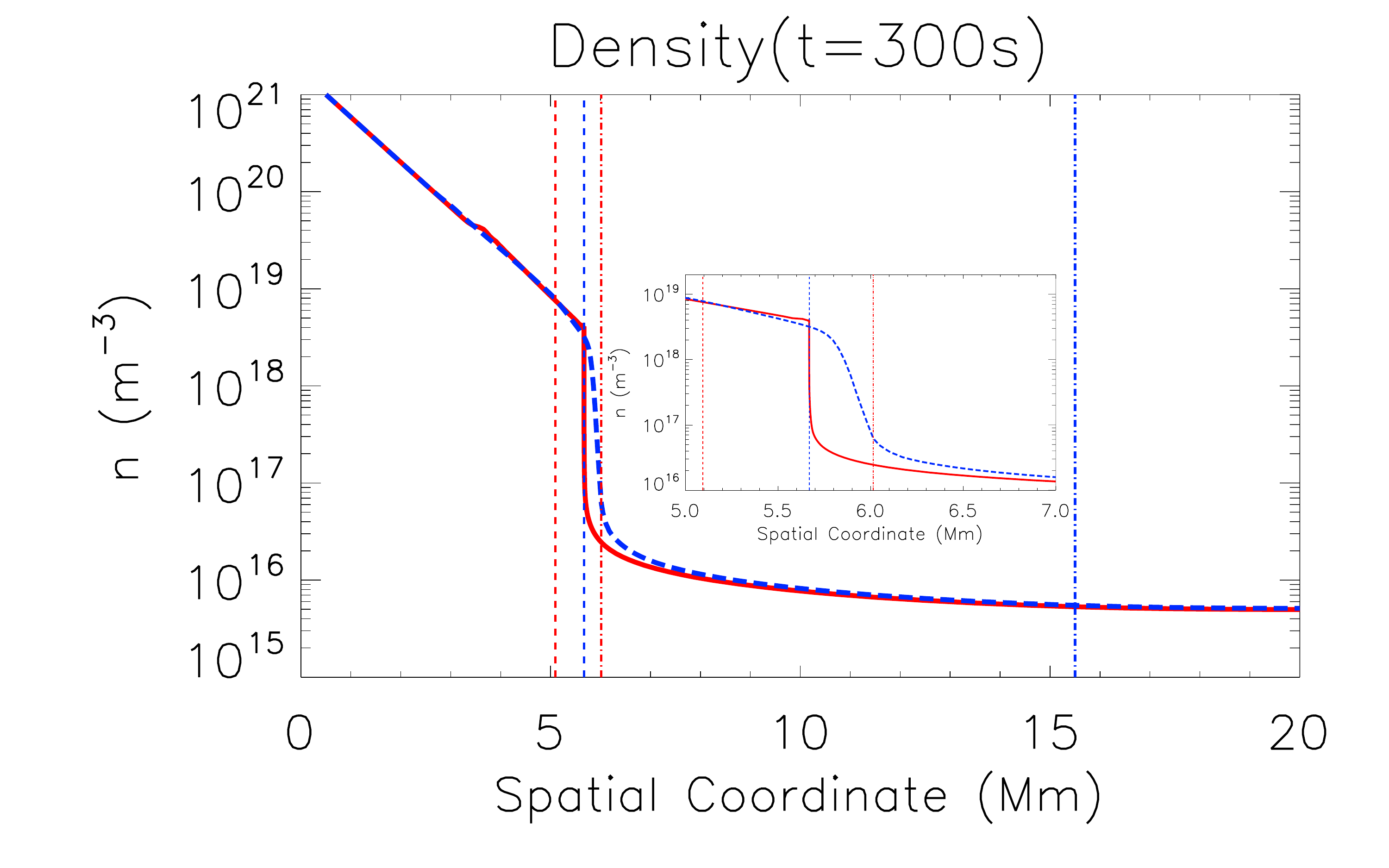}}
  \\[-6mm]
  \hspace*{-0.05\linewidth}
  \subfigure{\includegraphics[width=0.53\linewidth]
  {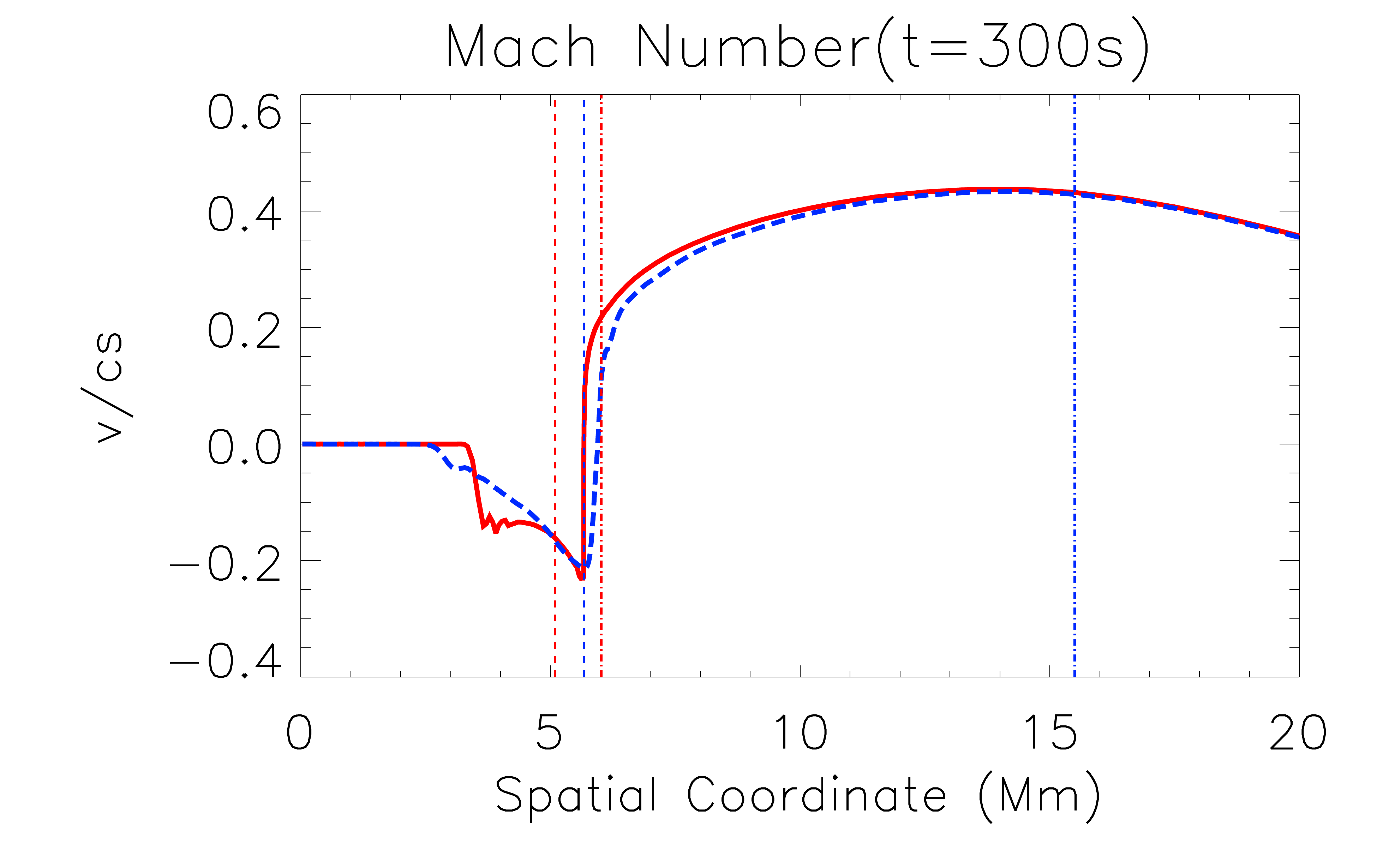}}
  \hspace*{-0.03\linewidth}
  \subfigure{\includegraphics[width=0.53\linewidth]
  {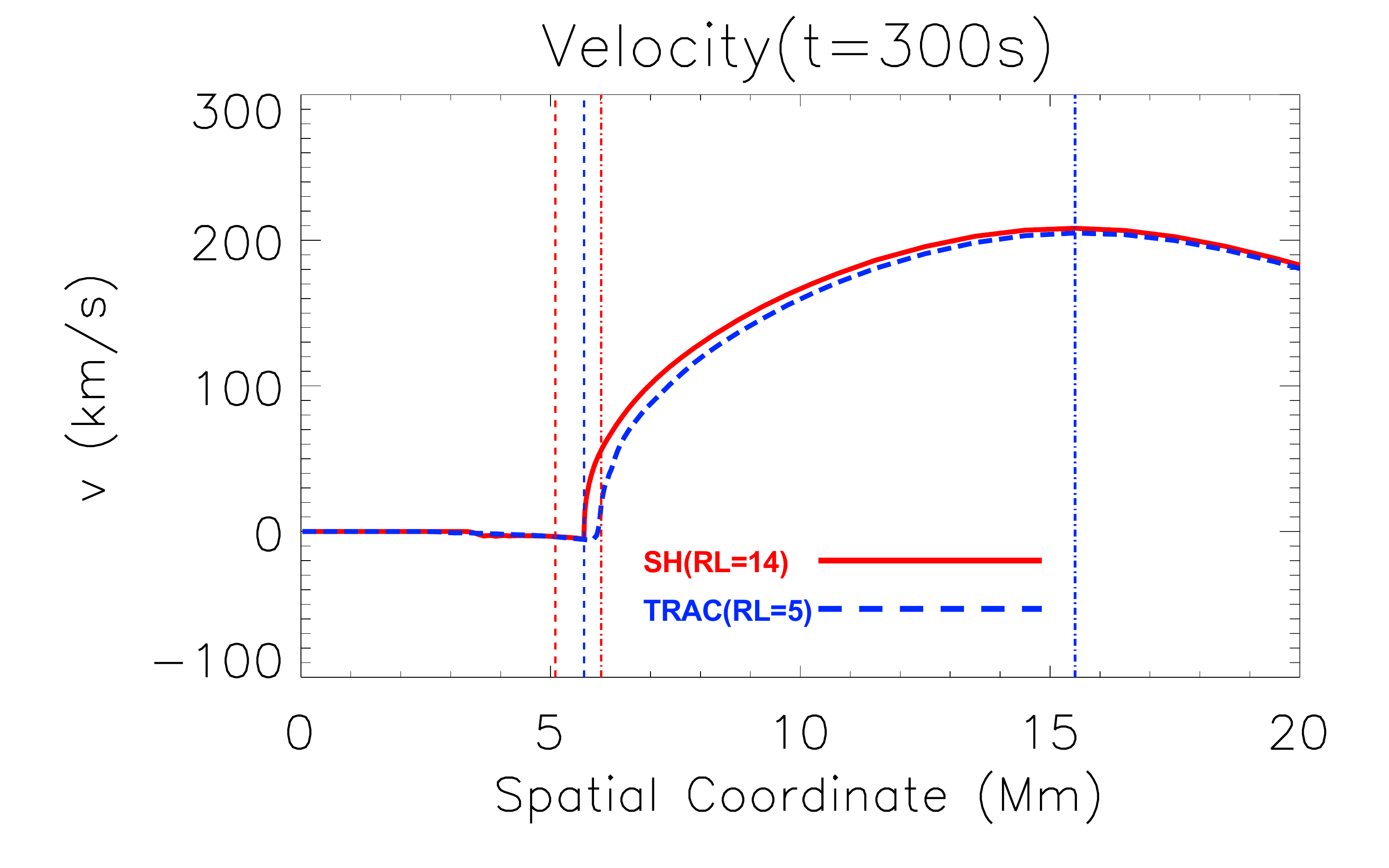}}
  \\[-6mm]
  \hspace*{-0.05\linewidth}
  \subfigure{\includegraphics[width=0.53\linewidth]
  {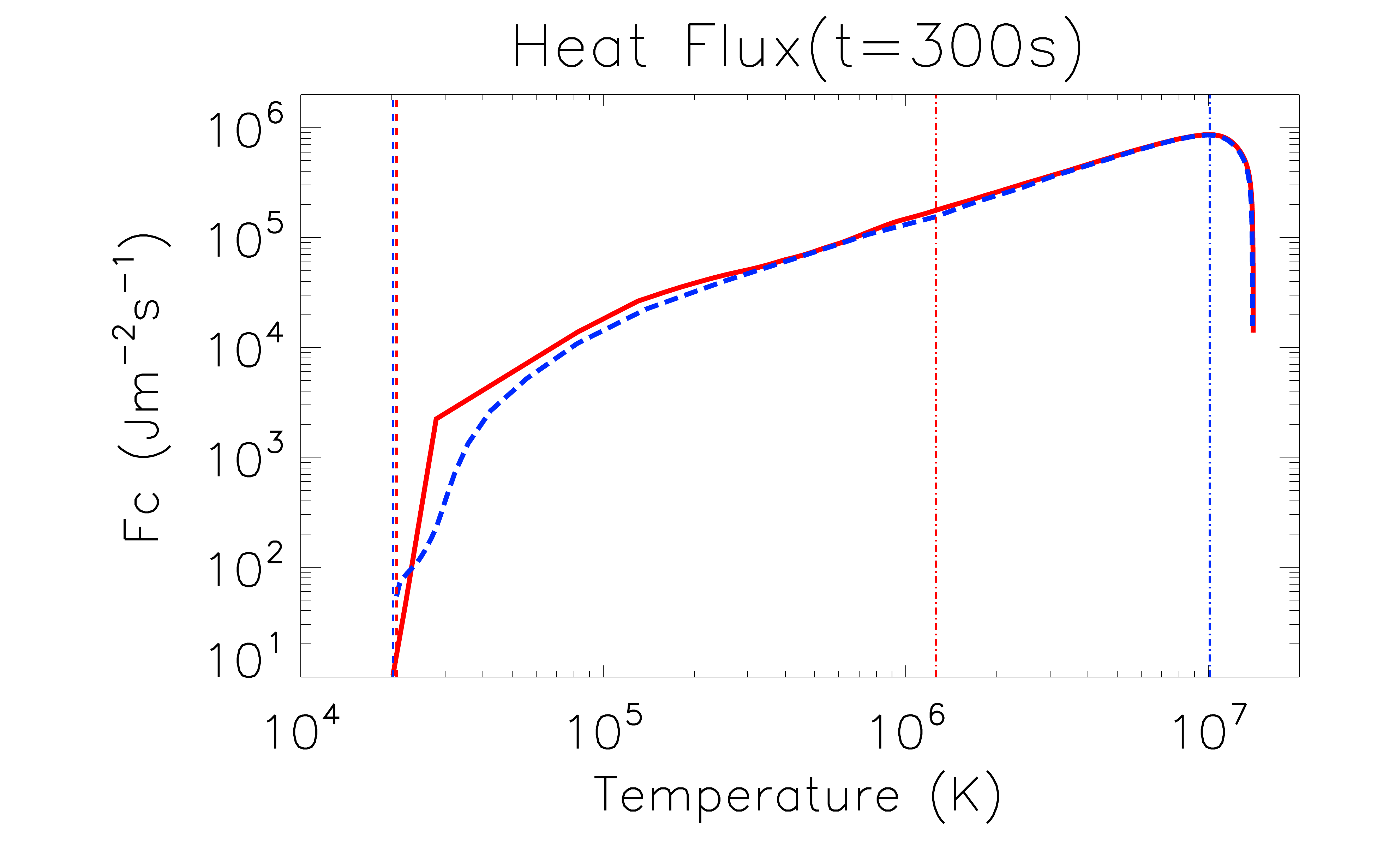}}
  \hspace*{-0.03\linewidth}
  \subfigure{\includegraphics[width=0.53\linewidth]
  {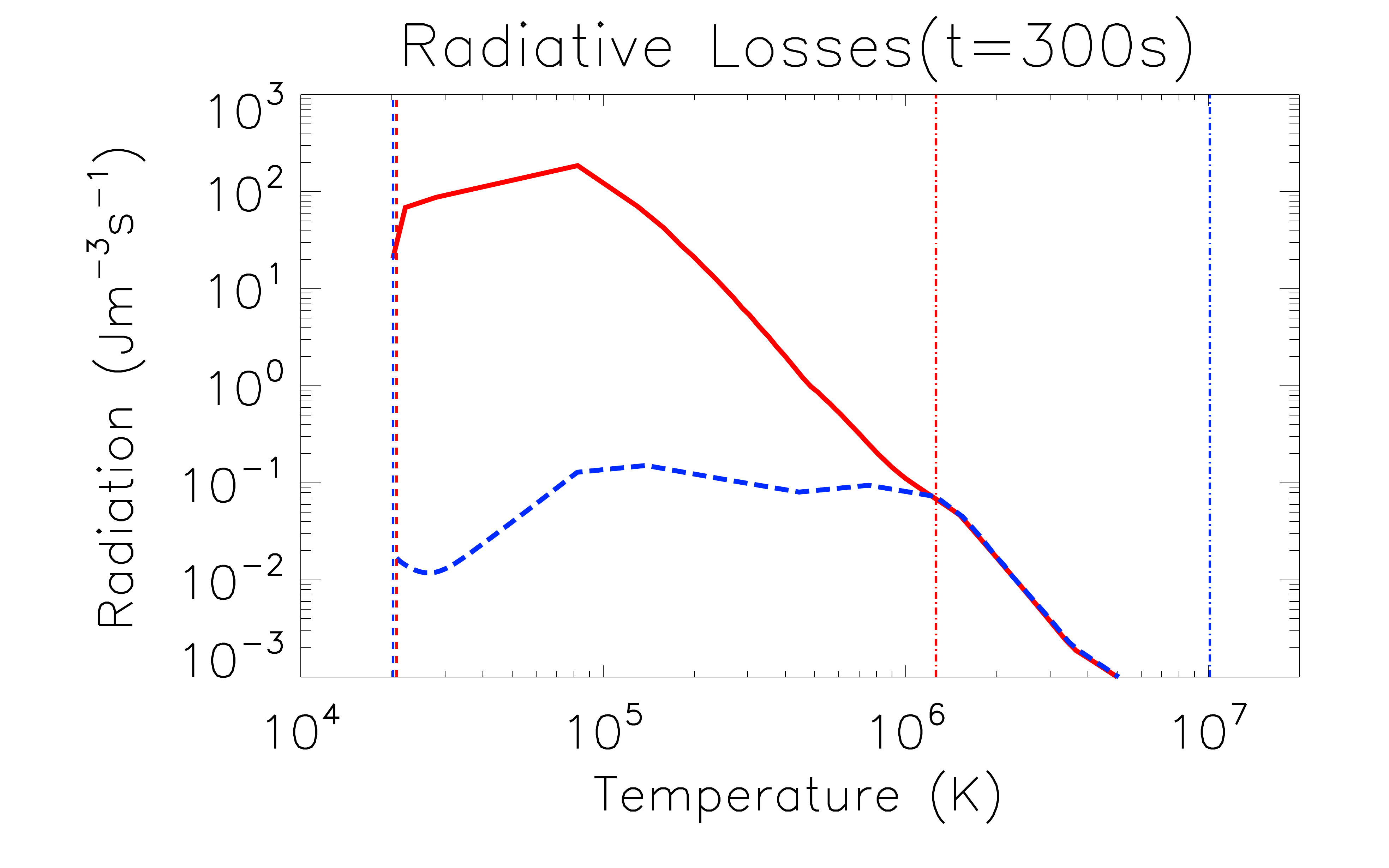}}
  \\[-6mm]
  \hspace*{-0.05\linewidth}
  \subfigure{\includegraphics[width=0.53\linewidth]
  {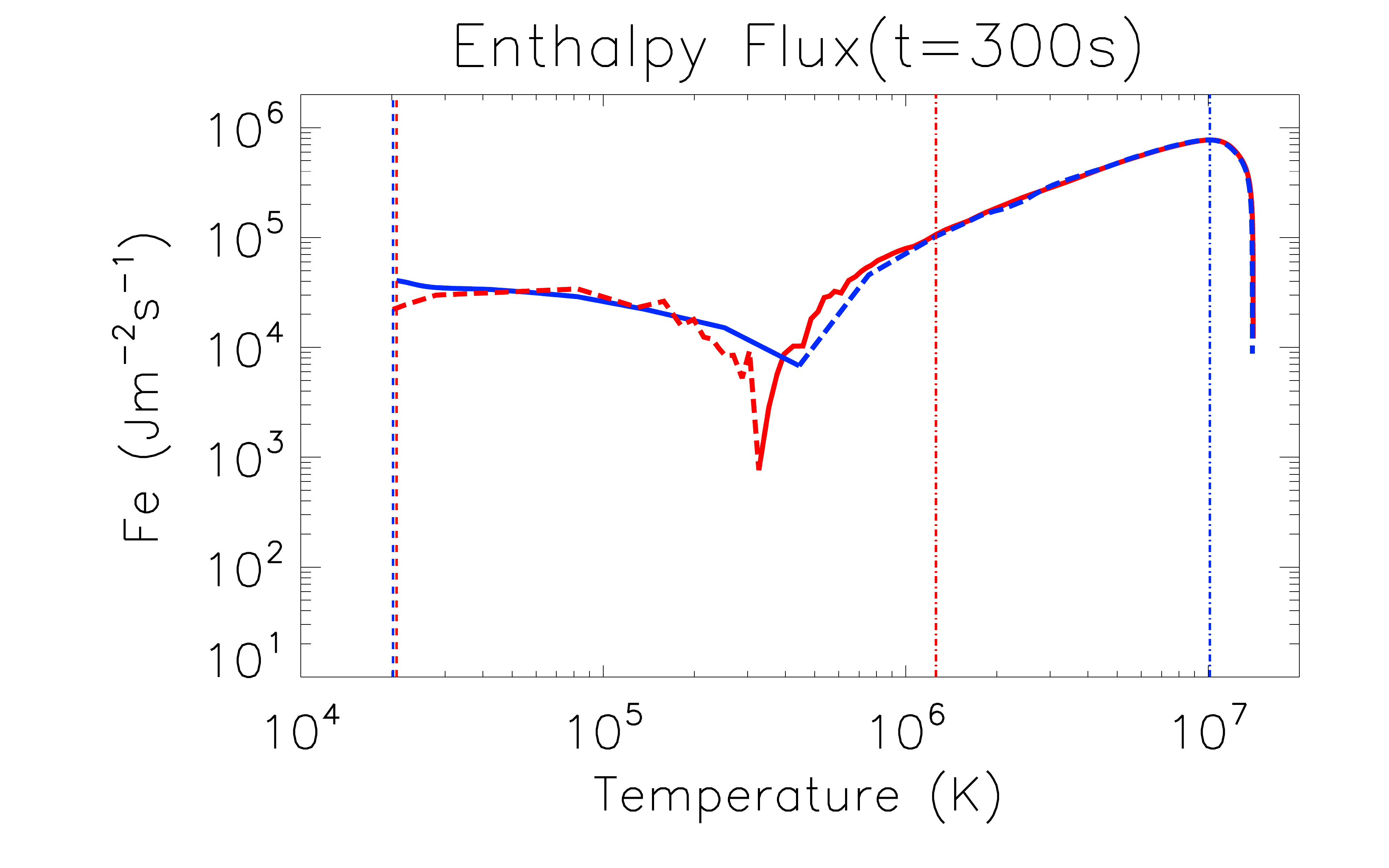}}
  \hspace*{-0.03\linewidth}
  \subfigure{\includegraphics[width=0.53\linewidth]
  {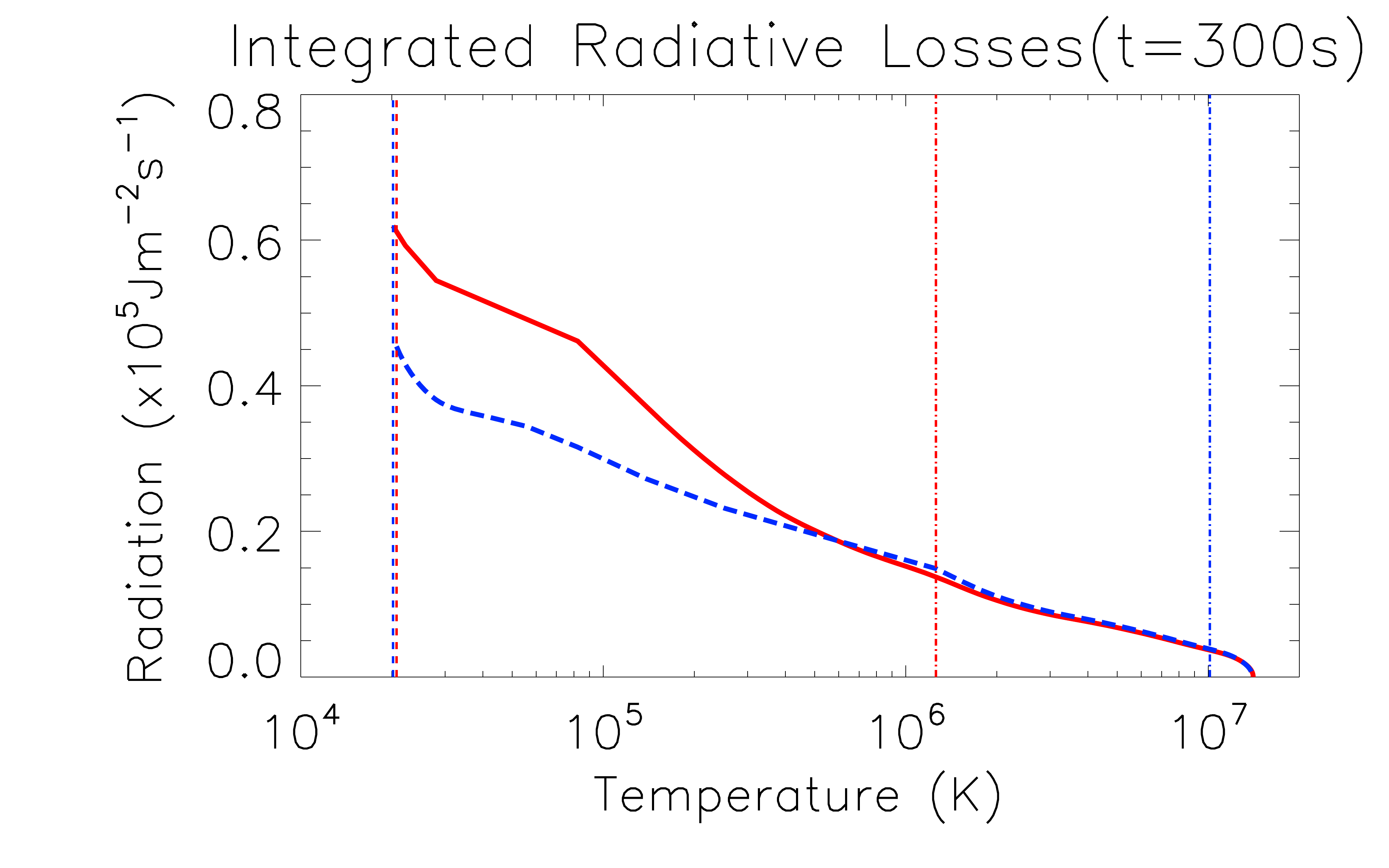}}
  \\[-9.0mm]
  \caption{Results for the 600~s heating pulse simulations
    (Section \ref{Sect:Long_pulse}).
    The upper four panels show the
    temperature, density, Mach number and velocity
    as functions of position along the loop,
    at the time of
    peak heating ($t=300$~s).
    The lower four panels show the heat flux, local
    radiative losses, enthalpy flux and
    integrated 
    radiative losses
    as functions of temperature.
    A change in line style
    indicates a change in sign of the enthalpy flux.
    The lines are colour coded in a way that reflects
    the conduction method used
    with dashed blue (solid red) representing the TRAC (SH)
    solution. 
    The dashed red (blue) vertical line
    indicates the
    base of the TRAC (SH) TR and the
    dot-dashed
    red (blue) vertical line the
    TRAC cutoff temperature, $T_c$, at the top of the 
    TRAC region
    (the SH temperature at the top of the TR).
  \label{Fig:600s_pulse_RL_5_300s_snapshots}
  }
\end{figure*}

\begin{figure*}
  \vspace*{-4.5mm}  
  \hspace*{-0.05\linewidth}
  \subfigure{\includegraphics[width=0.53\linewidth]
  {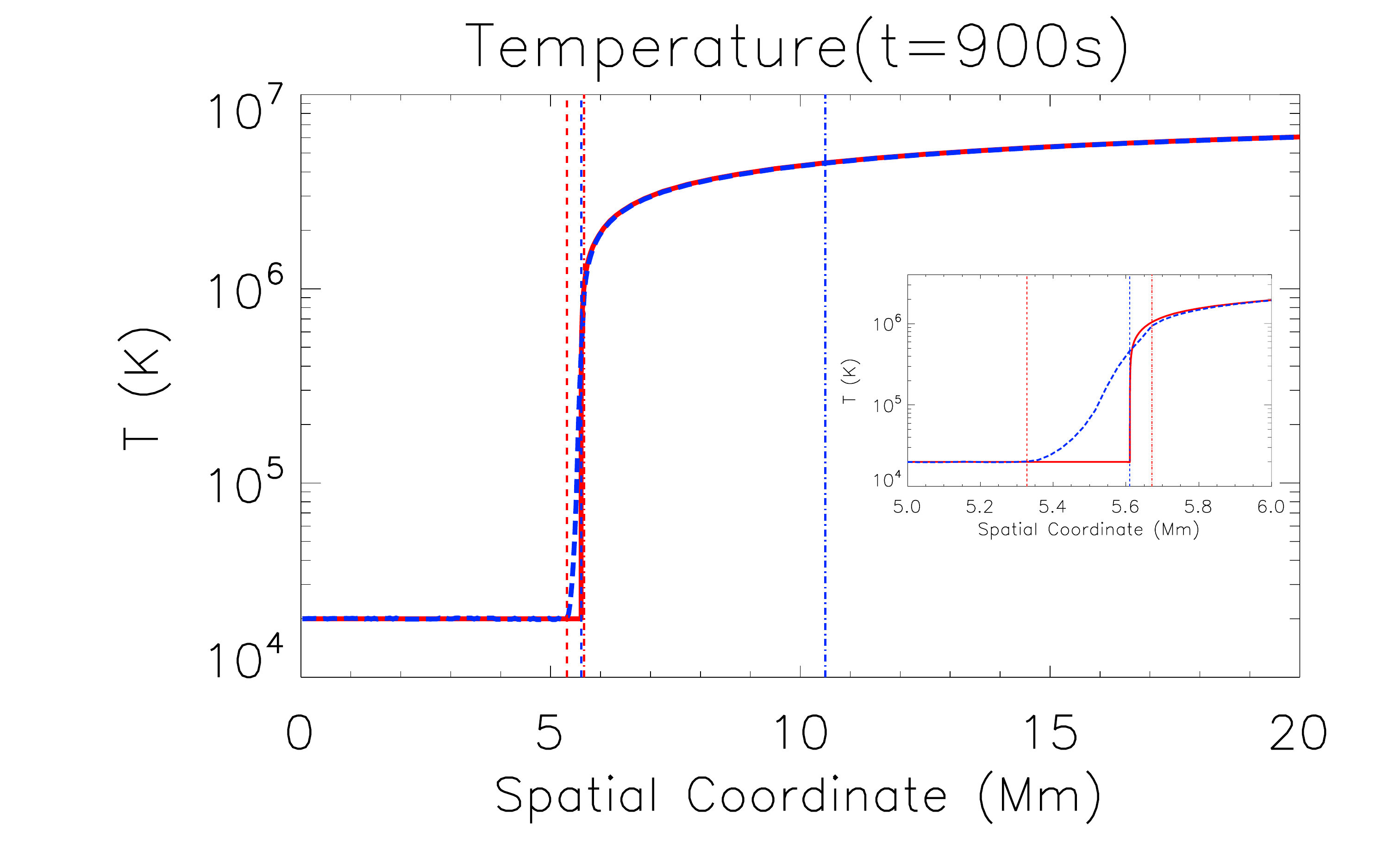}}
  \hspace*{-0.03\linewidth}
  \subfigure{\includegraphics[width=0.53\linewidth]
  {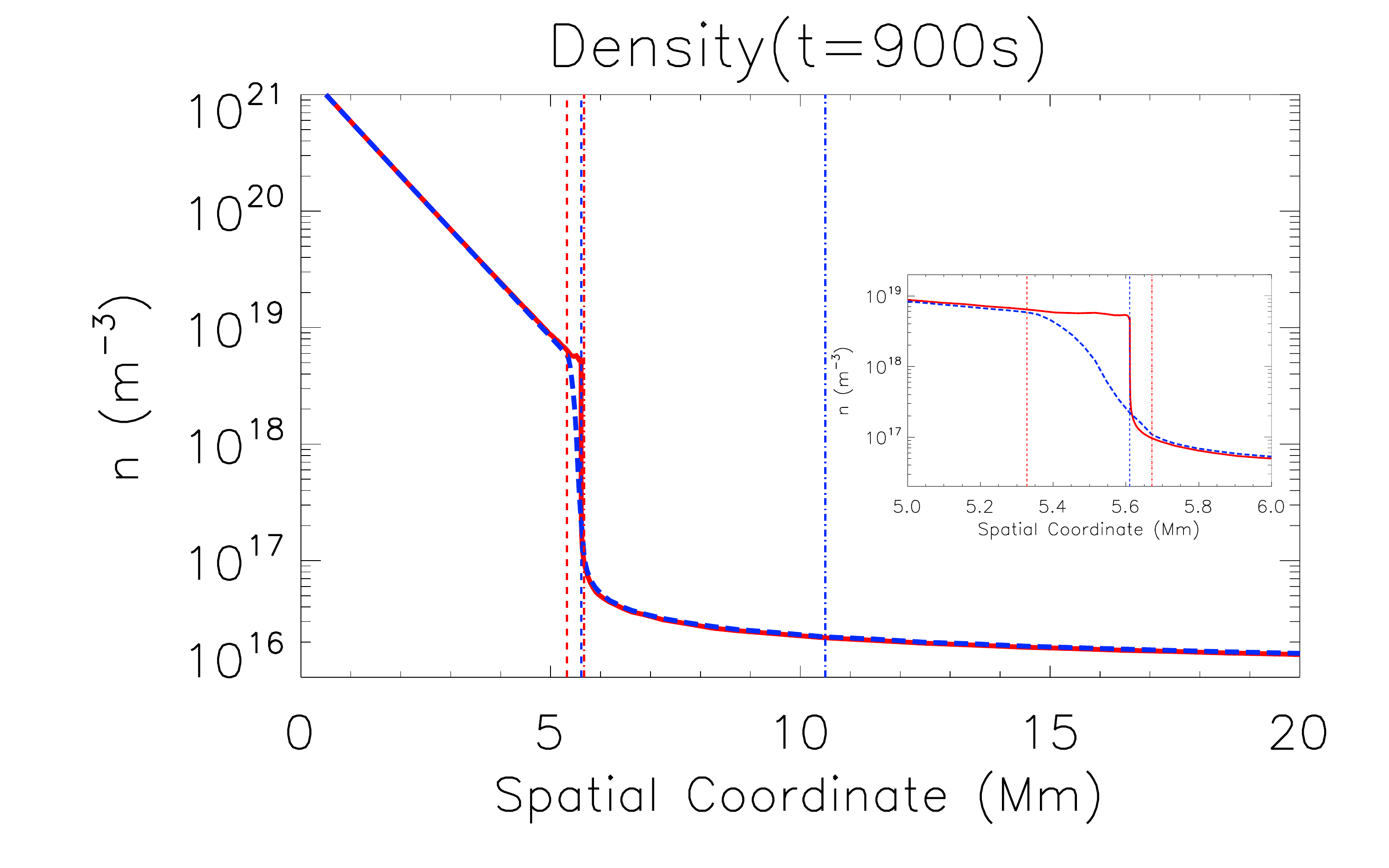}}
  \\[-6mm]
  \hspace*{-0.05\linewidth}
  \subfigure{\includegraphics[width=0.53\linewidth]
  {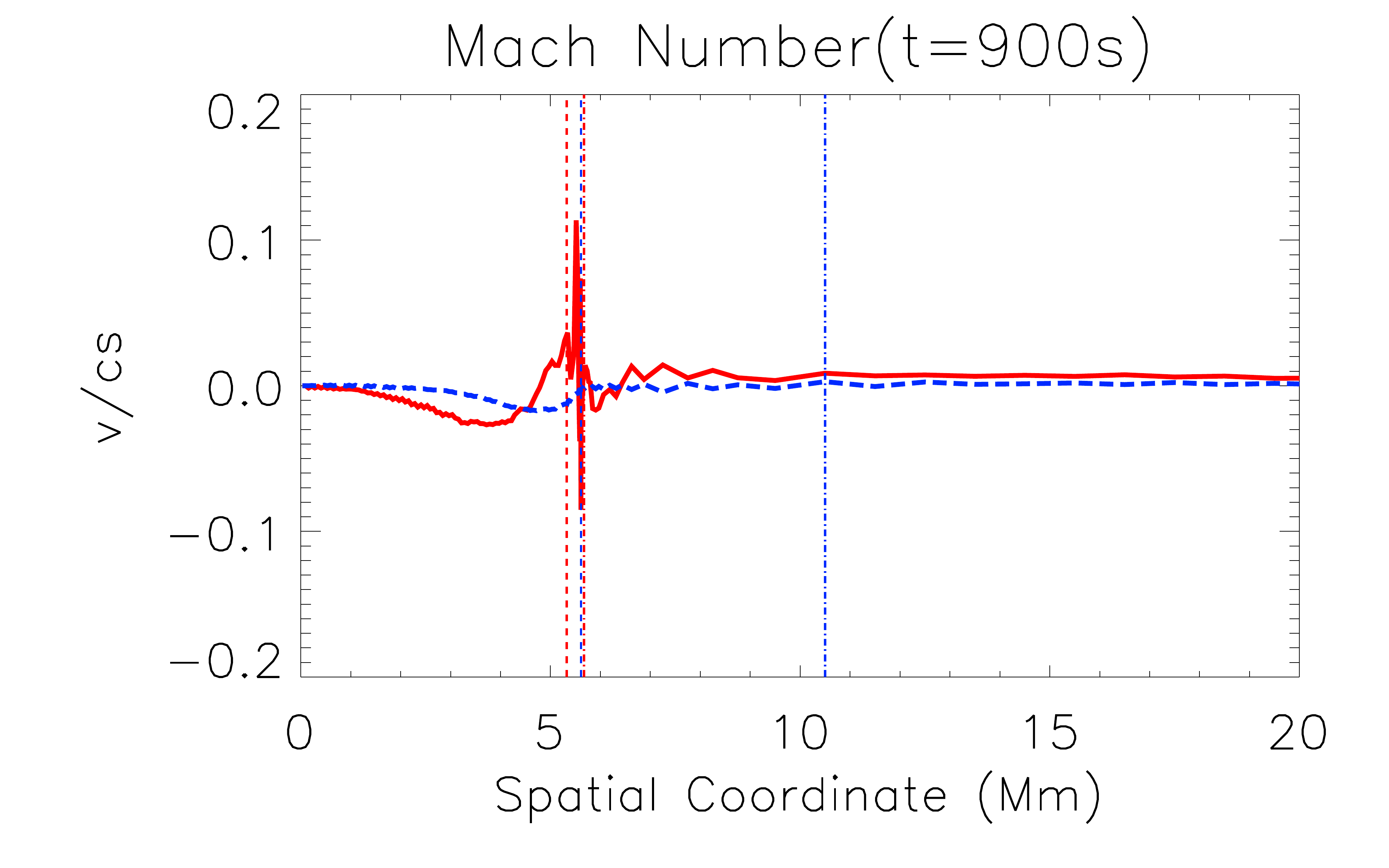}}
  \hspace*{-0.03\linewidth}
  \subfigure{\includegraphics[width=0.53\linewidth]
  {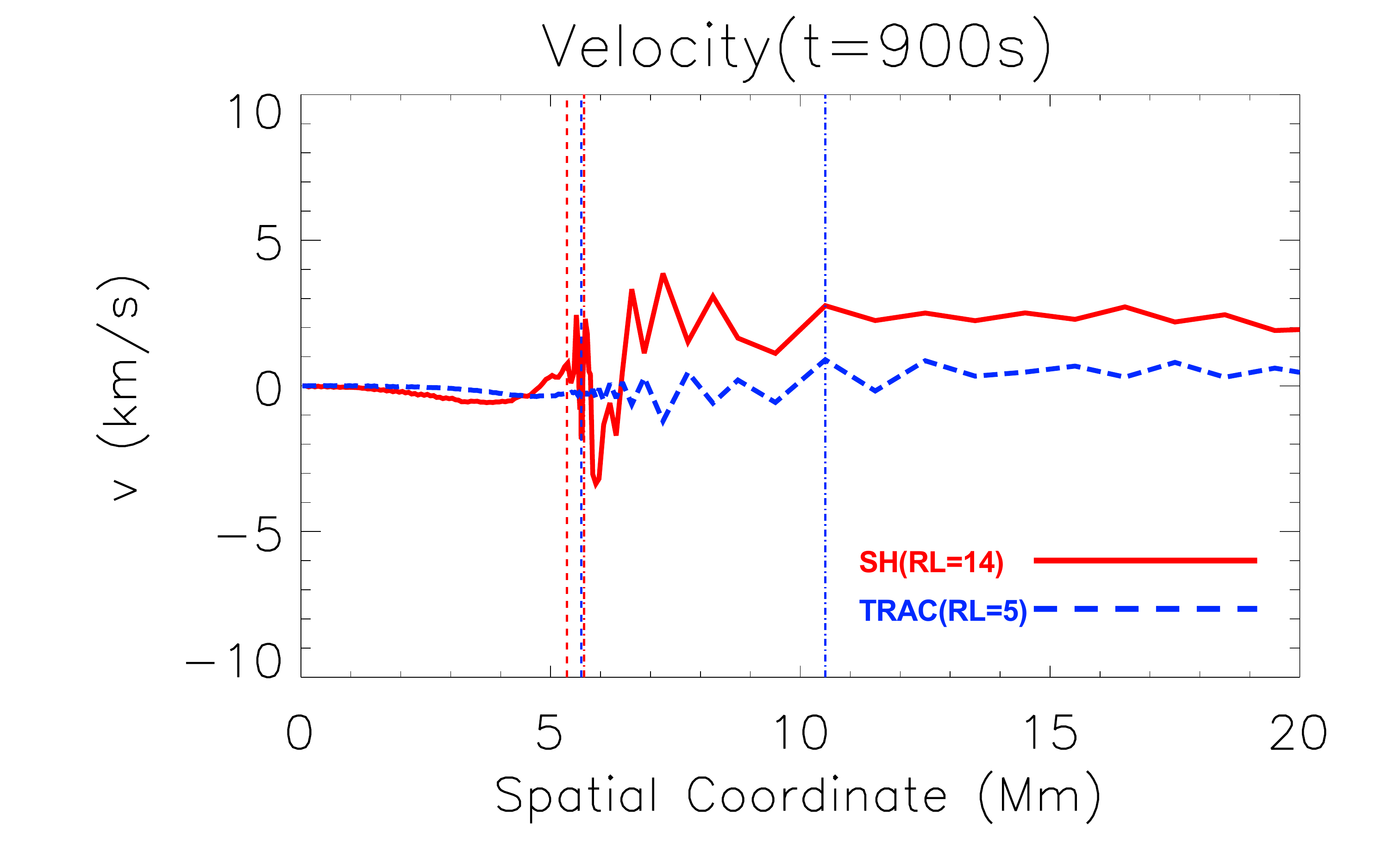}}
  \\[-6mm]
  \hspace*{-0.05\linewidth}
  \subfigure{\includegraphics[width=0.53\linewidth]
  {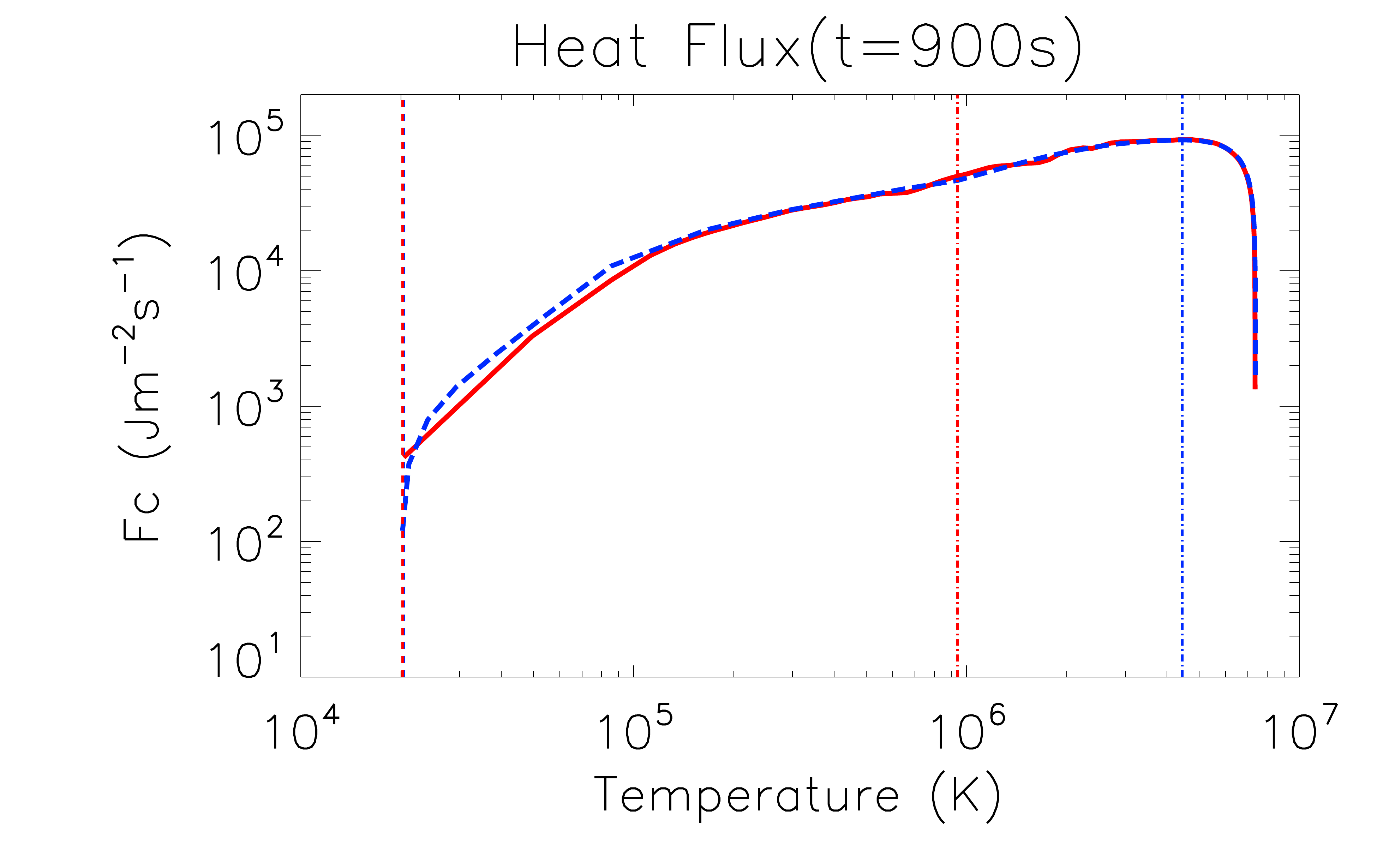}}
  \hspace*{-0.03\linewidth}
  \subfigure{\includegraphics[width=0.53\linewidth]
  {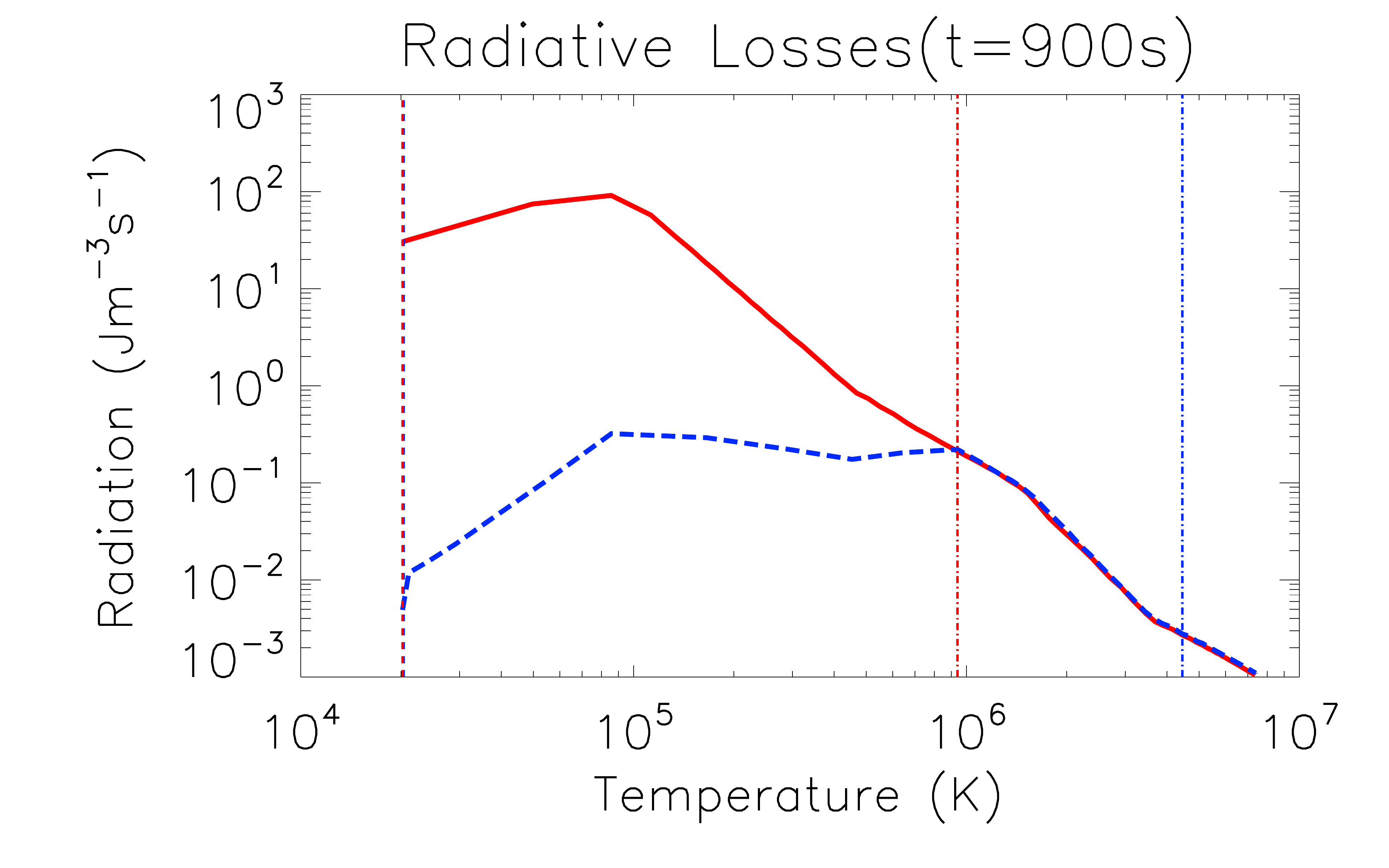}}
  \\[-6mm]
  \hspace*{-0.05\linewidth}
  \subfigure{\includegraphics[width=0.53\linewidth]
  {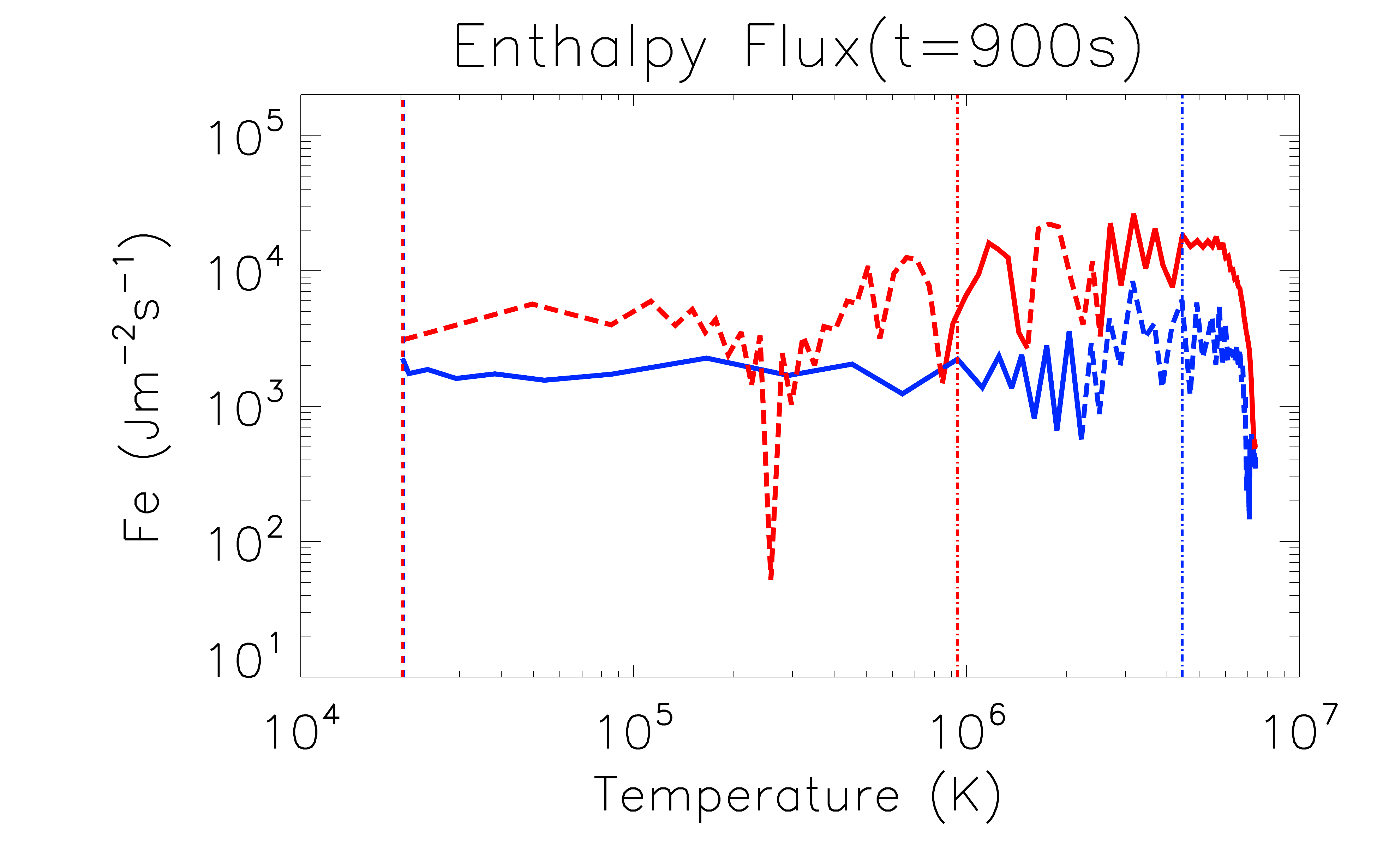}}
  \hspace*{-0.03\linewidth}
  \subfigure{\includegraphics[width=0.53\linewidth]
  {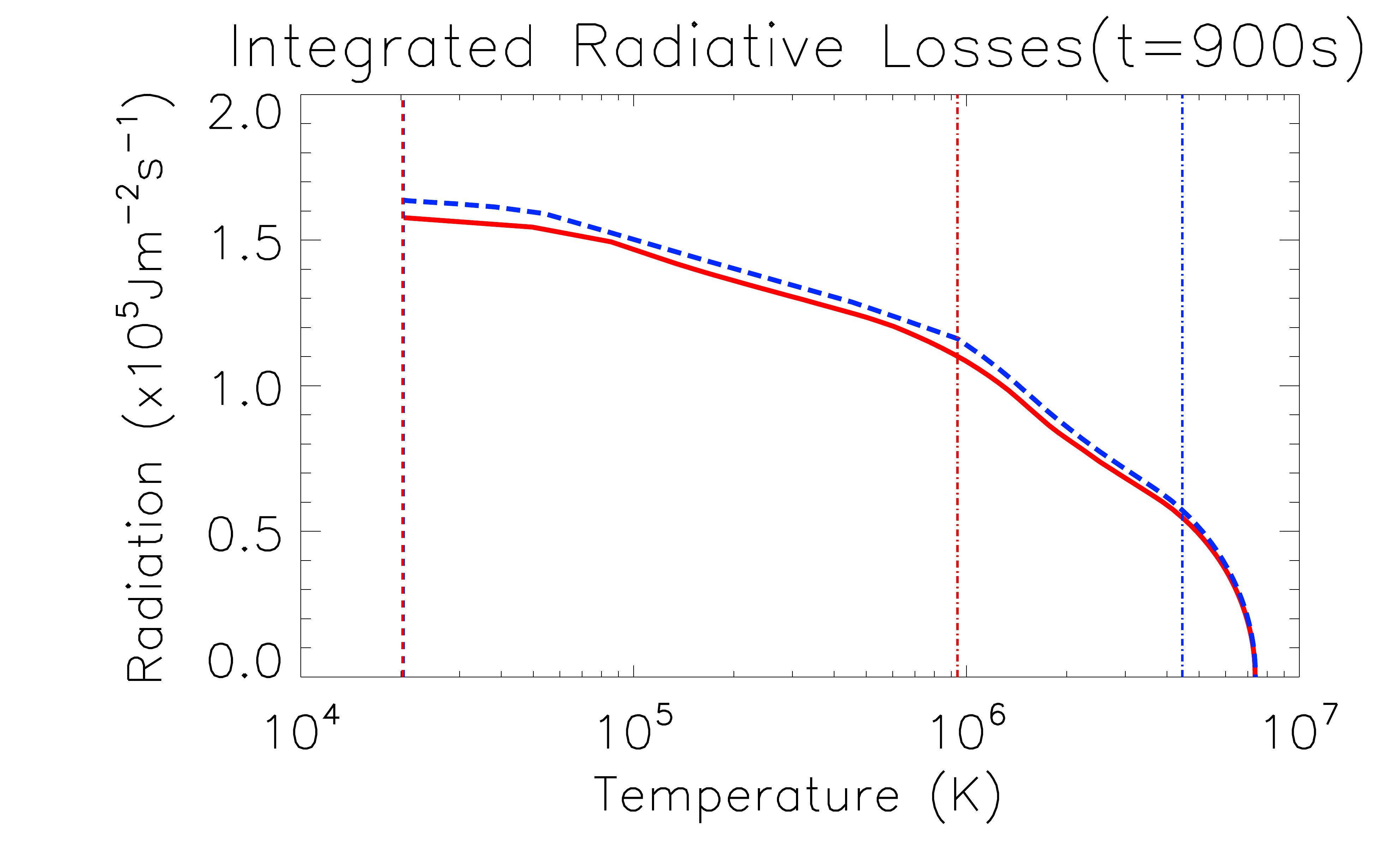}}
  \\[-8mm]
  \caption{Results for the 600~s heating pulse simulations
  (Section \ref{Sect:Long_pulse}).
  The panels show snapshots at the time of
  peak density ($t=900$~s).
  Notation is the same as that in Fig.
  \ref{Fig:600s_pulse_RL_5_300s_snapshots}.
  \label{Fig:600s_pulse_RL_5_900s_snapshots}
  }
\end{figure*}

\begin{figure*}
  \vspace*{-4.5mm}
  \hspace*{-0.05\linewidth}
  \subfigure{\includegraphics[width=0.53\linewidth]
  {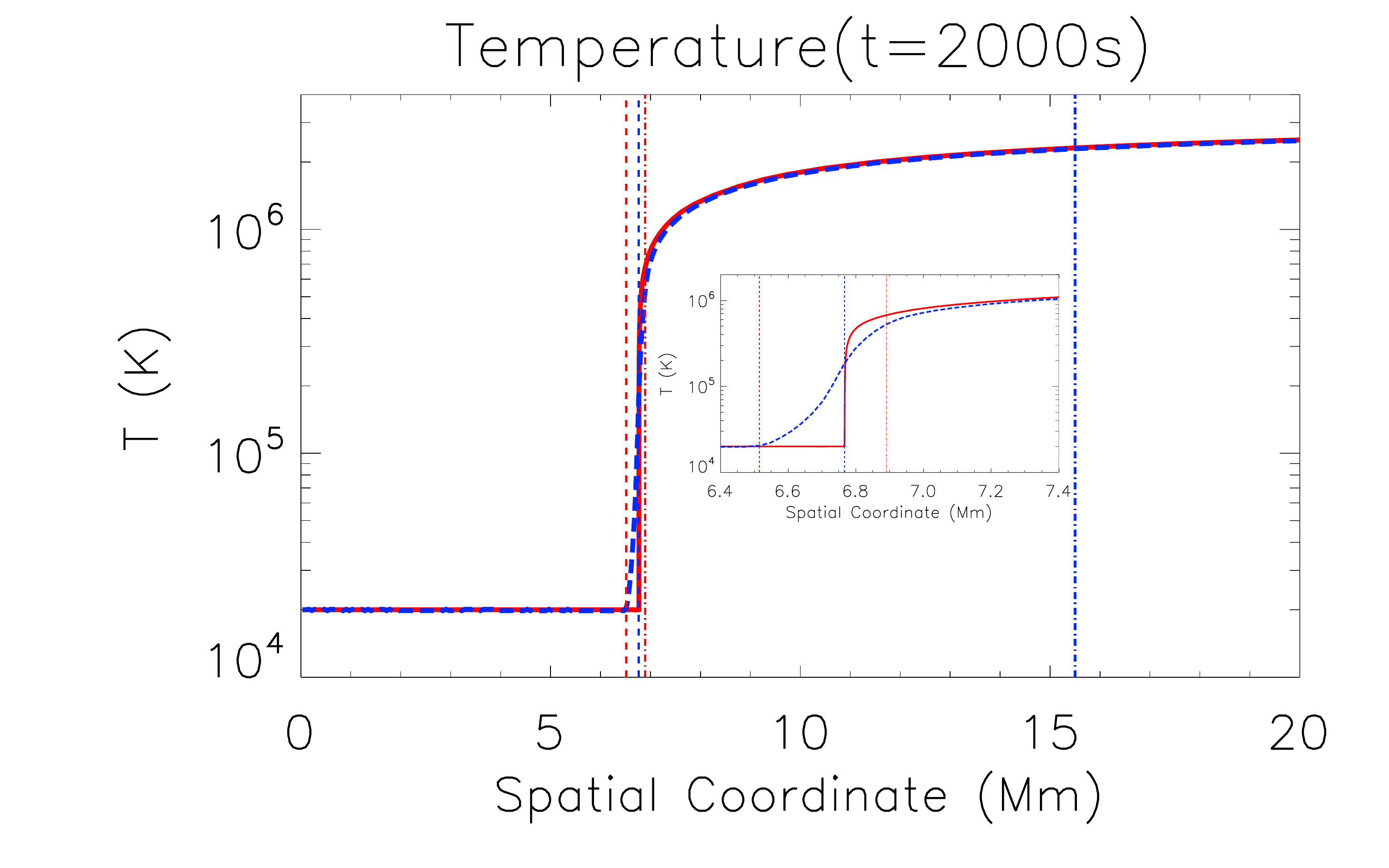}}
  \hspace*{-0.03\linewidth}
  \subfigure{\includegraphics[width=0.53\linewidth]
  {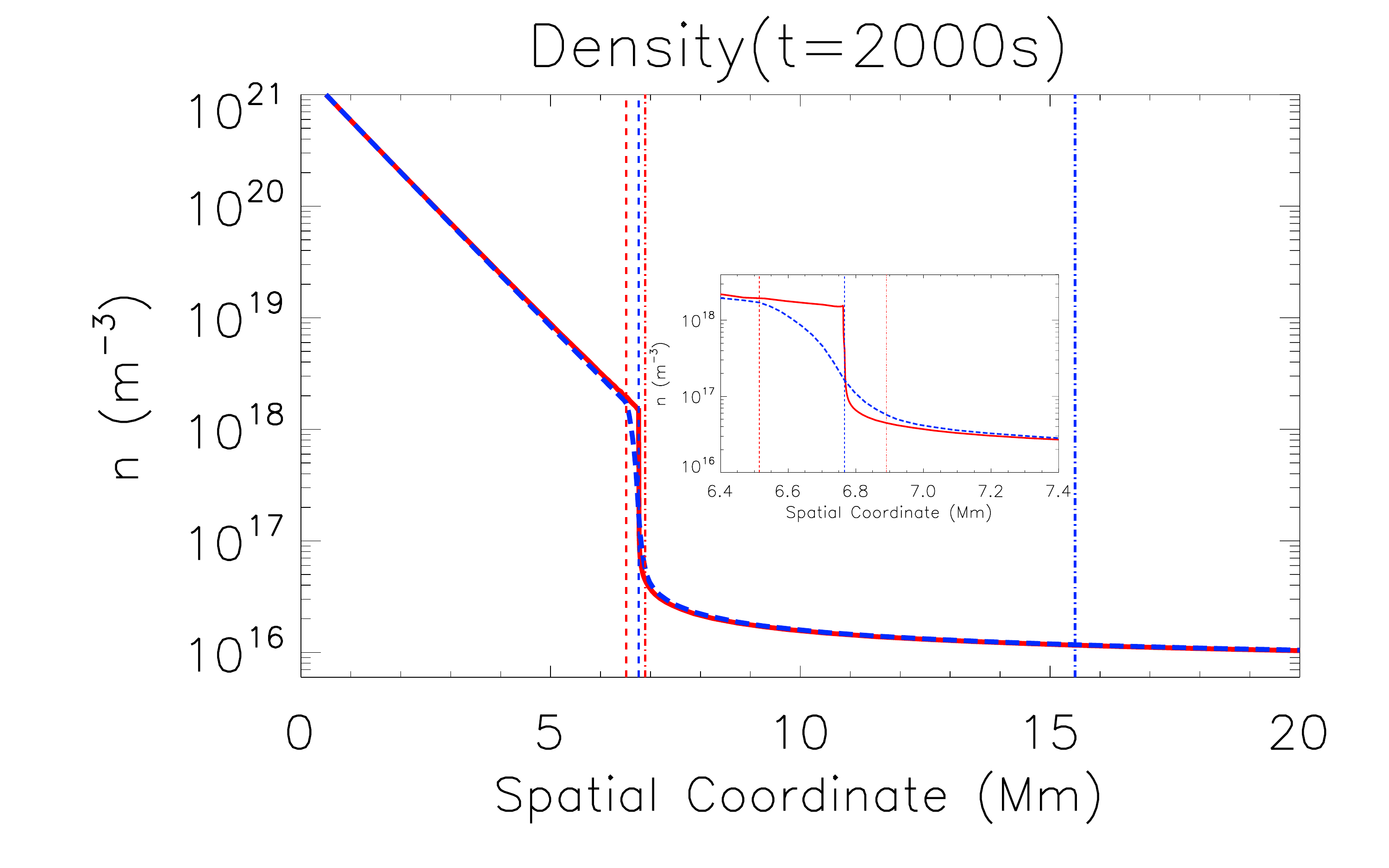}}
  \\[-6mm]
  \hspace*{-0.05\linewidth}
  \subfigure{\includegraphics[width=0.53\linewidth]
  {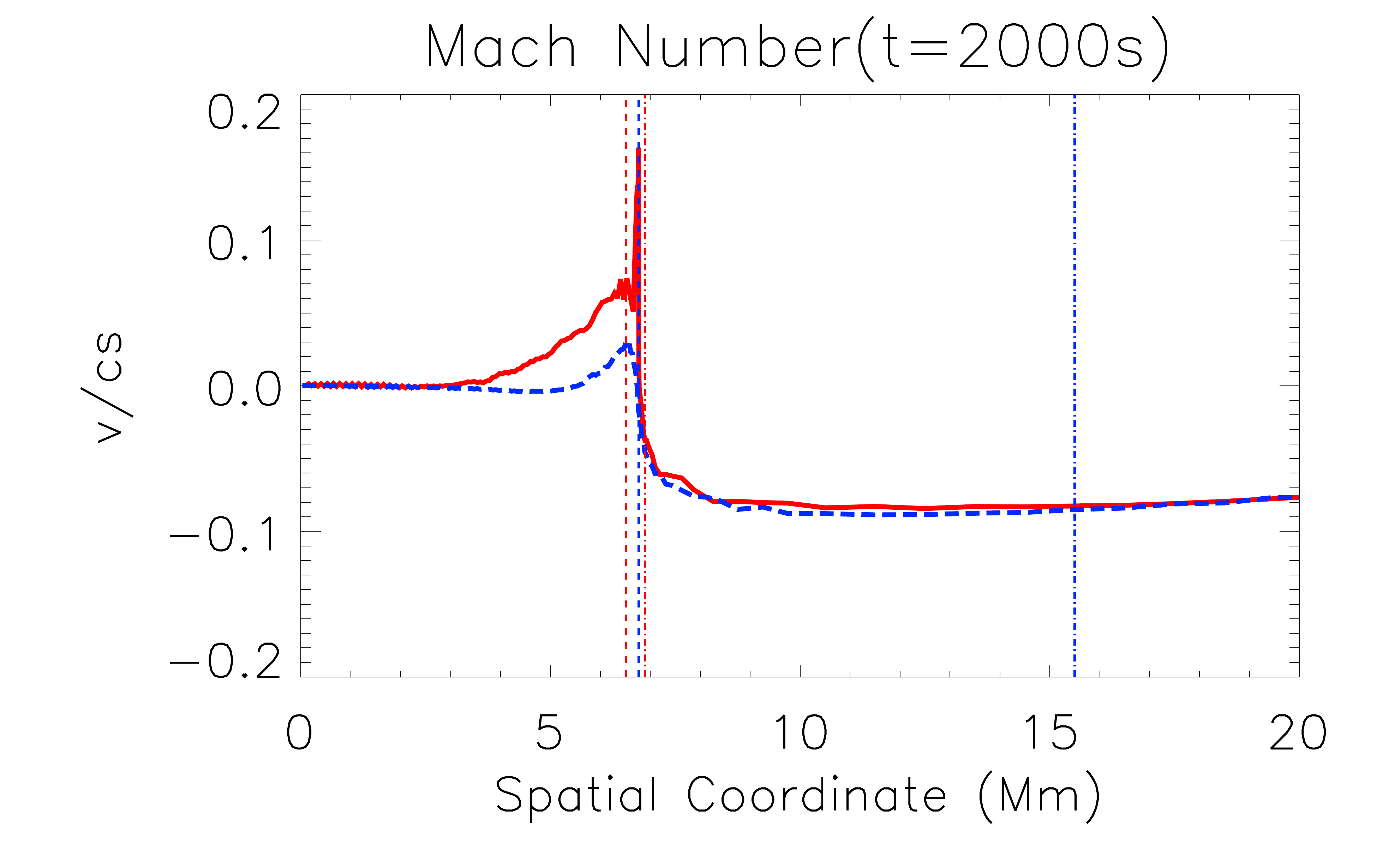}}
  \hspace*{-0.03\linewidth}
  \subfigure{\includegraphics[width=0.53\linewidth]
  {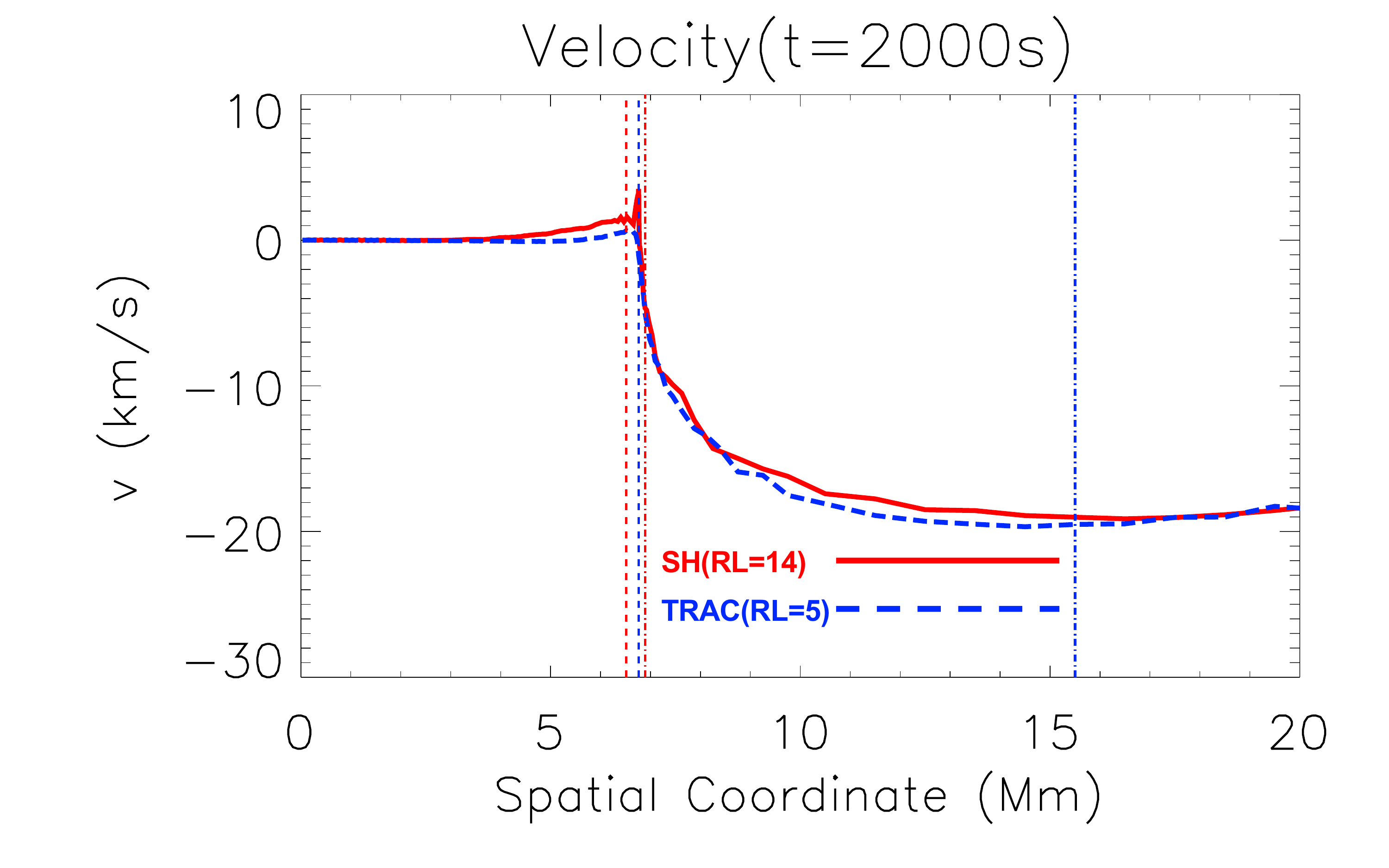}}
  \\[-6mm]
  \hspace*{-0.05\linewidth}
  \subfigure{\includegraphics[width=0.53\linewidth]
  {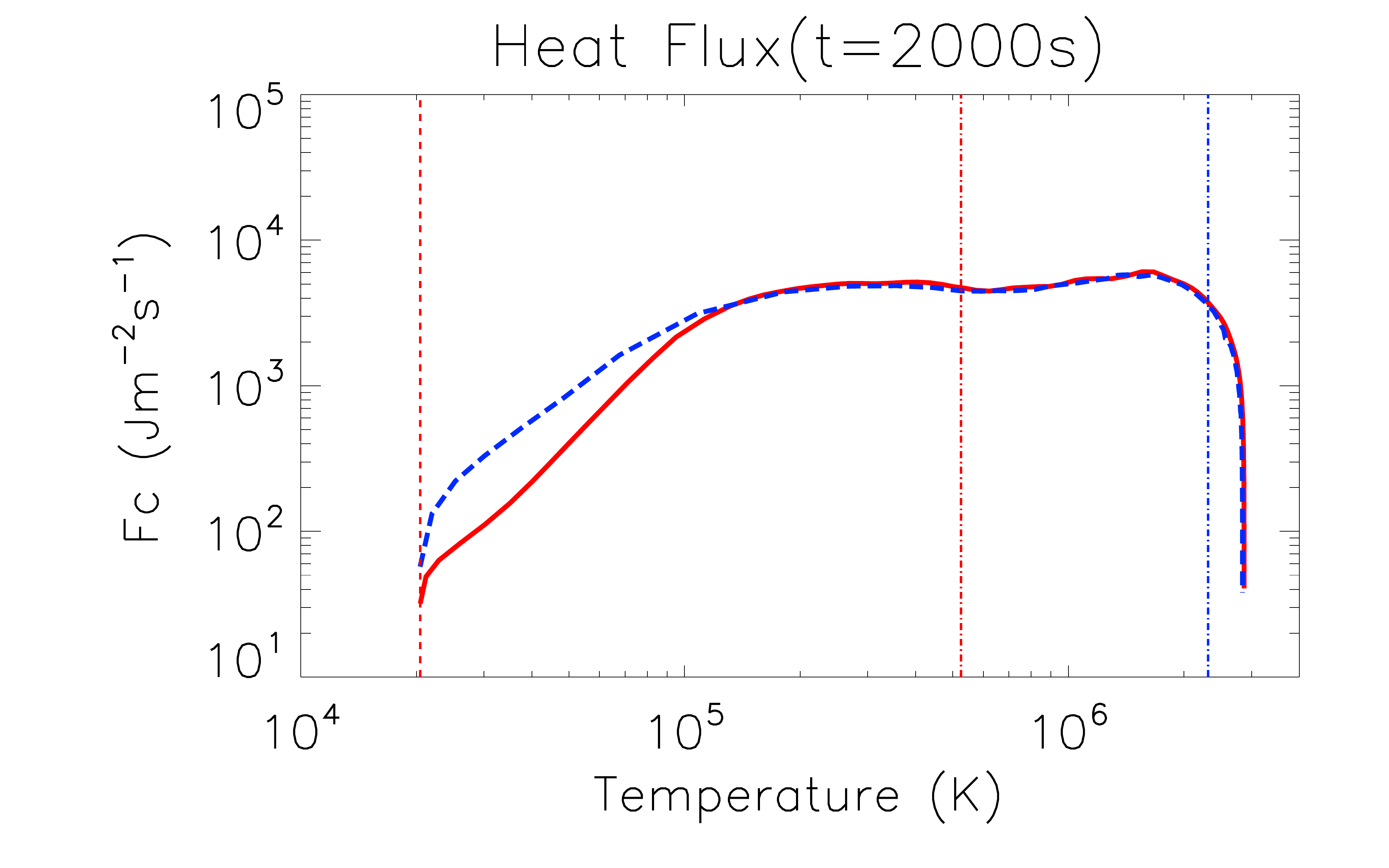}}
  \hspace*{-0.03\linewidth}
  \subfigure{\includegraphics[width=0.53\linewidth]
  {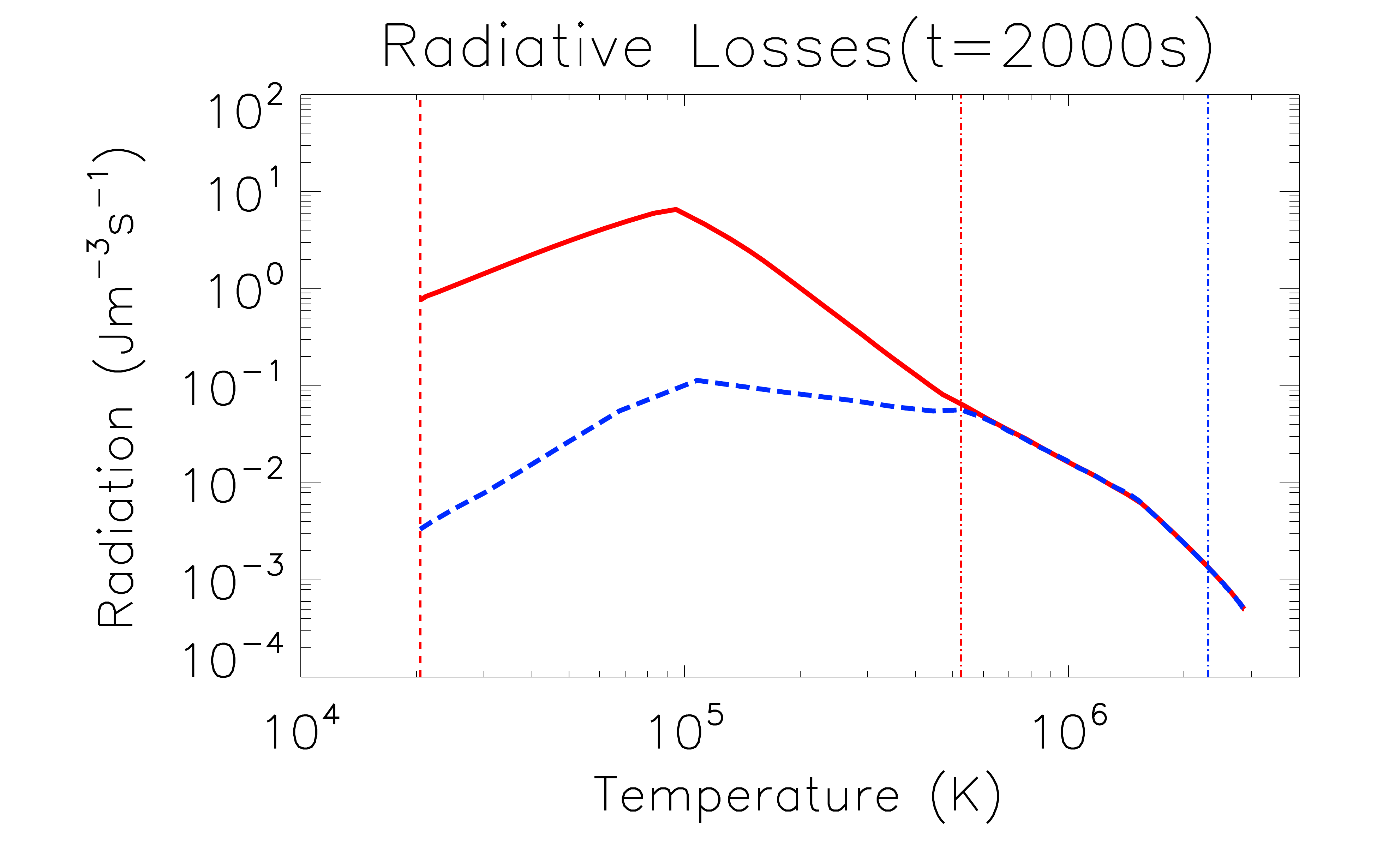}}
  \\[-6mm]
  \hspace*{-0.05\linewidth}
  \subfigure{\includegraphics[width=0.53\linewidth]
  {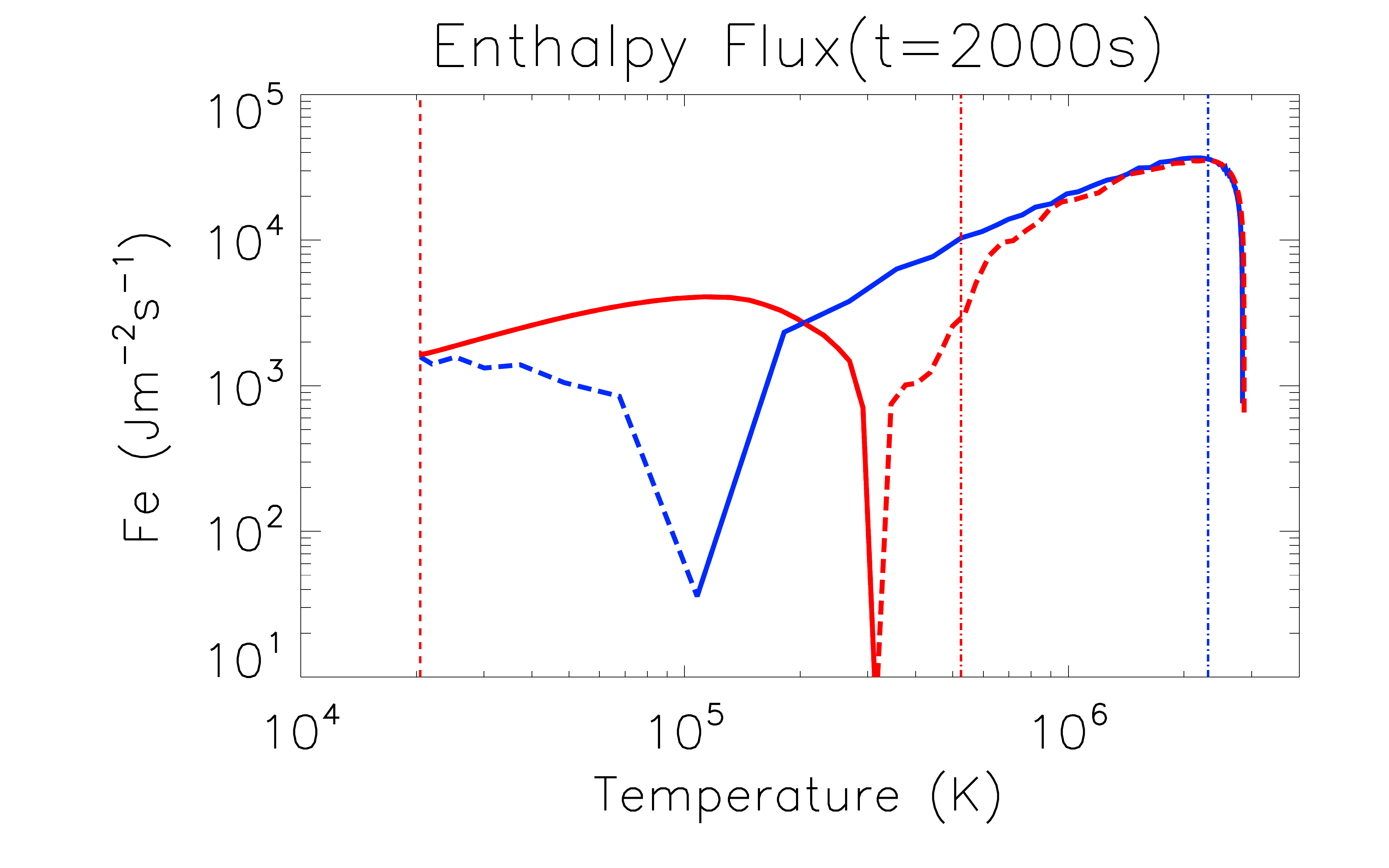}}
  \hspace*{-0.03\linewidth}
  \subfigure{\includegraphics[width=0.53\linewidth]
  {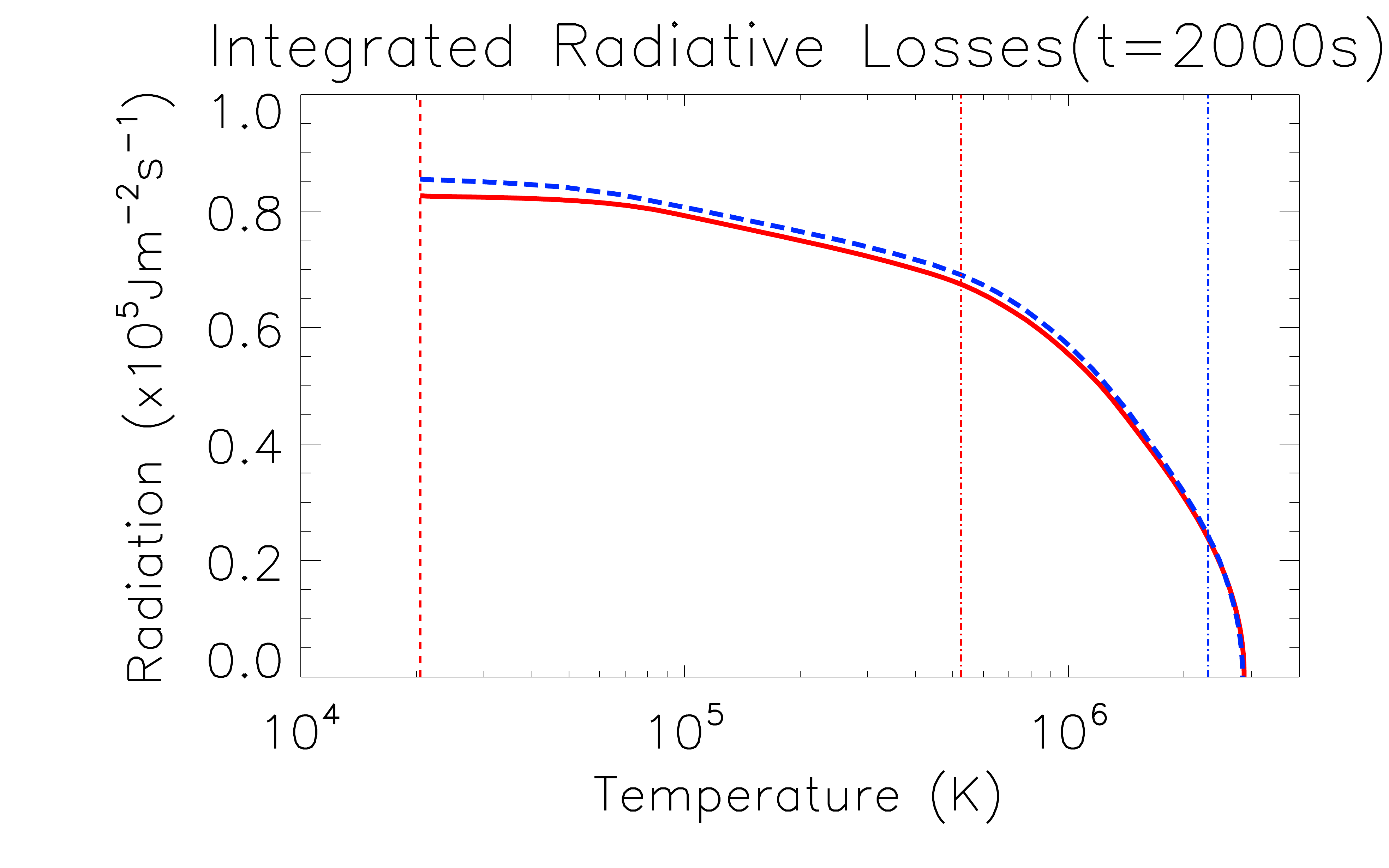}}
  \\[-8mm]
  \caption{Results for the 600~s heating pulse simulations
  (Section \ref{Sect:Long_pulse}).
  The panels show snapshots during the loop's decay
  phase ($t=2000$~s).
  Notation is the same as that in Fig.
  \ref{Fig:600s_pulse_RL_5_300s_snapshots}.
  \label{Fig:600s_pulse_RL_5_2000s_snapshots}
  }
\end{figure*}

  \indent
  Figs. \ref{Fig:600s_pulse_RL_5_300s_snapshots} - 
  \ref{Fig:600s_pulse_RL_5_2000s_snapshots}
  show 
  the evolution of a number of variables as a function
  of position along the loop, 
  in response to the 600~s heating pulse
  considered in Section \ref{Sect:Long_pulse},
  through a series of snapshots
  at three different 
  times: $t = 300$~s 
  (Fig. \ref{Fig:600s_pulse_RL_5_300s_snapshots}), 
  $t = 900$~s (Fig. \ref{Fig:600s_pulse_RL_5_900s_snapshots})
  and 
  $t = 2000$~s 
  (Fig. \ref{Fig:600s_pulse_RL_5_2000s_snapshots}). 
  These correspond to the time of maximum 
  heating, maximum density, and during the loop's draining 
  phase. 
  Therefore, the snapshots are representative
  of the three main phases for which it is valid to compare
  the global evolution of the loop with the 
  analytical predictions presented in Section 
  \ref{Sect:TRAC_analytical_assessment}.
  \\
  \indent
  In the upper four panels of these figures
  we focus on 20~Mm of the loop 
  around the TR, 
  showing the temperature, 
  density, local Mach number and velocity respectively. 
  An enlargement about the TRAC region is also
  shown inset on the temperature and density panels.
  In the 
  lower four panels, we show the heat flux, local radiative 
  loss, enthalpy flux and integrated radiative losses as a 
  function of temperature. The integrated losses are defined as 
  being from the top of the loop downwards
  to the base of the TR region and are shown on a 
  linear scale. In these panels, the red and blue lines are the 
  SH and TRAC solutions respectively. For SH (TRAC) solid 
  (dashed) lines indicate positive quantities and dashed 
  (solid)
  negative. Starting from the left, 
  the first 
  dashed red (blue) vertical line at the base of 
  the TR shows its location for TRAC (SH), and the next
  dot-dashed 
  red 
  line the top of the TRAC region ($T_c$). 
  The rightmost 
  vertical dot-dashed blue line is the top of the actual TR, 
  defined by 
  where the sum of the 
  downward conduction and enthalpy flux changes from 
  a loss to a gain 
  \citep[e.g.][]{paper:Veseckyetal1979,
  paper:Klimchuketal2008}. 
  \\
  \indent
  First we examine the temperature and density structure in the 
  lower TR region. The upper 
  two panels (row 1) in
  Figs. \ref{Fig:600s_pulse_RL_5_300s_snapshots} - 
  \ref{Fig:600s_pulse_RL_5_2000s_snapshots} 
  show the extent of 
  the TR broadening 
  that is associated with the TRAC method.
  As noted in Section
  \ref{Sect:Long_pulse_Tc_l}, the extent of the TRAC region 
  diminishes as the loop evolves, and this region extends the 
  TR both below and above the SH location, as was also shown 
  for a static loop by L09.
  \\
  \indent 
  The velocity and Mach number 
  results are also of interest. At
  $t = 300$~s, while the 
  mass flux out of the TRAC region is the same, the location of 
  maximum velocity is displaced upwards because of the
  effect the TR 
  broadening has on the density profile in this region. 
  The Mach number indicates that the peak 
  velocity is a significant fraction of the local sound speed 
  and, hence at this time,
  the subsonic approximation is only marginally valid. 
  This is the reason for the discrepancy 
  between the solutions presented in 
  Section \ref{Sect:TRAC_analytical_assessment} and the
  simulations as discussed shortly.
  \\
  \indent
  Again, 
  this is indicative that while the details of the plasma as a 
  function of temperature and density differ between 
  the SH and 
  TRAC models, 
  the result is to maintain good agreement between the 
  two in the corona.
  This can also be seen in the panel that shows the 
  local
  radiative losses as a function of temperature. For example, 
  the coronal properties of the two models converge 
  despite the differences in the local radiative losses
  below $T_c$.
  \\
  \indent
  The heat flux plots in 
  Figs. \ref{Fig:600s_pulse_RL_5_300s_snapshots} - 
  \ref{Fig:600s_pulse_RL_5_2000s_snapshots} 
  show that the top of the 
  TR is at $10.1$, $4.45$ and $2.31$~MK, 
  corresponding to
  roughly 75\%, 60\% and 80\% of the maximum loop temperature, 
  consistent with the detailed expectations of
  \cite{paper:Cargilletal2012a}.  
  In all cases $T_c$, as shown in Fig.
  \ref{Fig:600s_pulse_RL_5_time_evolution}g,
  is significantly 
  smaller than the temperature at the top of the TR, with 
  $T_c$ of order 1.25~MK at peak heating, declining to 0.94~MK 
  at peak density and 0.53~MK at 2000~s, so that $T_c$ just 
  follows the coronal average
  (but limited to 20\% of the maximum temperature). 
  The thickness of the TRAC region 
  is thus a small fraction of the TR thickness. 
  Furthermore, the SH 
  and TRAC temperature and density profiles converge a short 
  distance above the top of the TRAC region (see the upper two 
  panels (row 1) of
  Figs. \ref{Fig:600s_pulse_RL_5_300s_snapshots} - 
  \ref{Fig:600s_pulse_RL_5_2000s_snapshots} ), 
  significantly below the top of the TR,
  again suggestive of the limited 
  influence of the TRAC method on the corona.
  \\
  \indent
  The lower right plot of
  Figs. \ref{Fig:600s_pulse_RL_5_300s_snapshots} - 
  \ref{Fig:600s_pulse_RL_5_2000s_snapshots}  
  show that at $t = 900$ 
  and 2000~s, 
  the integrated radiative losses show good agreement 
  between the TRAC and SH methods, as predicted in 
  Section \ref{Sect:TRAC_analytical_assessment}. 
  However, the 
  agreement is less satisfactory at 
  $t = 300$~s. 
  While this is of 
  little consequence for the coronal behaviour since the 
  radiative losses are a relatively small fraction of the total
  energy budget at that time, 
  it is of interest to understand why this 
  occurs. 
  Analysis of the output shows that the subsonic 
  assumption is violated, as is the steady
  mass flux, implying a dynamic TRAC region. 
  The outcome is that the
  TRAC region
  integrated radiative losses in the simulation deviate
  from the analytical expression given in
  Eq. \eqref{Eqn:R_TRAC_strong_evaporation},
  during this short period.
  Further, the smaller integrated 
  TRAC losses are responsible for the slightly larger coronal 
  density at that time, as discussed above in Section
  \ref{Sect:Long_pulse_RL}.
  %
  %
  %
  %
  %%%%%%%%%%%%%%%%%%%%%%%%%%%%%%%%%%%%%%%%%%%%%%%%%%%%%%%%%%%%%    
  %
  % Fig:600s_pulse_RL_5_300s_snapshots
  %
  %
  %
  %
  %
  %%%%%%%%%%%%%%%%%%%%%%%%%%%%%%%%%%%%%%%%%%%%%%%%%%%%%%%%%%%%%    
  %
  % Fig:600s_pulse_RL_5_900s_snapshots
  %
  %
  %
  %
  %
  %%%%%%%%%%%%%%%%%%%%%%%%%%%%%%%%%%%%%%%%%%%%%%%%%%%%%%%%%%%%%    
  %
  % Fig:600s_pulse_RL_5_2000s_snapshots
  %
  \end{appendix}
\end{document}